\documentclass{aa}  
\usepackage{graphicx}

\usepackage{txfonts}
\usepackage{tabularx}
\usepackage{amsfonts}
\usepackage{bbold}
\usepackage{color}
\usepackage{transparent}
\usepackage{hyperref}
\usepackage{transparent}
\usepackage{rotating}
\usepackage{caption}
\usepackage{subcaption}

\usepackage{graphicx}	
\usepackage{amsmath}	
\usepackage{amssymb}	
\usepackage{tablefootnote}
\usepackage[flushleft]{threeparttable}
\usepackage{dblfloatfix} 
\usepackage{siunitx}
\usepackage{array}

\usepackage{physics}

\usepackage{multirow}
\usepackage{diagbox}
\usepackage[normalem]{ulem}

\usepackage{natbib}
\bibpunct{(}{)}{;}{a}{}{,} 

\usepackage{empheq}

\newcommand*{\hmxb}{HMXB\@\xspace}
\newcommand*{\hmxbs}{HMXBs\@\xspace}

\newcommand*{\bh}{BH\@\xspace}
\newcommand*{\bhs}{BHs\@\xspace}
\newcommand*{\eg}{e.g.,\@\xspace}
\newcommand*{\ie}{i.e.,\@\xspace}

\newcommand*\diff{\mathop{}\!\mathrm{d}}

\newcommand*{\msun}{$M_{\odot}$\@\xspace}

\newcommand*{\cyg}{\mbox{Cyg X-1}\@\xspace}
\newcommand*{\sgr}{\mbox{Sgr A$^*$}\@\xspace}
\newcommand{\m}[1]{M#1$^*$\@\xspace} 
\newcommand*{\gravity}{GRAVITY\@\xspace}

\newcommand*{\agns}{AGNs\@\xspace}
\newcommand*{\smbh}{SMBH\@\xspace}
\newcommand*{\smbhs}{SMBHs\@\xspace}
\newcommand*{\isco}{\text{ISCO}\@\xspace}
\newcommand*{\iscoo}{\text{ISCO}}
\newcommand*{\fido}{FIDO\@\xspace}
\newcommand*{\fidos}{FIDOs\@\xspace}
\newcommand*{\pic}{PIC\@\xspace}

\newcommand*{\ileyk}{\textcolor[rgb]{0,0,0}}
\newcommand*{\benoit}{\textcolor[rgb]{0,0,0}}
\newcommand*{\benjamin}{\textcolor[rgb]{0,0,0}}
\newcommand*{\kyle}{\textcolor[rgb]{0,0,0}} 

\usepackage[dvipsnames]{xcolor}

\makeatletter
\newcommand{\thickhline}{%
    \noalign {\ifnum 0=`}\fi \hrule height 1pt
    \futurelet \reserved@a \@xhline
}
\makeatother

\begin{document}


    \title{Spinning black holes magnetically connected to a Keplerian disk}
    \subtitle{Magnetosphere, reconnection sheet, particle acceleration \& coronal heating}
    
   \author{I. El Mellah
          \inst{1}
          \and
          B. Cerutti
          \inst{1}
         \and
          B. Crinquand
          \inst{1}          
          \and
          K. Parfrey
          \inst{2}   
          }

   \institute{Univ. Grenoble Alpes, CNRS, IPAG, 38000 Grenoble, France\\
              \email{ileyk.elmellah@univ-grenoble-alpes.fr}
         \and
         	School of Mathematics, Trinity College Dublin, Dublin 2, Ireland
    }
   \date{Received ...; accepted ...}

 
  \abstract
   {Accreting black holes (\bhs) may be surrounded by a highly magnetized plasma threaded by an organized poloidal magnetic field. Non-thermal flares and power-law spectral components at high energy could originate from a hot, collisionless and \kyle{nearly} force-free corona. The jets we often observe from these systems are believed to be rotation-powered and magnetically-driven.}
   {We study axisymmetric \bh magnetospheres where a fraction of the magnetic field lines anchored in a surrounding disk \ileyk{are connected to the event horizon of a rotating \bh}. We identify the conditions and sites of magnetic reconnection within 30 gravitational radii depending on the \bh spin.} 
   {With the fully \ileyk{general} relativistic particle-in-cell code \texttt{GRZeltron}, we solve the time-dependent dynamics of the electron-positron pair plasma and of the electromagnetic fields \ileyk{around the \bh}. The aligned disk is represented by a steady and perfectly conducting plasma in Keplerian rotation, threaded by a dipolar magnetic field.}
   {For prograde disks around Kerr \bhs, the topology of the magnetosphere is hybrid. Twisted open magnetic field lines crossing the horizon \kyle{power a Blandford-Znajek jet} while open field lines with their footpoint beyond a critical distance on the disk \ileyk{could} launch a magneto-centrifugal wind. In the innermost regions, coupling magnetic field lines ensure the transfer of significant amounts of angular momentum and energy between the \bh and the disk. From the Y-point at the intersection of these three regions, a current sheet forms where vivid particle acceleration via magnetic reconnection takes place. We compute the synchrotron images of the current sheet emission.}
   {Our estimates for jet power and \bh-disk exchanges match those derived from purely force-free models. \ileyk{Particles are accelerated at the Y-point which acts as a heat source for the so-called corona. It provides a physically motivated ring-shaped source of hard X-rays above the disk for reflection models.} Episodic plasmoid ejection might explain millisecond flares observed in Cygnus X-1 in the high soft state, but are too fast to account for daily non-thermal flares from \sgr. Particles flowing from the Y-point down to the disk could produce a hot spot at the footpoint of the outermost closed magnetic field line.}
   
   \keywords{acceleration of particles -- magnetic reconnection -- black hole physics -- radiation mechanisms: non-thermal -- methods: numerical}

   \maketitle


\section{Introduction}

Although their masses and luminosities span many orders of magnitude, stellar-mass black holes (\bhs) and supermassive black holes (\smbhs) share many similarities which emphasize how simple the intrinsic structure of a \bh is. Due to their high compactness, they can accrete ambient material and become intense sources of light visible up to cosmological distances. \kyle{The intensity, the spectrum, and the polarimetry of the radiation emitted by the plasma in the immediate surroundings of accreting \bhs grant us access to the two main properties of these universal beacons:} their mass and their spin.  

Multiple lines of evidence indicate that \bhs spin. In X-ray binaries hosting a \bh surrounded by an accretion disk, non-zero spins were measured using X-ray reflection spectroscopy and thermal continuum fitting \citep{Reynolds2020,Bambi2021}. Although these two diagnostics do not always yield consistent values and rely on many assumptions and degrees of freedom, they suggest that stellar-mass \bhs can be fast rotators (\eg GRS 1915+105, \citealt{Shreeram2020}, or \cyg, \citealt{Zhao2021}). X-ray reflection spectroscopy applied to \agns shows that \smbhs also have intrinsic angular momentum. Reasonable hopes for a direct measure of the spin of the \smbh at the center of the Milky Way, \sgr, via Lense-Thirring precession of the orbit of a nearby star can be sustained thanks to the efforts of the \gravity collaboration \citep{Abuter2018c}. The ubiquity of jets from accreting stellar-mass and \smbhs has also been interpreted as an indirect hint in favor of spinning \bhs.

Accretion onto X-ray active stellar-mass \bhs is thought to be partly mediated by an optically thick and geometrically thin accretion disk similar to the one first described by \cite{Shakura1973}. In semi-detached X-ray binaries, where the donor star fills its Roche lobe, an accretion disk forms owing to the high angular momentum of the stellar material provided at the inner Lagrangian point \citep[see \eg][]{Frank2002}. In detached high-mass X-ray binaries (\hmxbs), where the donor star is massive, the formation of a disk is still possible provided the UV-driven stellar wind fueling the \bh is slow and/or the orbital separation is small \citep{ElMellah2018,Sen2021}. \kyle{The properties of the accretion flow around \smbhs are less constrained}. The far infrared / radio images of the immediate vicinity of the \smbh \m{87} reported by \cite{Akiyama2019} are consistent with an underlying optically thin accretion disk where the magnetic field plays a major dynamic role \citep[the magnetically-arrested disk, or MAD,][]{Collaboration2021b}. This type of disks has been shown to be a robust bedrock for jet launching \`a la Blandford-Znajek \citep{Tchekhovskoy2011}. The presence of an accretion disk around \m{87} is thus corroborated by the \kyle{prominent} jet emitted by the \smbh \citep{Algaba2021}. For \sgr, numerical simulations by \cite{Ressler2020a} suggest that a MAD can form around the \smbh by capture of the stellar wind emitted by Wolf-Rayet stars orbiting within a few parsecs. Finally, the hot spot observed in orbit around \sgr by the \gravity collaboration was interpreted as the sign of an underlying disk \citep{Abuter2018b}. 

Accreting \bhs are thought to be also surrounded by a diffuse environment often called the corona. In the \hmxb \cyg, where a late O-type supergiant supplies material to an approximately 21\msun \bh \citep{Miller-Jones2021}, a high energy polarized component between 0.4 and 2MeV was found \citep{Laurent2011,Rodriguez2015}. It is present in the low-hard state but also in the high-soft state \citep[see][for spectral state classification in \cyg]{Grinberg2013}. In the high-soft state, radio emission from the jet is much dimmer and was long thought to be absent \citep{Remillard2006b,Rushton2012,Zdziarski2020}, which suggests that jet launching is quenched and is unlikely to be the source of this high energy emission. \kyle{\cite{Cangemi2021} showed that this high energy component was instead} compatible with a hybrid thermal/non-thermal Comptonization spectrum associated to a population of relativistic electrons. They also found that the magnetic field these electrons were embedded in could be high enough to accelerate particles up to relativistic Lorentz factors via magnetic reconnection in the corona \citep{Malzac2009,Poutanen2009}. These results were in agreement with \cite{DelSanto2013} who derived upper limits for the magnetic field magnitude in the corona between $10^5$ and $10^7$G. Similarly, the daily non-thermal flares from \sgr in near-infrared (NIR) have been interpreted as synchrotron emission from non-thermal relativistic electrons within a few 10 gravitational radii \citep{Witzel2018}, consistent with the presence of a hot corona around the \smbh. \cite{Witzel2020} carried an extensive multi-wavelength analysis of variability in \sgr, from sub-millimeter to X-rays, and ascribed the emission to high energy electrons in a magnetic field of $\sim$8-13G, in agreement with other diagnostics \citep{Moscibrodzka2009,Dexter2010,Ponti2017}. \ileyk{Noticeably, \cite{Ponti2017} showed during a flare, time evolution of spectral properties are coherent with a drop of the magnetic field, which suggests that magnetic reconnection could be at play. The recent polarization measurements performed by \cite{Collaboration2021a} in \m{87} point to a large-scale poloidal magnetic field of similar intensity as in \sgr threading a dilute plasma \citep{Collaboration2021b}.}

The aforementioned X-ray reflection spectroscopy diagnostics assume a pre-defined geometry for the corona which plays the role of a source of hard X-rays reprocessed by the underlying disk. In the common lamppost model, the corona is a point source located on the \bh spin axis, a few gravitational radii above the \bh \citep[see \eg][]{Kara2019}. However, both around accreting stellar-mass \bhs \citep{Zdziarski2021} and \smbhs \citep{Zoghbi2019}, evidence has emerged in favor of an extended corona \citep{Chainakun2019}. The actual geometry of the corona and the energy sources responsible for its formation however remain elusive, and so do the sites where the coronal heating takes place.

\ileyk{Configurations where magnetic field lines are all open and thread either the disk or the event horizon have been the object of much attention. In the equatorial plane within the \isco, a highly dynamic current sheet has been shown to form and be the stage of vivid magnetic reconnection \citep{Parfrey2019,Crinquand2021,Bransgrove2021,Ripperda2021}. Alternatively, the magnetosphere could contain closed magnetic field lines threading the event horizon but with a footpoint on the surrounding disk. It is the case extensively studied by \cite{Uzdensky2004,Uzdensky2005} and more recently, by \cite{Yuan2019b,Yuan2019c} and \cite{Mahlmann2018a} who worked in the purely force-free regime. They found steady solutions where disk-\bh coupling field lines in the innermost region co-exist with open field lines anchored further in the disk. The presence of magnetic loops formed in the disk and extending into the corona was first speculated by \cite{Galeev1979} and a stochastic model of their dynamics was designed by \cite{Uzdensky2008} but their formation remains elusive. Recent general relativistic magneto-hydrodynamics (GR-MHD) simulations by \cite{Chashkina2021} highlighted how these loops could be advected inwards in the disk. On the other hand, \cite{Parfrey2015}, \cite{Yuan2019c} and \cite{Mahlmann2020} used time-variable force-free simulations to show that efficient jet launching and dissipation in the corona could be sustained by the advection of successive magnetic loops of \kyle{varying} polarity.}

\ileyk{In this paper, we study the global axisymmetric structure of a \bh magnetosphere where the main physical ingredients are accounted for: a spinning \bh and a magnetic field threading both a highly magnetized hot pair-plasma corona and, in the \bh equatorial plane, a disk. In the hybrid magnetospheres we focus on, the topology of the magnetic field is not limited to field lines threading the \bh event horizon and extending to infinity. Initially, magnetic field lines are all anchored in the disk and cross the horizon, forming a single magnetic field loop which allows for exchanges of energy and angular momentum between the two components. We do not address the question of the origin of this configuration nor of the accretion of the loop itself but instead, we study the relaxation and time variability of this setup. We solve the dynamics of the plasma and of the electromagnetic fields in the corona specifically using an ab initio particle-in-cell approach, suitable to capture dissipative effects in highly magnetized plasmas. The disk is modeled as a steady and perfectly conducting plasma in Keplerian rotation. In the light of the major results obtained by previous authors, we believe this hybrid configuration can be a fruitful framework before studying more realistic magnetic field distributions on the disk. Our aim is to characterize the topology of the magnetic field and the possible reconnection sites in the corona, depending on the \bh spin, on the disk orientation (prograde or retrograde) and on the disk thickness.}

In Section\,\ref{sec:model}, we present the model of \bh magnetosphere we rely on and detail the numerical setup we designed to address it with the \texttt{GRZeltron} code. We present the results we obtained in Section\,\ref{sec:results} and discuss the implications for accreting \bhs, with an emphasis on \sgr, \m{87} and \cyg. Finally, we summarize our main conclusions in Section\,\ref{sec:Summary} and suggest future tracks to be explored.


\section{Model}
\label{sec:model}

\subsection{Kerr black hole magnetosphere}
\label{sec:Kerr_BH_magnetosphere}

\subsubsection{Kerr spacetime metric}
\label{sec:Kerr_spacetime_metric}


We consider a stationary and axisymmetric Kerr spacetime induced by a rotating \bh of mass $M$ and dimensionless spin $a$ \citep{Kerr1963}. We use the $3+1$ formalism in order to have a universal time coordinate $t$ and spatial hypersurfaces of constant $t$, mapped by the coordinates $x^i$ \citep{MacDonald1982}. It introduces a class of observers whose worldlines are orthogonal to these spatial hypersurfaces, the fiducial observers or \fidos. If we write $h_{ij}$ the spatial 3-metric on the hypersurfaces of constant $t$, the Kerr spacetime line element $\diff s^2$ can be written in the ADM form \citep{Arnowitt1962}:
\begin{equation}
    \diff s^2=-\alpha^2c^2\diff t^2+h_{ij}\left(\diff x^i+\beta^ic\diff t\right)\left(\diff x^j+\beta^jc\diff t\right)
\end{equation}
where $\alpha$ is the lapse function, $\boldsymbol{\beta}$ is the shift vector and $c$ is the speed of light. The former encapsulates the information on time \kyle{dilation} from coordinate time to the proper time measured locally by the \fido, while the latter is the 3-velocity of the \fido with respect to the coordinate grid.

\ileyk{The properties of interest of this metric for our study are the presence of an event horizon of radius $r_H=r_g\left( 1+\sqrt{1-a^2} \right)$ where $r_g=GM/c^2$ is the gravitational radius and $G$ is the gravitational constant, an ergosphere and an innermost stable circular orbit (\isco) \citep{Shapiro1983a}. At the event horizon, stationary observers are forced into rotation at an angular speed ${\omega_H=ac/2r_H}$ that we identify to the angular speed of a rotating \bh. Hereafter, we work in Kerr-Schild spherical coordinates $\left( r,\theta,\phi \right)$ in order to avoid the coordinate singularity at the event horizon which would arise in Boyer-Lindquist coordinates for instance.}

\subsubsection{Keplerian disk}
\label{sec:Keplerian_disk}

In our model, we rely on a simplified representation of a rotating and perfectly conducting disk entirely parametrized by the magnetic flux function $\Psi$ in the disk and the disk aspect ratio $\epsilon$. The latter is assumed to be uniform and is defined as the ratio of the scale height to the distance ${R=r\sin\theta}$ to the \bh projected on the disk plane. \kyle{We do not model accretion in our model since there is no radial speed in the disk}. The disk rotation axis is assumed to be aligned with the \bh spin axis such as the disk lies in the \bh equatorial plane. \ileyk{We account for the disk aspect ratio in the angular velocity profile by relying on the following approximate correction \citep[see \eg][]{Fromang2011}:} 
\begin{equation}
    \label{eq:Om_K}
    \Omega\left(R\right)=\Omega_K\sqrt{1-\epsilon ^2}
\end{equation}
where 
\begin{equation}
    \label{eq:omega_GR}
    \Omega_K\left(R\right)=\sqrt{\frac{GM}{r_g^3}}\frac{1}{\left(R/r_g\right)^{3/2}+a}
\end{equation}
is the relativistic counterpart of the Keplerian angular speed profile in Newtonian mechanics. \kyle{In this paper, we assume a dipolar magnetic field in the disk, centered on the \bh}. The influence of the disk magnetic flux distribution on the magnetic structure of force-free \bh magnetospheres was found to be modest by \cite{Yuan2019b} who considered, in addition to this dipolar case, asymptotically paraboloid and uniform vertical fields. In a follow-up paper, when exploring a more realistic setup where two successive poloidal loops of opposite polarity are accreted, \citep{Yuan2019c} found similar results when the inner loop was sufficiently weak with respect to the outer one. The magnetic field lines are frozen into the disk and rotate at the angular speed of their footpoint in the disk mid-plane. This magnetic flux function enables us to explore a configuration where initially, magnetic field lines threading the disk are all connected to the \bh by extending the $\Psi$ function to the whole simulation space (see Section\,\ref{sec:ICs}).

\subsubsection{Corona}
\label{sec:Corona}

Spectral fits of accreting stellar-mass \bhs and \smbhs motivate our choice to include in our model a hot collisionless electron/positron pair plasma to represent the corona. While the disk is kept \kyle{stationary}, we will study the time evolution of this corona. Although several mechanisms were invoked to explain its replenishment (\eg pair cascade, \citealt{Crinquand2020}, or magneto-centrifugal loading from the disk, \citealt{Blandford1982}), we adopt an empirical approach. Independently of the injection mechanism, we assume that the particle density in the corona is both low enough to ensure that the plasma is collisionless, and high enough to be in the force-free regime. In this framework, electromagnetic forces dominate the dynamics. The plasma flows along the poloidal magnetic field lines and screens any electric component along these lines. The force-free approximation requires a cold magnetization $\sigma \gg 1$, where $\sigma$ is defined as the ratio of the magnetic energy to the inertial mass energy of the particles:
\begin{equation}
    \sigma=\frac{B^2/4\pi}{\left(\gamma_+ n_+ + \gamma_- n_-\right)m_e c^2}
\end{equation}
with $B=\left| \boldsymbol{B} \right|$ the magnitude of the magnetic field measured by the \fido, $\gamma_{\pm}$ (resp. $n_{\pm}$) the Lorentz factor (resp. the number density) of the positrons/electrons and $m_e$ the mass of the electron. In these conditions, the plasma satisfies the force-free equation \citep[see \eg][]{Uzdensky2004} and \kyle{in the steady state,} each magnetic field line rotates at a fixed angular speed. In order to be achieved, this regime requires a net plasma charge density $\rho$ \kyle{above} the threshold set by the Goldreich-Julian charge density $\rho_{GJ}$ \citep{Goldreich1969}:
\begin{equation}
    \label{eq:GJ}
    \rho_{GJ}=-\frac{\boldsymbol{\Omega}\cdot\boldsymbol{B}}{2\pi c}
\end{equation}
where $\boldsymbol{\Omega}$ is the angular \kyle{velocity} vector of the magnetic field line. This charge density translates into a net number density $n_{GJ}=\rho_{GJ}/e$, with $e$ the charge of a positron ($e>0$). Hereafter, we assume that $\sigma \gg 1$ and that the plasma multiplicity $\kappa=\left(n_++n_-\right)/n_{GJ} \gtrsim 1$. We will detail in Section\,\ref{sec:Parameters} how we ensure that the number density $n=n_++n_-$ of electrons and positrons is high enough to be in a regime where the net charge density $\rho$ can be higher than $\rho_{GJ}$.

\subsection{Numerics}
\label{sec:numerics}

\subsubsection{The \texttt{GRZeltron} code}
\label{sec:GRZeltron}

\ileyk{We solve the dynamics of the plasma and of the electromagnetic fields in the whole corona, including in regions where the force-free approximation breaks down and where significant amounts of electromagnetic energy can be converted into particle kinetic energy. To do so, we use the general relativistic particle-in-cell (\pic) code \texttt{GRZeltron} first introduced in \cite{Parfrey2019} and based on the \texttt{Zeltron} code \citep{Cerutti2013}. Its principle is described in detail in \cite{Crinquand2021} and we only remind here the main aspects.}


\ileyk{We work in the $3+1$ formalism presented in Section\,\ref{sec:Kerr_spacetime_metric} and we note $\boldsymbol{B}$ and $\boldsymbol{D}$ the magnetic and electric fields locally measured by the \fido. These fields are advanced by solving Maxwell-Faraday and Maxwell-Ampère equations. In these equations intervene \kyle{the grid-based current (hereafter, the current) $\boldsymbol{J}=q\boldsymbol{v}$, where $q$ is the charge of the particles and $\vec{v}$ their 3-velocity on the spatial hypersurface measured in universal coordinate time. Maxwell-Faraday and Maxwell-Ampère equations also make use of the auxiliary fields $\boldsymbol{H}$ and $\boldsymbol{E}$ obtained from \citep{Komissarov2004a}:}}
\begin{empheq}[]{align}
    \label{eq:H}
    &\boldsymbol{H}=\alpha \boldsymbol{B}-\boldsymbol{\beta}\times\boldsymbol{D}\\
    \label{eq:E}
    &\boldsymbol{E}=\alpha\boldsymbol{D}+\boldsymbol{\beta}\times\boldsymbol{B}.
\end{empheq}
\kyle{We then solve the full equation of motion for the electrons and positrons, including the Lorentz force \citep[see \eg][]{Bacchini2018,Bacchini2019,Parfrey2019}.}
\ileyk{Although available in \texttt{GRZeltron}, we did not include any radiative drag force or cooling induced by synchrotron or Compton emission (see \citealt{Mehlhaff2020} and \citealt{Sridhar2021} for the impact of radiative feedback on particle acceleration). We now proceed to a description of the numerical setup we designed to address the problem described in Section\,\ref{sec:Kerr_BH_magnetosphere}.}

\subsubsection{Grid}
\label{sec:grid}

We perform global \pic simulations on a 2D $\left(r,\theta\right)$ grid in Kerr-Schild coordinates. Axisymmetry around the polar axis is assumed such that all quantities are independent of the $\phi$ coordinate. Although the grid is 2D, all the 3 components of the electromagnetic fields and of the particle velocities are accounted for. Furthermore, our model being symmetric with respect to the \bh equatorial plane, we only cover the region from $\theta_{min}=0$ up to $\theta_{max}=\pi/2$. The inner edge of the simulation space is set within the horizon, with $r_{min}=0.9r_H$, while the outer edge lies at $r_{max}=30r_g$, far enough to limit the impact of the outer boundary conditions detailed in Section\,\ref{sec:BCs} on the region of interest. To cover the whole simulation space with a uniform cell aspect ratio, we use a logarithmically stretched spacing in $r$. Unless stated otherwise, the results we present are for a grid with 2,048 cells in the $r$-direction and 1,120 cells in the $\theta$-direction, which ensures a cell aspect ratio close to 1. We argue in Section\,\ref{sec:Parameters} that given the parameters we explore, this resolution is enough to resolve the (non-relativistic) plasma skin depth $\delta_e=\sqrt{m_ec^2/4\pi ne^2}$ everywhere, a statement that we verify a posteriori in our simulations (see appendix\,\ref{sec:num_conv}). 

\subsubsection{Boundary conditions}
\label{sec:BCs}

The inner edge of the simulation space lies within the event horizon ($r_{min}<r_H$) so the dynamics above the event horizon is not impacted by the boundary conditions at $r_{min}$. At the pole ($\theta=0$), we apply boundary conditions based on the symmetry properties of the polar and axial electric and magnetic vectors. At the outer edge of the simulation space, we use an absorbing layer from $r=27r_g$ to $r=r_{max}$. In this region, resistive terms are added to Maxwell's equations in order to damp the electromagnetic fields and avoid spurious wave reflection \citep{Cerutti2015}. Without an ambient disk to hold the magnetic field, the initial magnetic field lines would be expelled from the \bh, in agreement with the no-hair theorem. \ileyk{We introduce a perfectly conducting disk from $\theta_d=\pi/2-\arctan\left(\epsilon\right)$ to $\theta_{max}=\pi/2$ where the magnetic field is frozen to its initial value (see Section\ref{sec:ICs}). For $r>r_{\iscoo}$, we enforce the magnetic field lines to co-rotate with the disk at the Keplerian angular speed of their footpoint in the equatorial plane, given by Equation\,\eqref{eq:Om_K}. \ileyk{Conversely, although the equatorial plane remains perfectly conducting up to the event horizon, we enforce a smooth transition from a Keplerian angular speed to zero angular speed at $r<r_{\iscoo}$}. In the disk, the electric field $\boldsymbol{E}$ is obtained from the assumption of a perfectly conducting environment.} Therefore, the electric field in the frame rotating with each disk annulus must vanish and we have:
\begin{equation}
    \boldsymbol{E}+\frac{\boldsymbol{V_{disk}}}{c}\times\boldsymbol{B}=\boldsymbol{0} 
    \label{eq:omega}
\end{equation}
in the disk, with ${\boldsymbol{V_{disk}}=\Omega\boldsymbol{\partial_{\phi}}}$. $\boldsymbol{H}$ and $\boldsymbol{D}$ in the disk are then deduced from the constitutive relations \eqref{eq:H} and \eqref{eq:E}.


\subsubsection{Initial conditions}
\label{sec:ICs}


We start from a dipolar magnetic field in vacuum defined by the following $\phi$-component of \ileyk{the 4-potential: 
\begin{equation}
    A_{\phi}=B_0 r_g^2\frac{\sin^2\theta}{r/r_g} 
    \label{eq:Aphi}
\end{equation}}
Constructing the magnetic field from the 4-potential enables us to enforce an initially divergence-free field, a feature which is maintained through the simulation. In this \ileyk{initial} configuration, all the magnetic field lines connect the disk to the event horizon. The further the footpoint of the magnetic field line on the disk, the higher the latitude of the point where this field line intersects the event horizon. 

It is the simplest divergence-free setup which connects the disk to the \bh but it introduces a set of possibly non-physical magnetic field lines in the equatorial plane within the \isco. Indeed, this region is thought to be orders of magnitude less dense than the disk \citep[see][though they focus on a case where magnetic fields are not dynamically important]{Potter2021}. In the highly magnetized regime we are interested in, such a density drop means that this region could be force-free but according to the no-ingrown-hair theorem, closed magnetic field lines anchored in the \bh can only maintain if they intersect or loop around a non force-free region such as the disk \citep{MacDonald1982,Gralla2014}. The behavior of the plasma in this region, which however vanishes as $a\rightarrow 1$, might thus be altered by this initial magnetic field configuration, especially because it is enforced as a boundary condition in the equatorial plane.

\subsubsection{Particle injection}
\label{sec:Particle_injection}

The corona is initially empty and needs to be populated in order to match the force-free conditions described in Section\,\ref{sec:Corona}. Then, since particles propagate at relativistic speeds and eventually escape from the simulation space, the corona has to be replenished in order to prevent it from emptying within a few ${10 r_g/c}$. In this paper, we remain agnostic on the physical mechanism which ensures that the corona is not empty and embrace an empirical approach. At each time step, we inject electron/positron pairs in cells where $n<3\Omega_{\iscoo}B/2\pi c e$ with a numerical weight given by:
\begin{equation}
    w=\frac{\Omega_{\iscoo}B}{2\pi c e}
\end{equation}
where $\Omega_{\iscoo}=\Omega_K\left(R=r_{\iscoo}\right)$ is the Keplerian angular speed in the disk mid-plane at the \isco, and $B=\left|\boldsymbol{B}\right|$ is the \kyle{magnitude of the local magnetic field}. The choice of $\Omega_{\iscoo}$ as a reference angular speed was found to be a good trade-off to load both magnetic field lines anchored in the disk and in the event horizon with enough plasma, for any \bh spin. Given that we start with a dipolar magnetic field, this weight guarantees that the number of particles per cell remains reasonably uniform through the simulation space since the $r^{-3}$ dependence of the magnetic field compensates the cell volume increase from $r_{min}$ to $r_{max}$. Particles are deleted when they enter the disk, the absorbing layer or leave the simulation space. They do not experience any radiative feedback in these numerical simulations.

\subsubsection{Parameters}
\label{sec:Parameters}


The main degrees-of-freedom of our model are the \bh spin $a$, the disk aspect ratio $\epsilon$ and the reference magnitude of the magnetic field $B_0$. In the simulations, the latter manifests through the ratio of the gravitational radius to the Larmor radius ${R_{L,0}=m_ec^2/eB_0}$. For the \bh spin, we explore the case of a non-rotating \bh ($a=0$), rotating \bhs with increasing spins from $a=0.6$ to $a=0.99$ surrounded by a prograde disk. At lower spin than $a=0.6$, critical features expected for $a>0$ like the separatrix (see Section\,\ref{sec:ff_region}) extend up to the outer edge of our grid at $r_{max}=30r_g$. In order to avoid numerical artifacts due to the outer boundary conditions, we limit our study of prograde disk to \bh spins $a\geq0.6$. We also investigate the case of a rotating \bh with a spin $\left|a\right|=0.8$ surrounded by a counter-rotating disk (\ie a=-0.8). We consider a geometrically thin disk with $\epsilon=5\%$ \ileyk{and obtain qualitatively similar results for a thicker disk with $\epsilon=30\%$ around a \bh with spin $a=0.8$.}

\begin{figure}
\centering
\includegraphics[width=0.99\columnwidth]{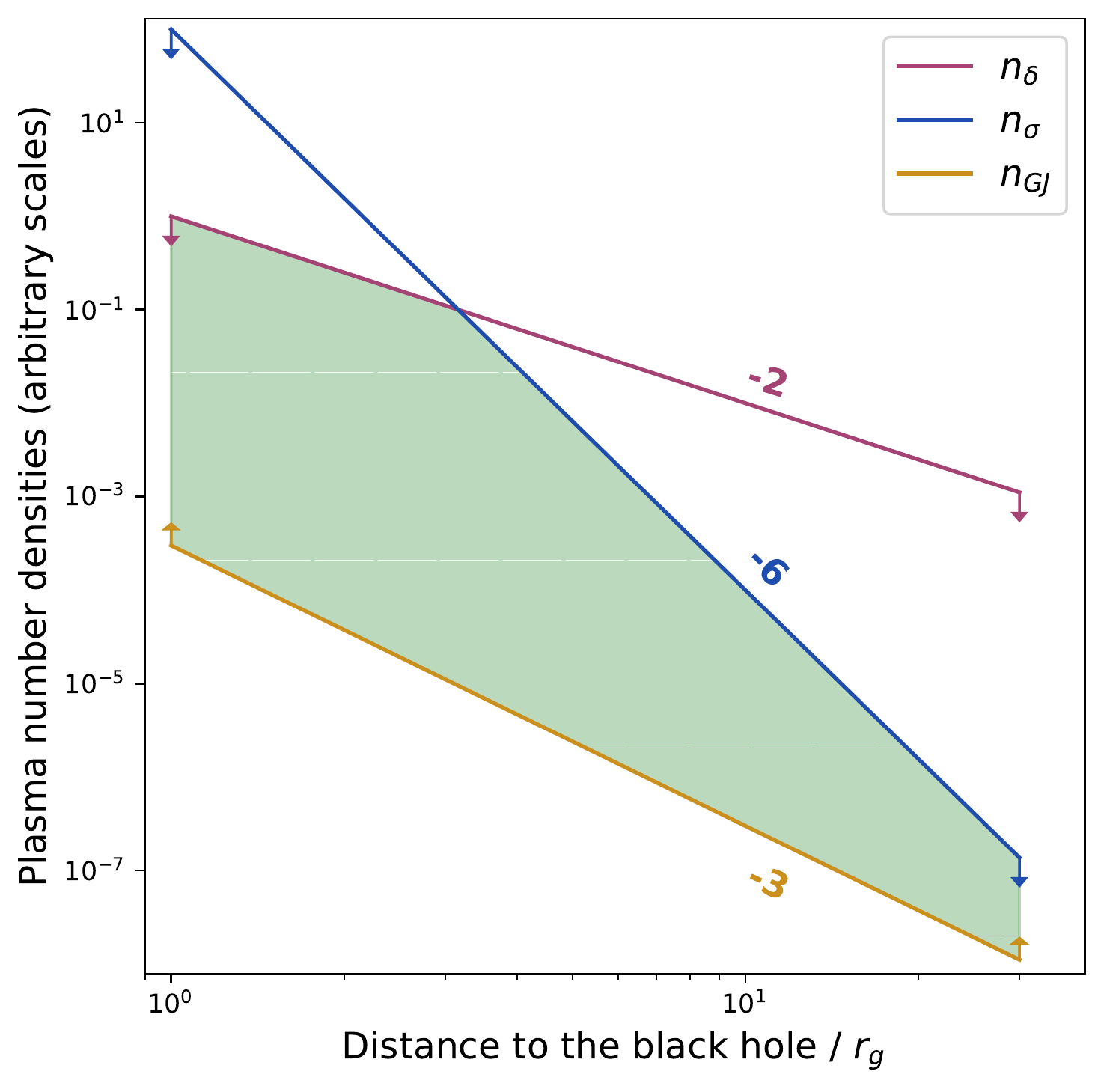}
\caption{Fiducial radial profiles for the critical densities $n_{\delta}$, $n_{\sigma}$ and $n_{GJ}$ defined in the text. The numbers are the power-law exponents and the arrows indicate the allowed regions. The green hatched region represents the region where the force-free regime can be achieved while still resolving the skin depth.}
\label{fig:B0_max}
\end{figure}

Realistic values of the dimensionless magnetic field ${\tilde{B}_0=r_g/R_{L,0}}$ are out of reach of the current computational capacities. Indeed, in \m{87} and \sgr, the magnetic field near the event horizon approximately ranges between 1 and 100G which corresponds to ${\tilde{B}_0\sim4\cdot 10^{8-10}}$ for \sgr and ${\tilde{B}_0\sim5\cdot 10^{11-13}}$ for \m{87}. In \cyg, $\tilde{B}_0$ is comparable since the gravitational radius, smaller by 5 to 7 orders of magnitude, is counter-balanced by a similar increase in the magnetic field magnitude \citep{DelSanto2013,Cangemi2021}. To remain in the force-free regime with such high magnetic fields, we would need a much higher particle number density according to Equation\,\eqref{eq:GJ}: a high magnetic field then sets a stringent upper limit on the skin depth which becomes so small that it cannot be resolved anymore. Reasoning in terms of number density, we must match the three following conditions in the whole simulation space:
\begin{itemize}
    \item the skin depth is resolved \ie $n<n_{\delta}=\frac{m_ec^2}{4\pi e^2}/\Delta r^2$, where $\Delta r$ is the radial extent of the cell and where we assumed a cell aspect ratio close to unity
    \item the magnetization $\sigma\gg 1$ \ie $n\ll n_{\sigma}=B^2/4\pi m_e c^2$
    \item the multiplicity $\kappa\gtrsim 1$ \ie $n\gtrsim n_{GJ}=\Omega_{\iscoo} B/2\pi c e$
\end{itemize}
The main numerical challenge of this setup lies in the combination of the dipolar magnetic field and the high ratio between the radius of the outer and inner edges of the simulation space since the aforementioned number densities vary by several orders of magnitude from $r_{min}$ to $r_{max}$. In Figure\,\ref{fig:B0_max}, we show representative radial profiles for $n_{\delta}$, $n_{\sigma}$ and $n_{GJ}$ assuming $B\propto r^{-3}$. The arrows indicate where the plasma number density must lie with respect to each profile according to the aforementioned conditions. The scaling of each profile is chosen such that the three conditions can be matched for any $r$ (green hatched region), which requires:
\begin{equation}
    2\left(\frac{r_{max}}{r_g}\right)^{3/2}\ll \tilde{B}_0 < \frac{1}{2} \frac{r_{min}^3}{r_g \Delta r^2\left(r=r_{min}\right)}
    \label{eq:parameters}
\end{equation}
With the spin, resolution and radial boundaries we use, it means that we must work with a dimensionless magnitude of the magnetic field which approximately verifies:
\begin{equation}
    320 \ll \tilde{B}_0 < 190,000
\end{equation}
In what follows, we set $\tilde{B}_0=10^5$, a value suitable to guarantee that from $r_{min}$ to $r_{max}$, the force-free regime is achievable for plasma densities at which the skin depth is resolved by our grid.

\section{Results}
\label{sec:results}

\subsection{Magnetic structure}
\label{sec:Magnetic_structure}

In Section\,\ref{sec:ff_region}, we first analyze the global structure of the corona in the light of properties derived in the force-free approximation and verify a posteriori the validity of our injection method. We then detail the properties of each of the three main regions identified in Sections\,\ref{sec:BH_Opened_magnetic_field}, \ref{sec:disk_Opened_magnetic_field} and \ref{sec:BH_Opened_magnetic_field}. In Section,\,\ref{sec:Toroidal_fields}, we describe the current sheet which separates the regions where magnetic field lines are open and anchored either in the disk or in the \bh.

\subsubsection{Force-free regions}
\label{sec:ff_region}

\begin{figure*}
\centering
\begin{subfigure}[b]{1.7\columnwidth}
   \includegraphics[width=0.99\columnwidth]{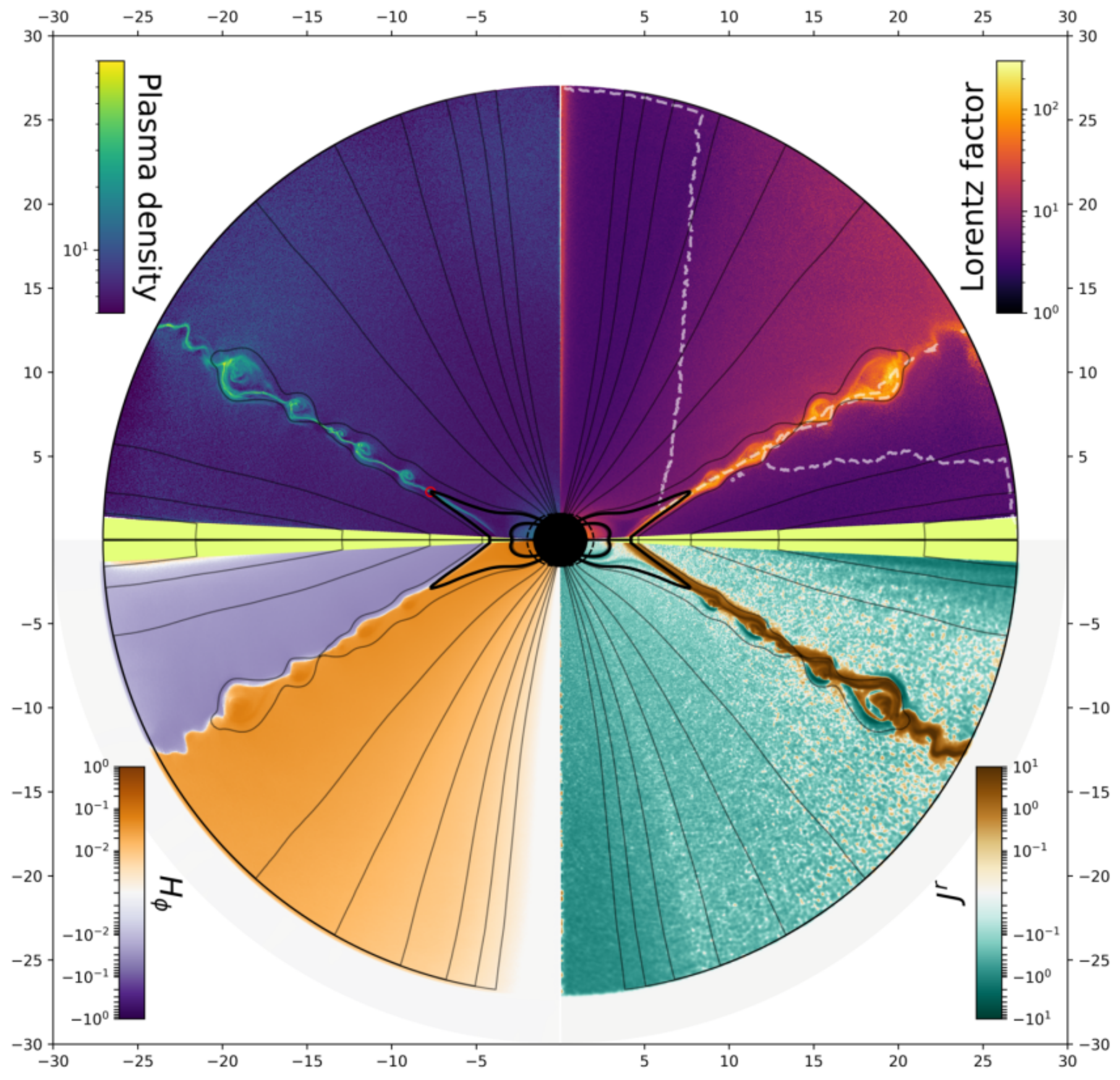}
\end{subfigure}
\begin{subfigure}[b]{1.7\columnwidth}
   \includegraphics[width=0.99\columnwidth]{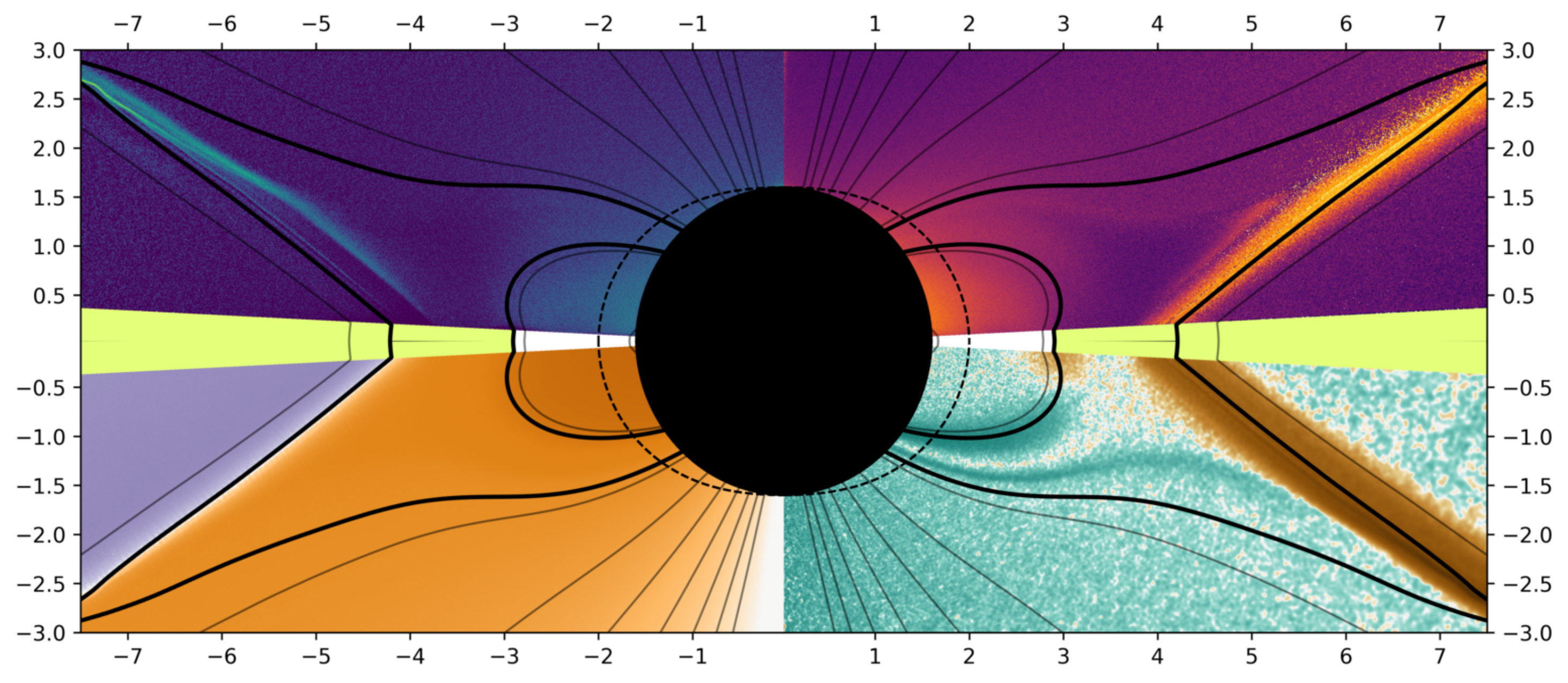}
\end{subfigure}
\caption{Poloidal magnetic field lines around the \bh (black solid lines), with the event horizon (black disk) and the ergosphere (black dashed line), for a fiducial snapshot from a simulation with $a=0.8$ and a thin disk (in \ileyk{Chartreuse yellow} in the equatorial plane). The thicker field lines delimit the region between the \isco and the separatrix coupling the \bh to the disk, visible in the zoom in the bottom panel. The colormaps represent the total plasma density $n$ in units of $n_{GJ}$ (top left), the mean Lorentz factor of the particles (top right), the toroidal component of the $\boldsymbol{H}$ field (bottom left) and the radial component of the \kyle{current $\boldsymbol{J}$} (bottom right). To compensate for spatial dilution, plasma density and current were multiplied by $r^2$. A smoothing Gaussian kernel of a few cells wide was applied to the current map. In the plasma density map, the red circle locates the Y-point. Distances to the \bh on the x and y-axis are given in units of $r_g$. The white line in the top right corner stands for the outer light surface.}
\label{fig:wheel}
\end{figure*}


In Figure\,\ref{fig:wheel}, we represented colormaps of 4 different quantities overlaid on top of the magnetic field lines, after the initial setup has relaxed. The \bh spin is 0.8, the disk is thin ($\epsilon=5\%$) and the results are qualitatively similar to what is found for other spins $\geq 0.6$. The two thick magnetic field lines correspond to the innermost and outermost coupling field lines anchored in the \bh and in the disk, respectively at the \isco and of magnetic potential $A_{\iscoo}$, and at $r=R_S$ and of magnetic potential $A_S$. Following \cite{Uzdensky2005}, we call the latter field line the separatrix. Space around the \bh is subdivided in 4 distinct regions delimited by those 2 critical field lines. 

In the innermost region, where $A>A_{\iscoo}$, magnetic field lines are anchored in the equatorial plane and do not rotate, in agreement with the boundary condition at their footpoint. As explained in Section\,\ref{sec:ICs}, it is an unphysical configuration due to the lack of non force-free medium in the region where $r<r_{\iscoo}$ and $\theta\in[\theta_d;\theta_{max}]$ (in white at the equator in Figure\,\ref{fig:wheel}). In reality, these magnetic field lines would be stretched in the equatorial plane due to the plasma quickly falling toward the \bh. It would form a current sheet where the force-free approximation would break up. In such a case, the magnetic field lines would either close within the horizon \citep{Komissarov2004a} or be fully open \citep{Crinquand2021,Bransgrove2021}. In the latter case though, no coupling between the \bh and the disk would exist since no magnetic field line would connect the two. In the perspective of a magnetically coupled configuration, it is likely that the lack of significant vertical magnetic field in the region where $r<r_{\iscoo}$ and $\theta\in[\theta_d;\theta_{max}]$ would lead to a field line anchored at the \isco which would penetrate the horizon at much lower latitude than in our simulations, resembling more the results obtained by \cite{Uzdensky2005} and \cite{Yuan2019b} who directly solved the relativistic Grad-Shafranov equation. The bloated region at the equator near the event horizon, visible in the zoom in bottom panel in Figure\,\ref{fig:wheel}, is thus probably the least accurate feature in these simulations. On the other hand, the 3 other regions, further described in the next sections, are both physically realistic and numerically robust. 

\begin{figure}
\centering
\begin{subfigure}[b]{0.99\columnwidth}
\hspace{0.7cm}\includegraphics[width=0.99\columnwidth]{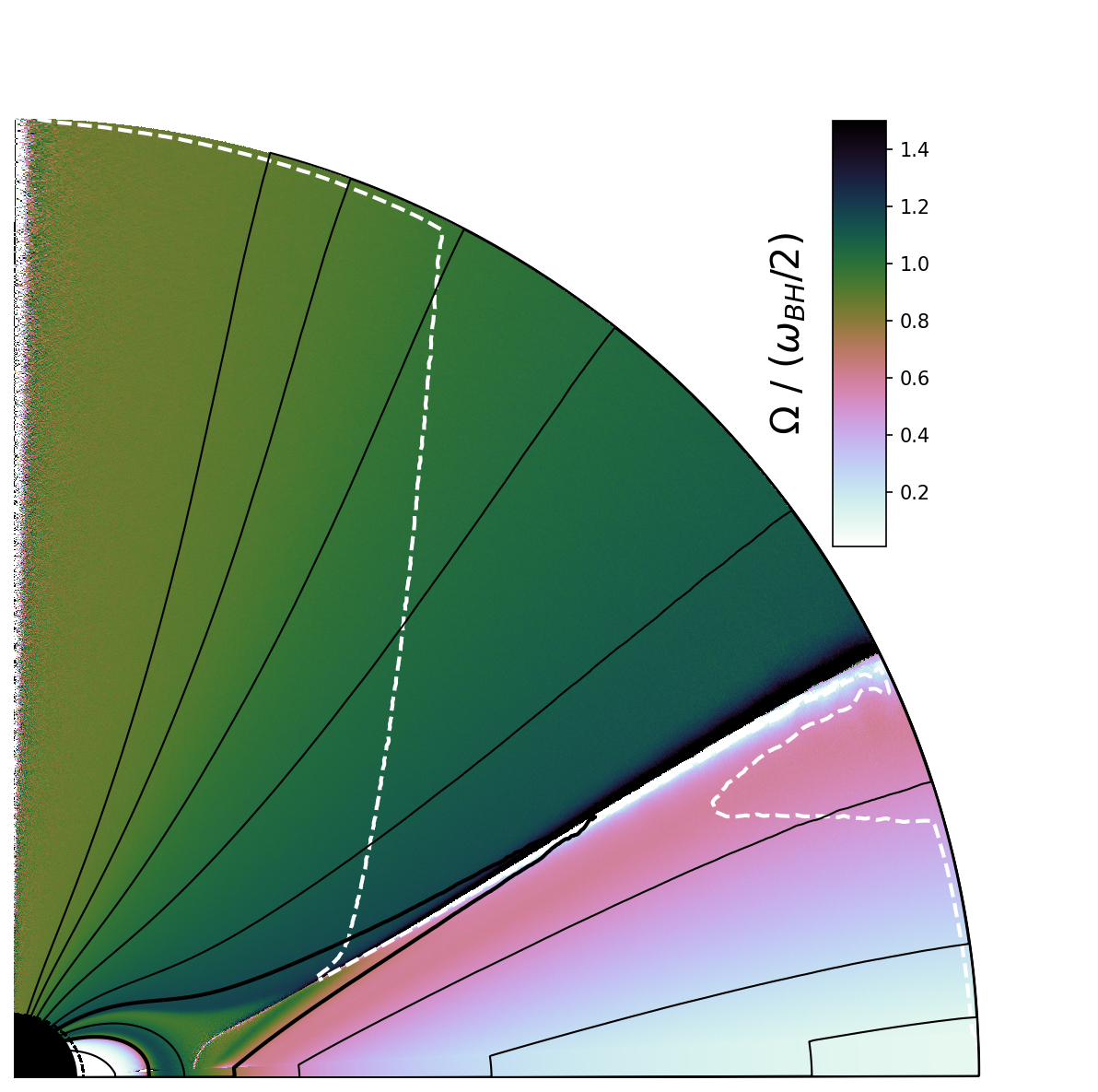}
\end{subfigure}
\begin{subfigure}[b]{0.99\columnwidth}
   \includegraphics[width=0.99\columnwidth]{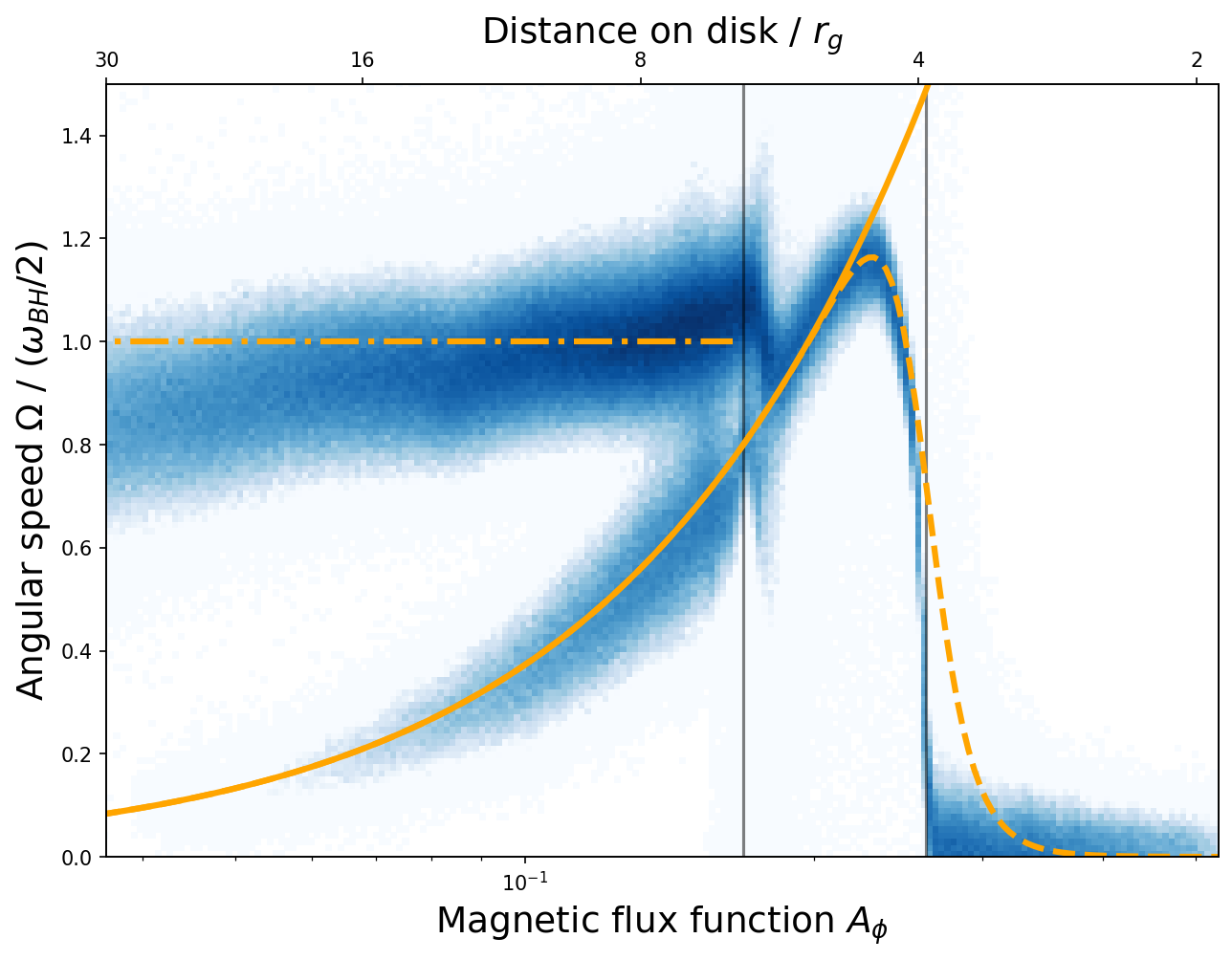}
\end{subfigure}
\caption{(top panel) Colormap of the time-averaged angular speed $\Omega$, with poloidal magnetic field lines overlaid, for a simulation with $a=0.6$ and $\epsilon=5\%$. The white dashed line is the outer light surface. (bottom panel) Angular speed of magnetic field lines as a function of their magnetic flux function $A_{\phi}$, with at the top the corresponding distance of the line footpoint on the disk. The color map is the histogram of the $(A_{\phi},\Omega)$ pairs measured in each cell of the simulation space at a fiducial time. The left (resp. right) vertical line stands for the separatrix footpoint (resp. the \isco). In orange are the Keplerian profile\,\eqref{eq:Om_K} without (solid) and with (dashed) \ileyk{the smooth cutoff}, and half of the \bh angular speed (dash-dotted).}
\label{fig:omega}
\end{figure}

Except in the current sheet, the electric field colinear to the magnetic field, \kyle{$\boldsymbol{D}\cdot\boldsymbol{B}/|\boldsymbol{B}|^2$}, remains marginally small, with values typically below a few 10$^{-4}$, which is indicative of a corona close to force-free. All over the grid, the initial conditions of high magnetization and plasma pair multiplicity above 1 described in Section\,\ref{sec:Parameters} are respected. The skin depth is also resolved everywhere. We plotted in Figure\,\ref{fig:omega} the histogram of $\left(A_{\phi},\Omega\right)$ values in each cell of the simulation space. The scale of the colormap is logarithmic, with a cutoff at its lower end fixed at 1\% of the maximum value of the histogram. The local angular speed $\Omega$ of the magnetic field lines is estimated from Equation\,\eqref{eq:omega}:
\begin{equation}
    \Omega=-\frac{E_{\theta}}{\sqrt{h}B^r}
\end{equation}
with $h$ the determinant of the spatial 3-metric. When overlaid on top of poloidal magnetic field lines, the $\Omega$ iso-contours match (see top panel in Figure\,\ref{fig:omega}), which indicates that each magnetic field line rotates at a fixed angular speed. They depart from each other only in the immediate vicinity of the current sheet, where $B^r$ is close to zero. The two vertical lines in the bottom panel in Figure\,\ref{fig:omega} stand for the outermost magnetic field line (left line) and for the magnetic field line anchored at the \isco (right line). The top $x$-axis indicates the footpoints on the disk of the magnetic field lines corresponding to a given magnetic potential $A_{\phi}$ according to Equation\,\eqref{eq:Aphi}, such as the distance to the \bh increases toward the left. At these footpoints, the local Keplerian angular speed profile\,\eqref{eq:Om_K} is enforced (orange solid line), with a cutoff at the \isco (orange dashed line). The field lines which remain anchored in the disk follow this profile, whether they are open (for $A_{\phi}<A_S$, lower branch) or coupled to the \bh (for $A_{\phi}>A_S$). The apparent fading at low $A_{\phi}$ and low $\Omega$ is due to the lower number of cells of the mesh at large distance from the \bh. The opening of the magnetic field lines for $A<A_{\phi}$ leads to the set of open field lines anchored in the \bh. They span the same range of $A_{\phi}$ values as their counterparts anchored in the disk but they rotate at an angular speed close to $\omega_H/2$, the optimal value for rotational energy extraction from the \bh according to the force-free Blandford-Znajek process (upper branch for $A<A_S$ in Figure\,\ref{fig:omega}).

\begin{figure}
\centering
\begin{subfigure}[b]{0.99\columnwidth}
   \includegraphics[width=0.99\columnwidth]{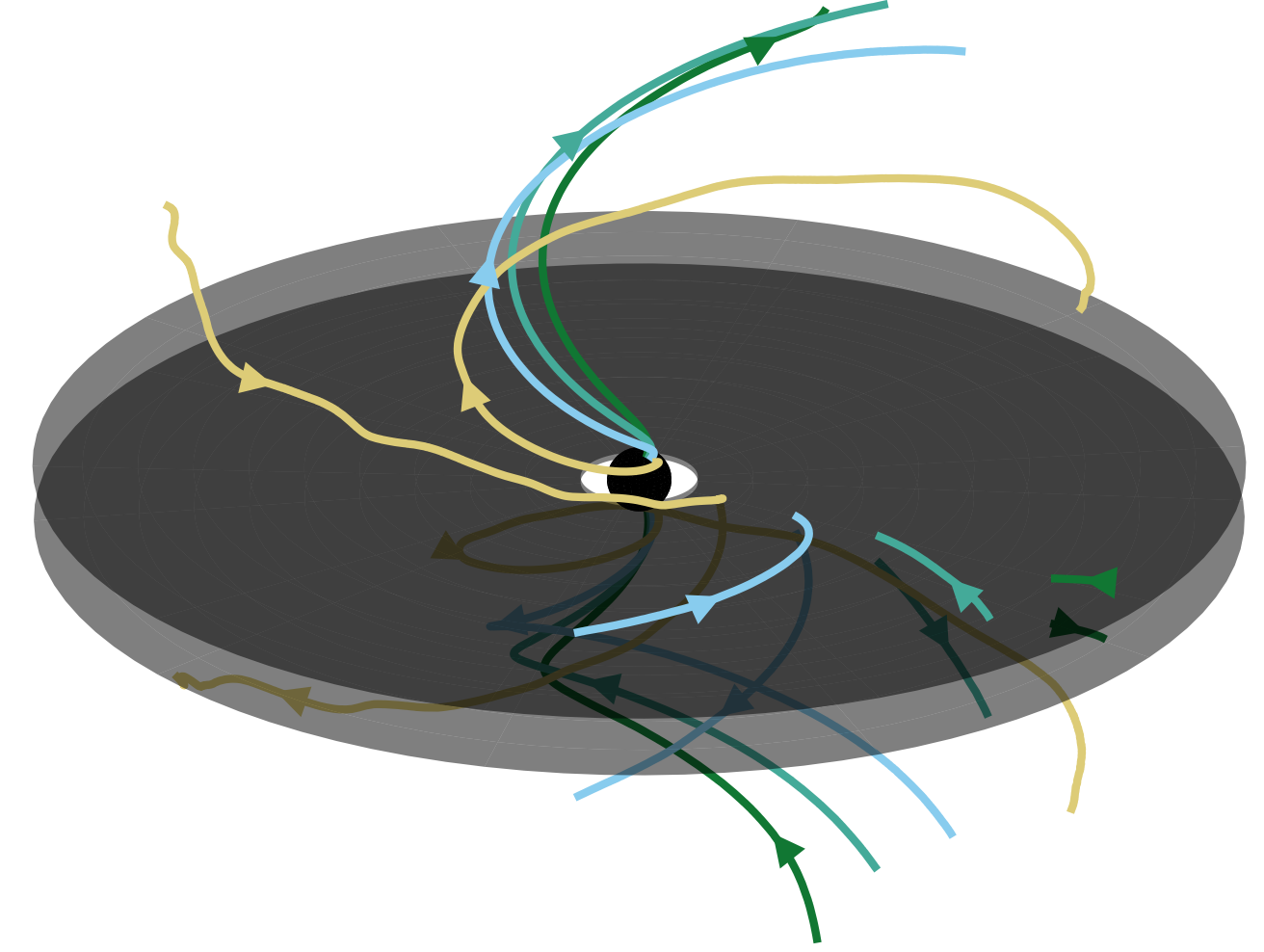}
\end{subfigure}
\begin{subfigure}[b]{0.99\columnwidth}
   \includegraphics[width=0.99\columnwidth]{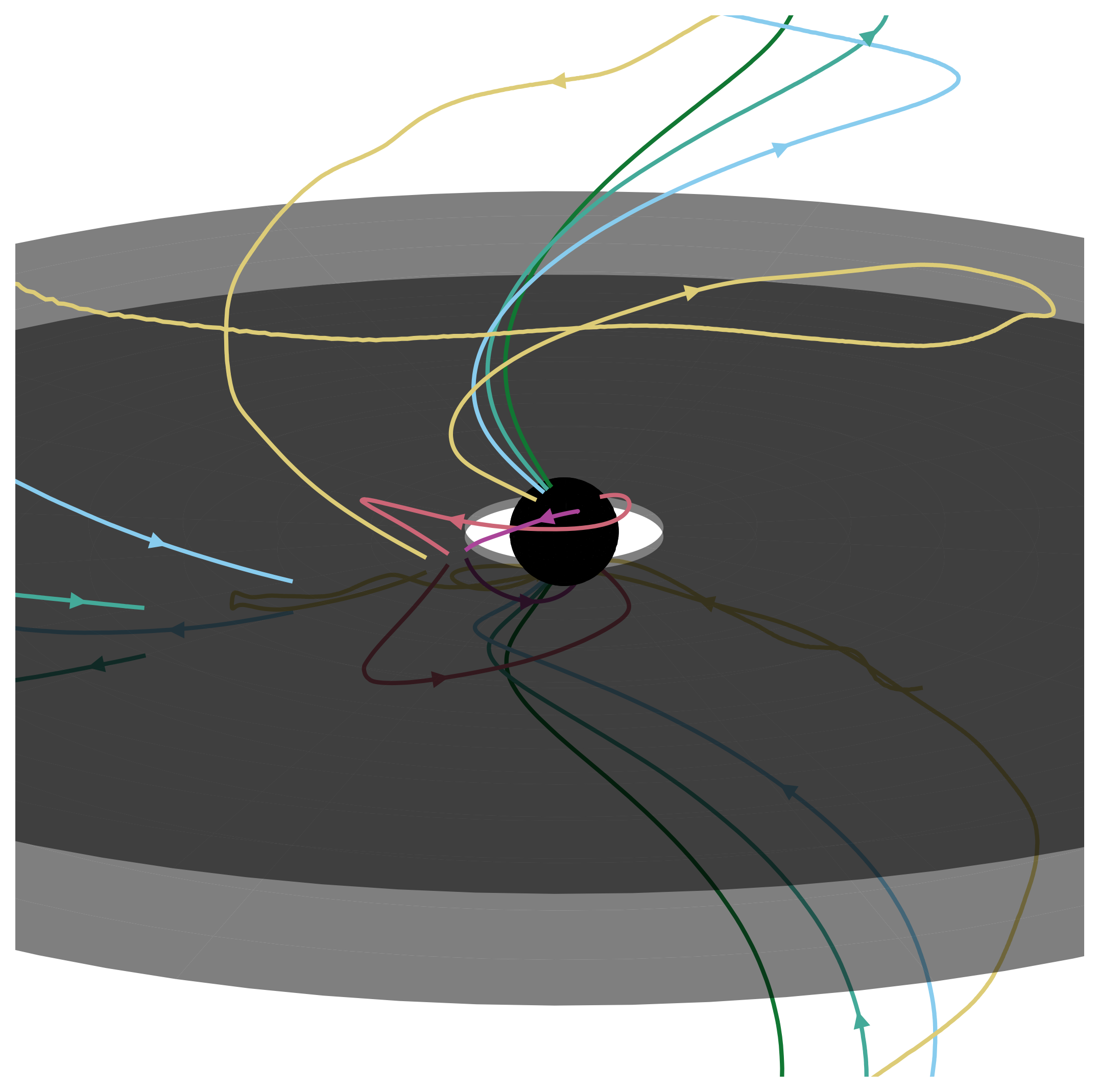}
\end{subfigure}
\caption{Three-dimensional representation of the magnetic field lines around the \bh (central black sphere), with the disk upper and lower surfaces in semi-transparent black. Field lines are color-coded from low magnetic potential (dark green) to high magnetic potential (magenta). Two additional field lines are represented in the zoom in the bottom panel.}
\label{fig:3D_B}
\end{figure}

Although the numerical grid we work with is two-dimensional, we use the axisymmetric assumption to provide a three-dimensional representation of the magnetic field lines in Figure\,\ref{fig:3D_B} for a \bh spin $a=0.8$ and a disk thickness $\epsilon=5$\%. The upper and lower surfaces of the disk are semi-transparent to make the field lines below the disk visible. They extend down to the \isco and the central black sphere is the event horizon of the \bh. The orientation of each field line is specified and their color stands for their magnetic potential $A_{\phi}$, from dark green for low values (\ie footpoint of the magnetic field line far on the disk at initial state) to magenta for high values (\ie footpoint near the \isco). Near the \bh spin axis, we find a set of open magnetic field lines anchored in the event horizon, reminiscent of the configuration considered by \cite{Blandford1977}. They correspond to the four twisted field lines threading the event horizon near the pole, visible in both panels in Figure\,\ref{fig:3D_B} (dark green, light green, blue and yellow field lines). We show in Section\,\ref{sec:BH_Opened_magnetic_field} that given their properties, these field lines do form the magnetic backbone of a relativistic collimated jet. In the outermost regions, above the disk, there are open magnetic field lines anchored in the disk. They formed after the closed magnetic field lines of the initial dipole splitted beyond the separatrix footpoint. A subset is visible in the upper and lower panels in Figure\,\ref{fig:3D_B} and they have the same colors as the field lines crossing the event horizon since they have the same magnetic potential $A_{\phi}<A_S$. Finally, the field lines with a magnetic potential between $A_S$ and $A_{\iscoo}$ remain anchored both in the \bh and in the disk, forming a closed region which couples these two components. In the bottom panel in Figure\,\ref{fig:3D_B}, they are the two innermost magnetic field line, represented in coral and magenta. In-between the two regions where the magnetic field lines are open, a current sheet forms. Magnetic reconnection kicks in and plasmoids form at the point where these two regions meet with the closed part of the magnetosphere, called the Y-point (red circle in the upper left panel in Figure\,\ref{fig:wheel}).

\ileyk{For a geometrically thick disk ($\epsilon=30\%$) such as the one which is believed to be present around \m{87}, the fundamental mechanism responsible remains effective. We carried out a simulation of a thick disk in prograde rotation around a \bh with spin $a=0.8$ and found qualitatively similar results, with the same regions of closed and open magnetic field lines, a Y-point at the end of the separatrix and a tearing unstable current sheet. The location of the footpoint of the separatrix at the surface of the disk does not change significantly with respect to simulations with a lower aspect ratio. The plasma density in the current sheet is more dilute and even if plasmoids still form, magnetic reconnection is less vivid with marginally lower rates of electromagnetic energy dissipated. Although these results suggest that our conclusions might hold in a thick disk around a supermassive \bh, it must be acknowledged that the biases introduced by the steady disk are enhanced in this configuration, where the disk has a larger spatial extent. \kyle{In this case, a magnetic diffusivity could be set in the disk as boundary conditions \citep{Parfrey2017b}. Alternatively, a fluid approach where kinetic effects are encapsulated in parametrized magnetic diffusivity coefficients would be insightful to derive a physically accurate magnetic structure in the disk \citep{Ripperda2020}. In what follows, we detail the results we obtain with thin disks only.}}

\subsubsection{Opened magnetic field lines anchored in the black hole}
\label{sec:BH_Opened_magnetic_field}

In the polar regions of the \bh, magnetic field lines thread the event horizon at their basis and form a flux tube around the spin axis. \kyle{In some previous studies devoted to magnetic configurations where the \bh is connected to a surrounding Keplerian disk, there were no open magnetic field lines threading the horizon (hereafter the jet launching region) because of numerical constrains \citep{Uzdensky2005,Yuan2019b}. On the contrary, we find that the jet launching region co-exists with magnetic field lines which couple the disk to the \bh.} Indeed, \cite{Uzdensky2005} rely on numerical relaxation techniques to find a steady solution of the relativistic Grad-Shafranov equation, with both the \bh spin $a$ and the footpoint of the separatrix $R_S$ as degrees-of-freedom \citep[see also][]{Uzdensky2002}. These methods require an initial guess for the magnetic potential $A_{\phi}$ but due to the lack of non-ideal effects in the force-free framework, the topology of the magnetic field is frozen as $A_{\phi}$ successively converges toward the solution. The regularity of the solution at the event horizon \citep{Nathanail2014} and the fact that the Grad-Shafranov equation is singular on the light surfaces drove \cite{Uzdensky2005}, \cite{Mahlmann2018a} and \cite{Yuan2019b} to consider solutions where a separatrix subdivides magnetic field lines in two sets: the open ones anchored in the disk, beyond the separatrix, and the ones coupling the \bh to the disk, with a footpoint on the disk closer than $R_S$. Our approach does not suffer from this limitation since we solve a more fundamental set of equations where magnetic reconnection is allowed and susceptible to modify the topology of the initially prescribed magnetic field. \kyle{In the intermediate approach of time-dependent force-free simulations, where the topology of the magnetic field can change due to numerical or physical dissipation, a hybrid closed-open magnetosphere is also found, although with different initial and boundary conditions \citep{Parfrey2015,Yuan2019c,Mahlmann2020}}.


\ileyk{The open magnetic field lines threading the horizon are dragged by the rotation of the \bh, which manifests as a twisting of the magnetic field lines in Figure\,\ref{fig:3D_B}. \ileyk{A significant toroidal component of the magnetic field on the grid, $H_{\phi}$, develops.} Since the magnetic moment of the initial dipole is aligned along the same direction as the angular momentum of the \bh, $H_{\phi}$ is positive (resp. negative) in the lower (resp. upper) hemisphere, as visible in the bottom left panel in Figure\,\ref{fig:wheel}. The upper branch for $A<A_S$ in Figure\,\ref{fig:omega} shows that most of these field lines rotate at an angular speed close to $\omega_H/2$ within 20\%. The field lines near the spin axis (\ie at low $A_{\phi}$) tend to rotate slower than $\omega_H/2$. It is probably an effect of the outer edge of the simulation space since these field lines are precisely those which do not cross the outer light surface inside the simulation space.}

We now comment on the magnetic flux and the electromagnetic power carried in this region. We represented these quantities measured for different \bh spin values in the middle panel in Figure\,\ref{fig:rdcd}. We compute the magnetic flux $\Phi_{jet}$ through a solid angle from the pole up to the separatrix. Since the magnetic field is divergence-free, we can compute this flux through a fraction of a sphere at any distance from the \bh. We average the values obtained over radii from $r_H$ up to the sphere which intersects the Y-point, in order to avoid the turbulent current sheet beyond. In Figure\,\ref{fig:rdcd}, we compare it to the magnetic flux threading the full upper hemisphere of the \bh at the event horizon, $\Phi_H$, given by its initial value for a magnetic dipole:
\begin{equation}
    \Phi_H=\oiint\boldsymbol{B}\cdot \boldsymbol{\diff \Sigma}=2\pi A_{\phi}\left(r_H,\pi /2 \right)=2\pi B_0 r_g^2 \frac{r_g}{r_H}
\end{equation}
The measured fraction $\Phi_{jet}/\Phi_H$ carried by the open magnetic field lines threading the \bh event horizon increases from 30\% at $a=0.6$ up to 45\% for $a=0.9$ and it plateaus beyond. In parallel, we estimate the power $P_{jet}$ carried by the jet by computing the flux of the Poynting vector as seen from an observer at infinity:
\begin{equation}
    \boldsymbol{S}=\frac{c}{4\pi}\boldsymbol{E}\times\boldsymbol{H}.
    \label{eq:poynting_flux}
\end{equation}
This flux hardly varies with the distance to the \bh, which indicates that electromagnetic energy is seldom dissipated in this region, in agreement with the force-free regime. We also checked that the flux of kinetic energy of the particles is negligible compared to the Poynting flux in this region: the basis of the jet is essentially devoid of mass and is controlled by the electromagnetic fields. The jet power quickly increases as the spin increases from 0.6 to 0.99. In order to provide an empirical characterization of this evolution, we compared $P_{jet}$ to a formula inspired by the results of \cite{Tchekhovskoy2011}:
\begin{equation}
    P_{th}=\frac{k}{4\pi}\frac{\omega_H^2}{c}\Phi_{jet}^2 f\left(\omega_H\right)
    \label{eq:jet}
\end{equation}
where $f$ is a correcting factor at large spin found by \cite{Tchekhovskoy2011} and given by:
\begin{equation}
    f ( \omega_H )=1+1.38\left(\frac{\omega_Hr_g}{c}\right)^2-9.2\left(\frac{\omega_Hr_g}{c}\right)^4
\end{equation}
The $k$ factor encompasses the information on the geometry of the magnetic field lines and does not depend on the \bh spin. For a split monopole (resp. a parabolic) geometry, $k\sim 0.053$ (resp. $k\sim 0.044$). Since we consider a different magnetic configuration, we reinjected the magnetic flux $\Phi_{jet}$ we measure into Equation\,\eqref{eq:jet} and fitted for the $k$ factor. We obtained the blue crosses in the middle panel in Figure\,\ref{fig:rdcd} \ileyk{corresponding to:
\begin{equation}
    k=0.072\pm 0.002 
\end{equation}}
We retrieve that the correcting factor $f$ is necessary to capture the dependence of $P_{jet}$ on $\Phi_H$ for spin values $a\geq$0.95.

\subsubsection{Opened magnetic field lines anchored in the disk}
\label{sec:disk_Opened_magnetic_field}

Among the magnetic field lines anchored in the Keplerian disk, those beyond the separatrix do not remain connected to the \bh and open up. The footpoint of the separatrix on the disk is located at a distance $R_S$ from the \bh which depends essentially on the \bh spin (and weakly on the disk thickness). The existence of this outermost closed magnetic field line at a finite distance around spinning \bhs was first highlighted by \cite{Uzdensky2005}. The key argument is that in an initial configuration where magnetic field lines all thread the disk and the event horizon, the field lines will slip along the horizon at different angular speed owing to the different locations of their footpoints on the Keplerian disk. This longitudinal shearing of the field lines induces a toroidal component. For $a>0$, near enough from the \bh pole, this component will induce a magnetic pressure too high to be confined by the tension of the polodial component at large distances. It leads to the opening up of the magnetic field lines we observe in the relaxation phase of our simulations.

\begin{figure}
\centering
\includegraphics[width=0.99\columnwidth]{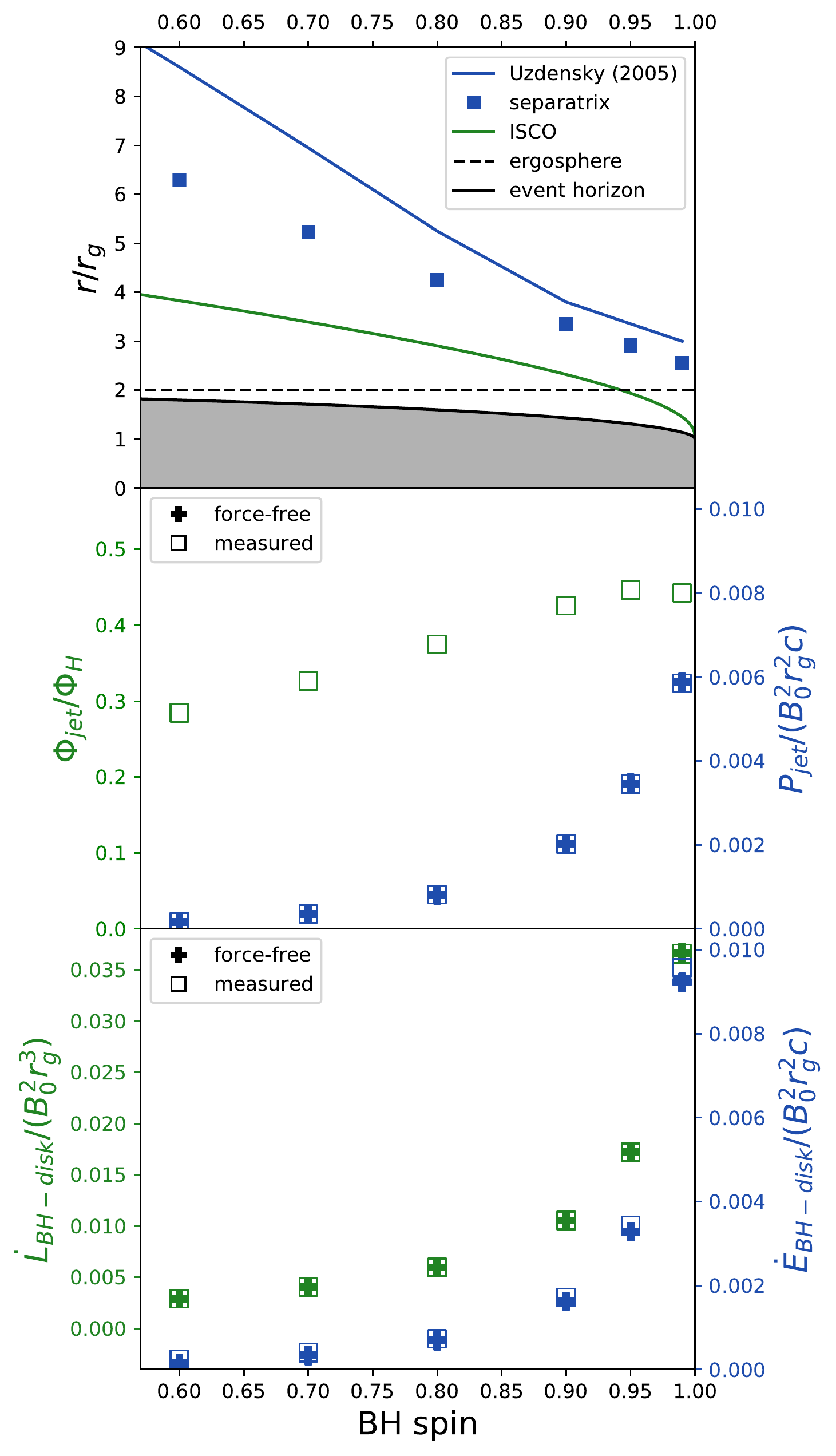}
\caption{(Top panel) \isco (green solid line), event horizon (black solid line) and ergosphere at the equator (dashed solid line) as a function of the \bh spin. The blue line indicates approximately the location of the footpoint of the separatrix found by \cite{Uzdensky2005} while the blue squares are the values measured in our simulations. (Middle panel) Magnetic flux $\Phi_{jet}$ of the open magnetic field lines threading the event horizon (green, left axis). In blue (right axis) are the jet electromagnetic power measured (squares) and predicted by \cite{Tchekhovskoy2011} (crosses). (Bottom panel) Angular momentum (left axis) and energy (right axis) transferred from the \bh to the disk per unit time, with and without the force-free approximation (crosses and squares).}
\label{fig:rdcd}
\end{figure} 

\cite{Uzdensky2005} estimated the maximal extent of the coupling part of the magnetosphere on the disk by looking for a steady solution of the relativistic Grad-Shafranov equation at a given \bh spin $a$ and location of the separatrix $R_S$. For each spin, \cite{Uzdensky2005} showed that there is a maximal value of $R_S$ beyond which the solver could not converge. We represented the upper limit \cite{Uzdensky2005} found in the upper panel in Figure\,\ref{fig:rdcd} (blue line). The grey shaded region corresponds to the region within the event horizon (black solid line). The green line is the inner edge of the Keplerian disk, set at the \isco, and lies within the ergosphere (black dashed line) for $a$ greater than approximately 0.95. We locate the separatrix by using the change in sign of the $H_{\phi}$ variable which happens in the current sheet where the magnetic field reconnects (see bottom left quarter in Figure\,\ref{fig:wheel}). Once the simulations have relaxed, the position of the footpoint remains stable within 5\%. \ileyk{The method we use yield robust estimates of $R_S$ which represent lower limits since transient closed magnetic field lines are observed up to a footpoint $\sim$20\% higher than $R_S$ (\eg in Figure\,\ref{fig:wheel} where a closed magnetic field line anchored at $\sim$4.7$r_g$ on the disk extends up to $\sim23r_g$).}

Finally, we notice that the open magnetic field lines anchored in the disk are very inclined, with a maximum inclination $i$ with respect to the disk plane approximately given by the inclination of the current sheet. It is visible both in Figure\,\ref{fig:wheel} and in the three-dimensional representation in Figure\,\ref{fig:3D_B}. As $a$ varies between 0.6 and 0.99, we find that $i$ remains approximately between 30$^{\circ}$ and 35$^{\circ}$. In the magneto-centrifugal model introduced by \cite{Blandford1982}, the maximum inclination angle to launch an outflow is 60$^{\circ}$ which is $>i$ for any $a\geq 0.6$. It means that matter from the disk is susceptible to slide out along the magnetic field lines and provide a possible source of matter to ensure force-free conditions in the corona. \cite{Yuan2019b} studied this effect in great detail and showed that when accounting for relativistic effects, the range of inclinations suitable to launch a disk wind is even wider, especially at higher spins.

\subsubsection{Coupling magnetic field lines}
\label{sec:Coupling_field_lines}

As the \bh spin increases, the separatrix moves closer to the \bh, and so does the \isco. The net effect is a smaller disk surface connected to the \bh at larger spins, but the connected region lies deeper in the gravitational potential and probes stronger magnetic fields. We now quantify $\dot{L}_{BH-disk}$ and $\dot{E}_{BH-disk}$, the amount of angular momentum and energy exchanged per unit time respectively. To do so, we use the flux $\boldsymbol{S}$ of energy density at infinity given by Equation\,\eqref{eq:poynting_flux} and the flux $\boldsymbol{S_L}$ of angular momentum \ileyk{in an axisymmetric configuration} given by \citep{Blandford1977,MacDonald1982,Komissarov2004a}:
\begin{equation}
    \boldsymbol{S_L}=\frac{1}{4\pi}\left(-E_{\phi}\boldsymbol{D}-H_{\phi}\boldsymbol{B}\right),
    \label{eq:ang_gen}
\end{equation}
where positive values denote transfers from the \bh to the disk. We integrate these fluxes over both sides of the disk, from the \isco up to the separatrix, to obtain $\dot{L}_{BH-disk}$ and $\dot{E}_{BH-disk}$, displayed as green and blue squares respectively in the bottom panel in Figure\,\ref{fig:rdcd}. We compare these values to what would be expected in a fully force-free configuration, where the infinitesimal angular momentum and energy exchanged per unit time along a magnetic flux tube delimited by field lines of magnetic potential $A_{\phi}$ and $A_{\phi}+\diff A_{\phi}$ would be given by \citep{Blandford1977}:
\begin{empheq}[]{align}
    \label{eq:ang_ff}
    &\diff \dot{L}_{ff}=H_{\phi}\diff A_{\phi}\\
    \label{eq:ene_ff}
    &\diff \dot{E}_{ff}=H_{\phi}\Omega\diff A_{\phi}.
\end{empheq}
For each spin, we integrate the quantities $H_{\phi}$ and $H_{\phi}\Omega$ over the magnetic potential $A_{\phi}$ on a wedge just above the disk, from the \isco to the separatrix, and obtain the green and blue crosses in the bottom panel in Figure\,\ref{fig:rdcd} for the exchange rates of angular momentum and energy respectively. 

The good match between the quantities obtained from the angular momentum and energy fluxes\,\eqref{eq:ang_gen} and \eqref{eq:poynting_flux} and the ones derived in the idealized force-free case with equations\,\eqref{eq:ang_ff} and \eqref{eq:ene_ff} indicate that only a small fraction of the energy funneled along the coupling magnetic field lines is dissipated between the \bh and the disk. Except in the immediate vicinity of the Y-point, the closed part of the magnetosphere is essentially force-free and little electromagnetic energy is dissipated along the magnetic field lines coupling the disk to the event horizon. 

We notice that $\dot{L}$ is positive for any $a\geq0.6$ which can be understood in terms of co-rotation radii. At $a\sim 0.36$, the co-rotation radius is approximately located at the \isco so for higher spins, the angular speed from the \isco to the separatrix is lower than the \bh angular speed $\omega_H$. Thus, the coupling magnetic field lines can only spin down the \bh and transfer angular momentum to the disk, which is consistent with the positive values we measure. The rate at which rotational energy is extracted from the \bh and deposited onto the disk is comparable to the jet power for $a\geq0.6$.


\subsubsection{Current sheet}
\label{sec:Toroidal_fields}


\begin{figure}
\centering
\includegraphics[width=0.99\columnwidth]{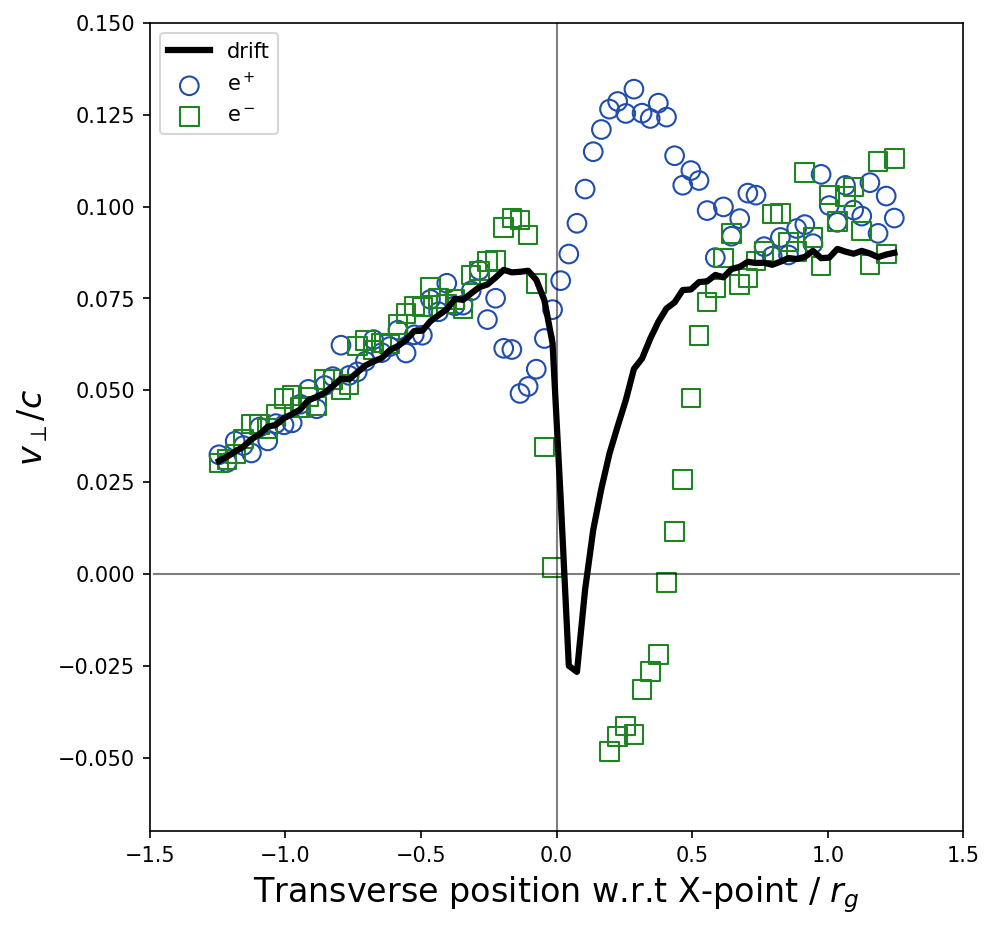}
\caption{Profiles along a line transverse to the current sheet and passing through an X-point. The components of the drift velocity (solid black line) and of the positron and electron velocities (resp. blue circles and green squares) normal to the current sheet are represented.}
\label{fig:cs}
\end{figure} 


\begin{figure*}[!b]
\centering
\includegraphics[width=1.9\columnwidth]{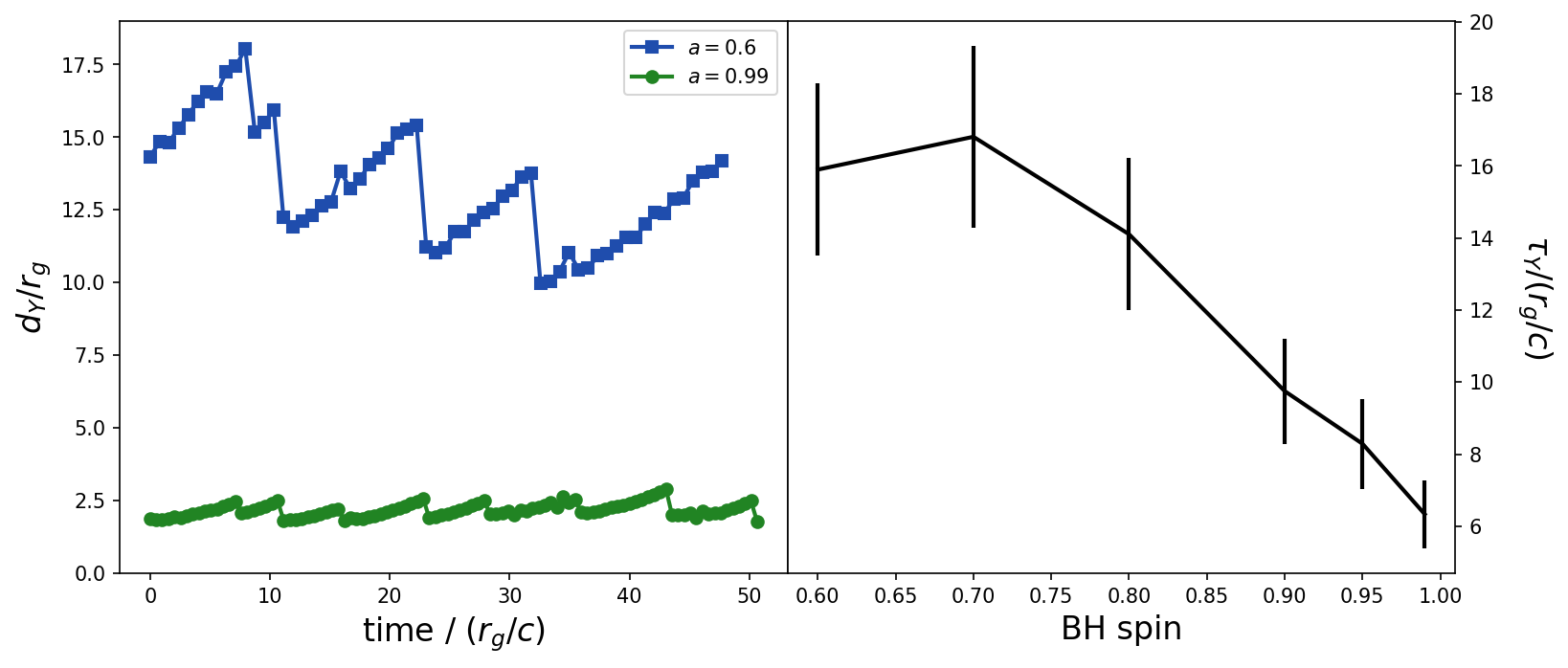}
\caption{(Left panel) Distance $d_Y$ between the Y-point and the footpoint of the separatrix, as a function of time, for different \bh spin values. The origin of time is arbitrary but over these time lapses, the simulations have reached a numerically relaxed state. (Right panel) Estimates of the time periodicity $\tau_Y$ of the motion of the Y-point, as a function of the \bh spin.}
\label{fig:dY}
\end{figure*}

For any \bh spin $a\geq 0.6$ and for both disk thickness $\epsilon=5$\% and $\epsilon=30$\%, a current sheet forms between the two regions of opened magnetic field lines. As mentioned in Section\,\ref{sec:disk_Opened_magnetic_field}, the $H_{\phi}$ component displays a clear change of sign at the current sheet, visible in the bottom left quarters in Figure\,\ref{fig:wheel}. It is in agreement with the toroidal motion of the magnetic field lines which rotate with the \bh and with the prograde disk (see Figure\,\ref{fig:3D_B}). Due to the opening of the magnetic field lines, the polarity reverses at the current sheet and \ileyk{the rotational drag of the field lines by the ergoregion or by the disk yields} a toroidal component of $\boldsymbol{H}$ of opposite sign on both sides of the current sheet. In the global steady state we reach, Maxwell-Amp\`ere's equation shows that a jump in $H_{\phi}$ along the longitudinal direction corresponds to a local peak in $J^r$. \ileyk{The current sheet hosts a strong current flowing away from the \bh visible in the bottom right quarters in Figure\,\ref{fig:wheel}}. This positive current is essentially carried by positrons and the associated electric circuit closes in the disk. Magnetic reconnection takes place in this current sheet, mostly at the Y-point but also further along the sheet, at local X-points. With the magnetic field amplitude we can reach (see Section\,\ref{sec:Parameters}), the sheet is resolved with 3 to 10 grid points and has a typical thickness $\delta\sim 5\cdot 10^{-2} r_g$, thin enough to be tearing unstable \citep{Zenitani2007}. The current sheet breaks up which leads to the formation of a chain of magnetic islands where particles gather into plasmoids \citep{Loureiro2007}, visible in the upper left quarters in Figure\,\ref{fig:wheel}. The distance between plasmoids vary with the \bh spin. 


\benjamin{We now estimate the reconnection rate by looking at the physical properties of the upstream flow near the current sheet. In Figure\,\ref{fig:cs}, we report the transverse profiles of the components of the positrons and electrons velocities normal to the current sheet. In a simulation with an intermediate \bh spin ($a=0.8$) and a thin disk ($\epsilon=5$\%), we worked at the innermost X-point on the current sheet, formed just after a plasmoid detached from the Y-point. The box around the current sheet we consider is 3$r_g$ long (x-axis in Figure\,\ref{fig:cs}), centered on the X-point and 0.5$r_g$ wide so as to work with averaged quantities and reduce the numerical noise. \ileyk{Figure\,\ref{fig:cs} displays the poloidal component of the particle velocity normal to the current sheet, $v_{\perp}$, with $v_{\perp}>0$ (resp. $v_{\perp}<0$) if particles are above the current sheet and move towards it or if they are below the current sheet and move away from it (resp. if they are above the current sheet and move away from it or if they are below the current sheet and move towards it).} The current sheet is approximately along a radius ($\theta\sim\text{constant}$) but since the plasma tends to flow along the magnetic field lines, $v^{r}\gg v^{\theta}$ in general and both velocity components contribute to $v_{\perp}$. The exact shape of the $v_{\perp}$ profiles depends on the inclination angle of the current sheet we use to perform the projection but the glitches at the X-point (\ie at the vertical grey line) subsist for any reasonable inclination. The profiles have a net non-zero offset which corresponds to a bulk instantaneous transverse motion of the current sheet, possibly due to the drift kink instability \citep{Zenitani2007,Barkov2016}. Once corrected for this offset, a change of sign at the X-point is measured with a step $\Delta v_{\perp}/c\sim5$\% for positrons and $\Delta v_{\perp}/c\sim15$\% for electrons. In the highly magnetized regime where we work, $\sigma\gg 1$ upstream, the Alfven speed is comparable to the speed of light and the reconnection rate $\beta_{rec}$ is given by:\ileyk{ 
\begin{equation}
    \beta_{rec}\sim\frac{\Delta v_{\perp}/2}{c}\sim 5\%.
    \label{eq:rec_rate}
\end{equation}}
An alternative way to compute the reconnection rate relies on the electromagnetic fields and the associated drift velocity $\boldsymbol{v_d}$ defined as:
\begin{equation}
    \frac{\boldsymbol{v_d}}{c}=\frac{\boldsymbol{E}\times\boldsymbol{B}}{B^2},
    \label{eq:drift_vel}
\end{equation}
and $\Gamma=|\boldsymbol{v_d}|/c$ is the Lorentz factor of the plasma bulk motion. In Figure\,\ref{fig:cs}, we show the transverse profile of the component of $\boldsymbol{v_d}$ normal to the current sheet component of the drift speed (solid black line). It is coherent with the particle velocity profiles and the step in velocity is intermediate between the values obtained from the positron and electron profiles. We compute $\beta_{rec}\sim 5\%$ which is similar to what \cite{Crinquand2021} measured in PIC simulations of magnetic reconnection in the equatorial plane \ileyk{near the event horizon}.}

These values are of the same order of magnitude as what was computed in PIC simulations of magnetic reconnection in collisionless pair plasma \citep{Kagan2015,Sironi2014,Werner2018}. They are approximately an order of magnitude higher than the ones derived from GR-MHD simulations of \bh magnetosphere where the finite volume approach induces a numerical diffusivity at small scales \citep[see][for a comparison of magnetic reconnection in PIC and GR-MHD simulations]{Bransgrove2021}. Furthermore, it must be noticed that this reconnection configuration presents some specific features which differ from idealized situations. First, magnetic reconnection is asymmetric here since the plasma properties on the disk and jet sides (\eg density, velocity and multiplicity) differ \citep[see][for recent simulations of asymmetric reconnection in the relativistic regime]{Mbarek2021}. A persistent signature of this asymmetry manifests in the left-right asymmetry of the transverse profiles in Figure\,\ref{fig:cs}, where the profiles significantly depart from a simple step function. In the poloidal plane, the bulk motion of the plasma is faster above than below the current sheet (see Figure\,\ref{fig:cs}). This shear of the plasmoids can be seen in the top left quarters in Figure\,\ref{fig:wheel} and can trigger Kelvin-Helmholtz instability and enhance the fraction of electromagnetic energy dissipated in the process \citep{Sironi2021}. Second, the angular speed of the magnetic field lines threading the plasma varies, with magnetic field lines above the current sheet rotating faster since they cross the event horizon (see Figure\,\ref{fig:omega}). It creates an additional shear in the toroidal direction, in addition to the strong twisting of the magnetic field lines visible in Figure\,\ref{fig:3D_B}.

\subsection{Particle acceleration}
\label{sec:Particle_acceleration}

\subsubsection{Y-point}
\label{sec:Y-point}

At the furthest point on the separatrix, the outermost closed magnetic field line, the current sheet connects to the closed magnetosphere, forming a Y-point. Below, most of the plasma flows to the disk at relativistic speed (see upper right insert of bottom panel in Figure\,\ref{fig:wheel}). It hits the disk at the point where $r\sim R_S$, where the separatrix is anchored. The Y-point is located near the outermost light surface, as visible in the upper right quarter of the upper panel in Figure\,\ref{fig:wheel}. It moves back and forth along the current sheet as plasmoids form and flow away. In the left panel in Figure\,\ref{fig:dY}, we show the distance $d_Y$ between the Y-point and the footpoint of the separatrix it belongs to as a function of time, for $a=0.6$ (blue squares) and $a=0.99$ (green circles). It shows how the separatrix progressively stretches (\ie when $d_Y$ increases) until magnetic reconnection occurs at a point lower on the separatrix and a plasmoid detaches. This lower point becomes the new Y-point, which corresponds to the sharp decreases in the left panel in Figure\,\ref{fig:dY}, and the separatrix stretches again, closing the loop. We can see that for higher \bh spin values, the extent of this stretching is more limited and the Y-point spans a smaller region. A higher spin also means more frequent plasmoid formation and shorter time intervals $\tau_Y$ between successive tearing of the separatrix near the Y-point.

\begin{figure}
\centering
\includegraphics[width=0.99\columnwidth]{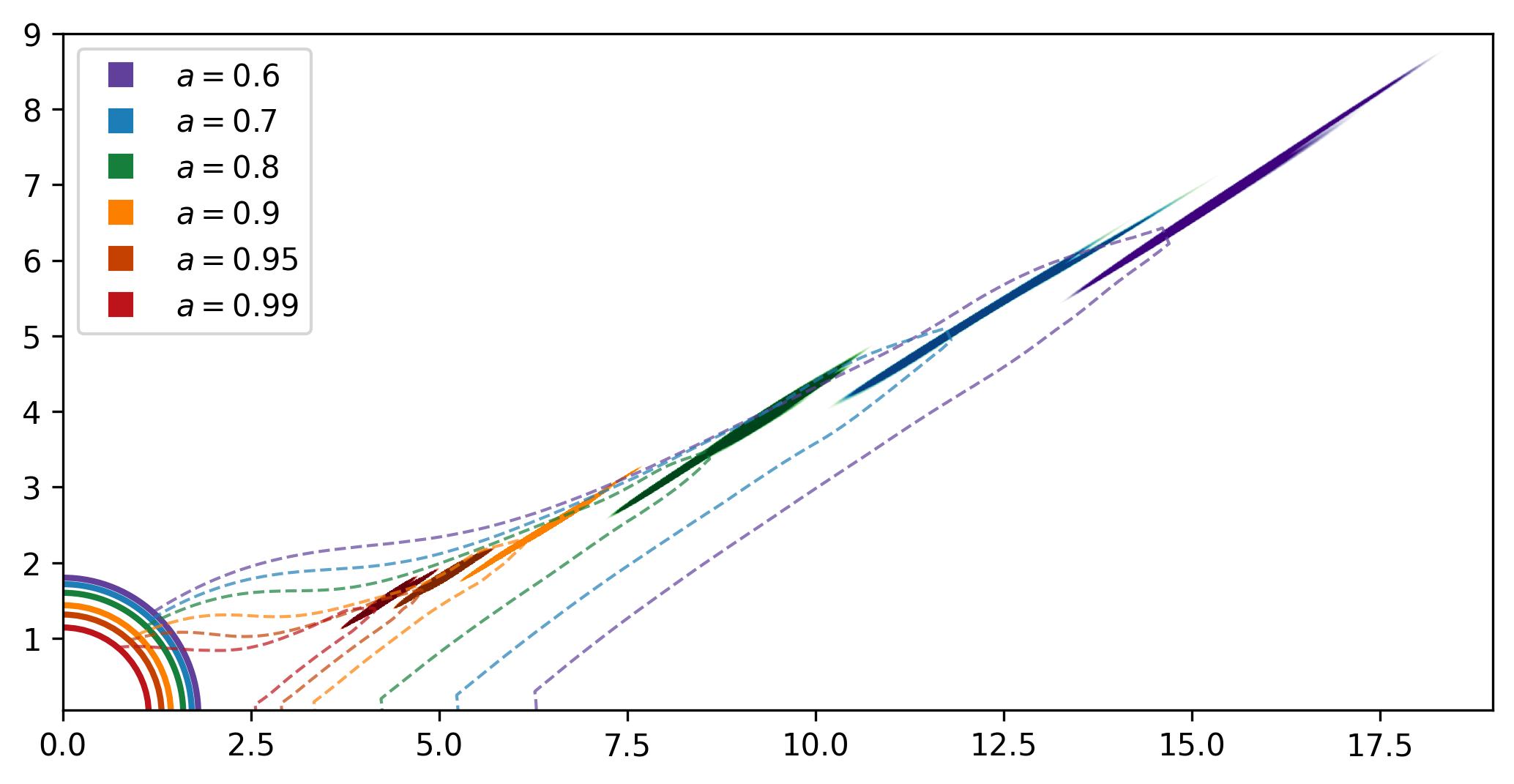}
\caption{Density maps of the locations of the Y-point for different \bh spin values. The inner solid lines correspond to the event horizons and the dashed lines are the separatrix. Distances to the \bh are given in $r_g$.}
\label{fig:Y_location}
\end{figure}

In the right panel in Figure\,\ref{fig:dY}, we represented estimates of $\tau_Y$ as a function of the \bh spin, with error bars. It illustrates the aforementioned trend, with $\tau_Y$ decreasing from $\sim 9 r_g/c$ to less than 5$r_g/c$ for $a$ increasing from 0.6 to 0.99. These values are related to the reconnection rate $\beta_{rec}$ measured in Section\,\ref{sec:Toroidal_fields} using the upstream particle speed and the electromagnetic fields near an X-point in the current sheet. Indeed, the drops in $d_Y$ happen each time a new magnetic island is formed, which depends on how fast the magnetic field lines reconnect. The most unstable mode of the tearing instability corresponds to a spacing $L$ between X-points of $2\pi\sqrt{3}\delta$ \citep{Zenitani2007}. Consequently, if the tearing instability controls the rate of formation of plasmoids at the Y-point, we can estimate a reconnection rate $\beta_{rec,Y}$ as:
\begin{equation}
    \beta_{rec,Y}=\frac{L}{c\tau_Y}
\end{equation}
which ranges between 6 and 11\% for $\tau_Y$ from 9 to 4.8$r_g/c$. The good match with the order of magnitudes derived in Section\,\ref{sec:Toroidal_fields} suggests that the tearing instability is the dominant mechanism at play near the Y-point. \ileyk{Also, the decreasing $\tau_Y$ we measure as the \bh spin increases can be interpreted in the light of the spatial extent of the closed magnetosphere: as the spin increases, the Y-point moves closer to the \bh and the distance $L$ covered by the Y-point as the line stretches shrinks, as visible in Figure\,\ref{fig:Y_location}. In the highly magnetized regime we work in, the reconnection rate hardly varies \citep[see \eg][]{Werner2018} and the characteristic time scale $\tau_Y$ between the formation of two plasmoids at the Y-point thus shrinks. }

\begin{figure}
\centering
\includegraphics[width=0.99\columnwidth]{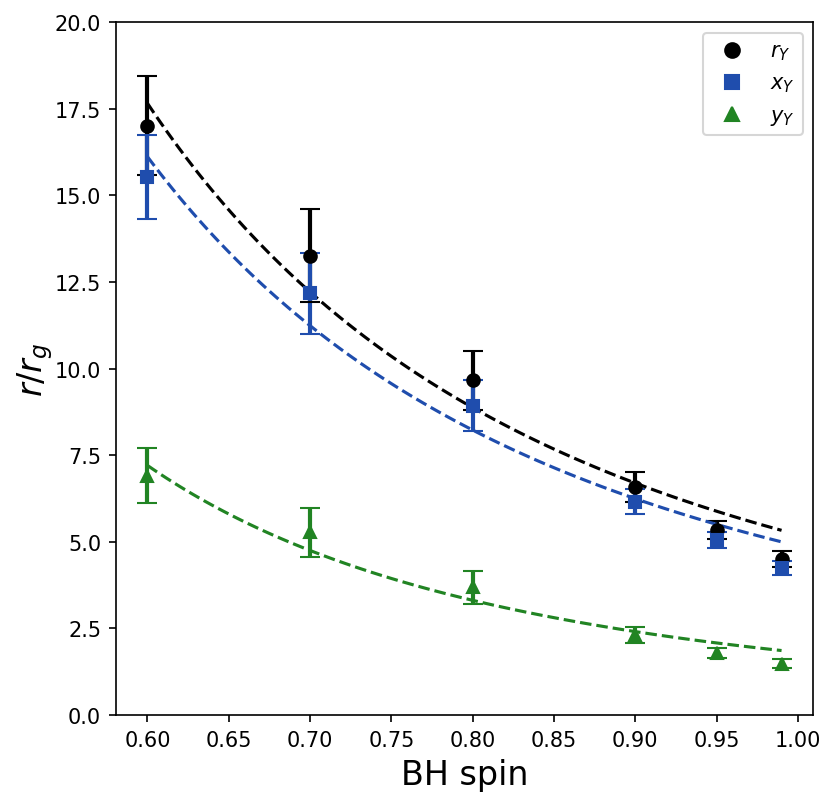}
\caption{Coordinates $(x_Y,y_Y)$ and distance $r_Y$ to the \bh center of the Y-point as a function of the \bh spin. The dashed lines are best fit power laws (see Appendix\,\ref{app:best_fit}).}
\label{fig:rY}
\end{figure}

\ileyk{In order to better appreciate the typical coordinates of the Y-point above the disk, we represented in Figure\,\ref{fig:Y_location} the spatial occurrence rate of the Y-point for different spins from 0.6 (purple) to 0.99 (red). The dashed lines stand for the time-averaged separatrix and the solid lines for the event horizon. We retrieve that the Y-point is closer from the separatrix footpoint for higher \bh spins and that it moves within a smaller region. Figure\,\ref{fig:rY} shows how the height above the disk $y_Y$ and the projected distance to the \bh $x_Y$ of the Y-point decrease when the spin increases (resp. green and blue points), along with the distance $r_Y$ to the \bh (black points). The error bars represent the 1$\sigma$-variation of these quantities around their median value and the dashed lines are the best fit obtained based on power laws (see Appendix\,\ref{app:best_fit}).} 

In our axisymmetric framework, this Y-point unfolds into a 3D ring above and below the disk. Once the quasi-periodic motion of the Y-point along the current sheet is accounted for, it manifests in 3D maps integrated over durations $>\tau_Y$ as a section of a cone between two planes normal to the \bh spin axis. In Section\,\ref{sec:Synthetic_images}, we compute the images of the synchrotron emission of the particles. We discuss in Section\,\ref{sec:discussion} the possible links of this geometry with the lamppost model \citep{Ross2005} and compare the physical values of $\tau_Y$ to the typical duration between flares in \cyg, \sgr and \m{87}. 

\begin{figure*}
\centering
\begin{subfigure}[b]{0.9\columnwidth}
   \includegraphics[width=0.99\columnwidth,height=13cm]{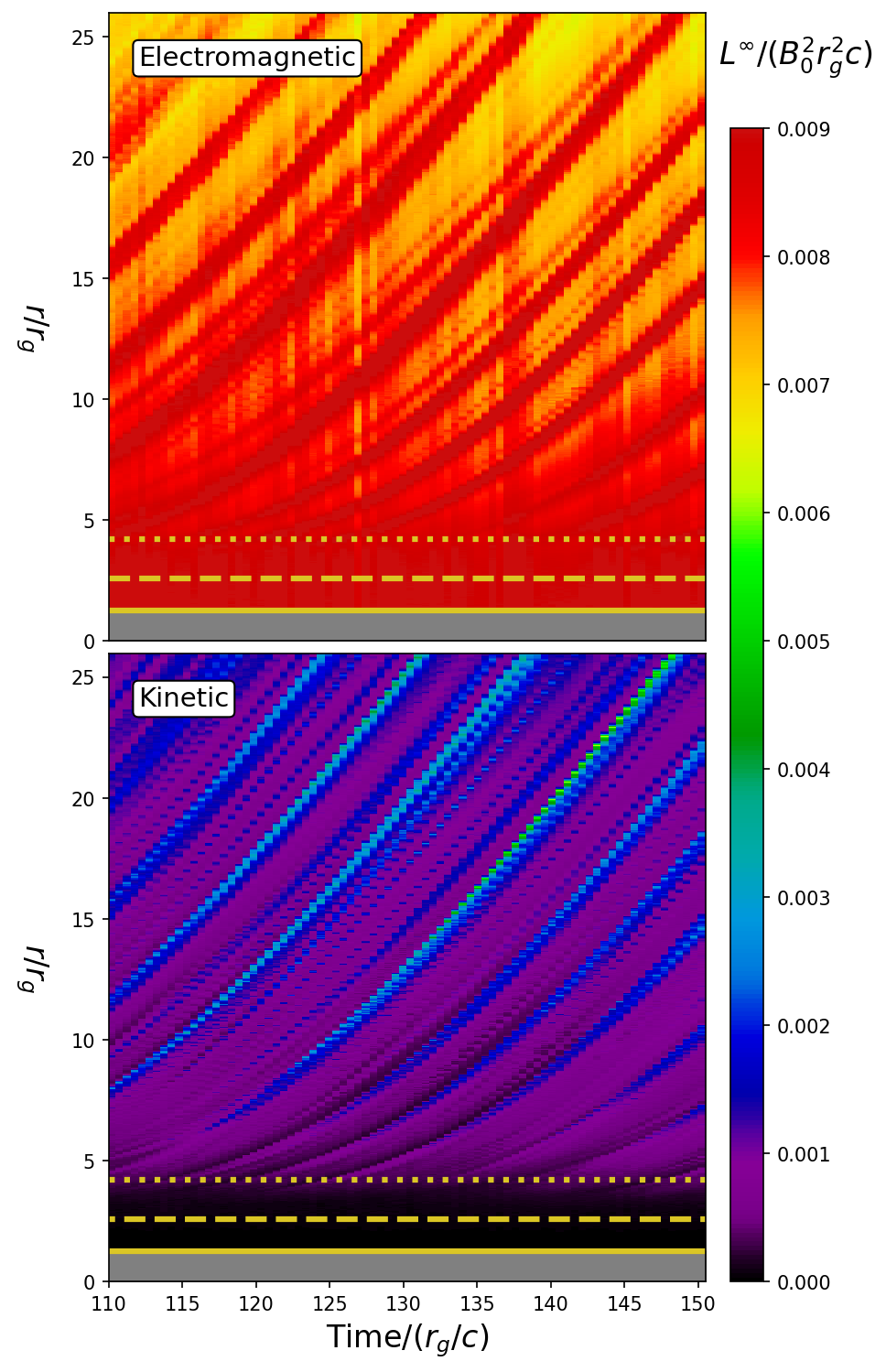}
\end{subfigure}
\begin{subfigure}[b]{1.04\columnwidth}
   \includegraphics[width=0.99\columnwidth,height=13cm]{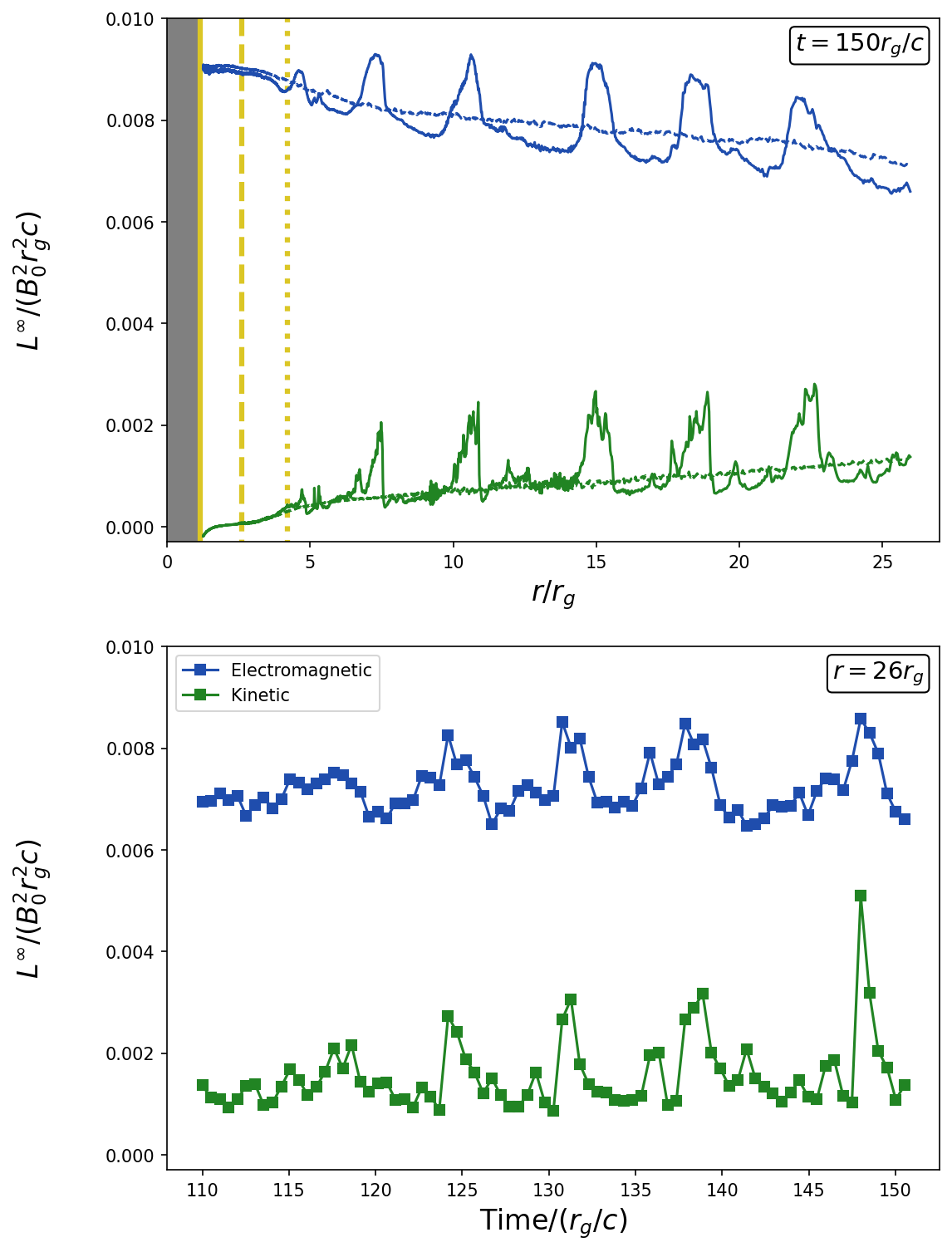}
\end{subfigure}
\caption{(left column) Electromagnetic (top panel) and particle kinetic (bottom panel) energy flux through a surface extending up to a radius $r$ (left axis), as a function of time and for a thin disk with a \bh spin $a=0.99$. (right column) Slices at constant time $t=150r_g/c$ (top panel) and constant radius $r=26r_g$ (bottom panel) of the fluxes. The dashed lines in the upper panel are the time-averaged profiles. The solid, dashed and dotted yellow lines show the event horizon, the distance to the separatrix footpoint and to the Y-point respectively.}
\label{fig:energy_budget}
\end{figure*}

\subsubsection{Energy budget}
\label{sec:Energy_fluxes}

Magnetic reconnection at the Y-point and beyond emphasizes the importance of particle acceleration in these simulations. For the most dynamic simulation, where $a=0.99$, we monitor the evolution of the fluxes of electromagnetic and particle kinetic energy, respectively given by:
\begin{equation}
    L_{EM}^{\infty}=\iint_{\Sigma} \boldsymbol{S} \cdot \diff \boldsymbol{\Sigma}
\end{equation}
and
\begin{equation}
    L_{K}^{\infty}=\iint_{\Sigma} n e_{\infty} \boldsymbol{v} \cdot \diff \boldsymbol{\Sigma}
\end{equation}
with $e_{\infty}=-u_t$ the energy at infinity of the electrons and positrons. Since there are non-zero longitudinal components of $\boldsymbol{S}$ and $\boldsymbol{v}$ on the wedge surface just above the disk, we need to consider an integration surface $\Sigma$ which covers two sides in order to account for (i) the fluxes through a fraction of a sphere at constant $r$, \ileyk{from the pole up to $\theta_{d}$, and (ii) the fluxes through the disk surface at constant $\theta=\theta_{d}$ from the event horizon up to a given $r$. The fluxes through the pole tend to zero by symmetry and due to the vanishing surface element. We compute the sum of the fluxes through these 2 surfaces for each $r$ from $r_h$ up to $r=26r_g$,} and for each time snapshot between $t=110r_g/c$ and $t=150r_g/c$, once the initial conditions have relaxed. The results are presented in Figure\,\ref{fig:energy_budget} which illustrates how the electromagnetic energy is progressively transferred to the particles.

The space-time diagrams in the left column show these fluxes for $L_{EM}^{\infty}$ (top panel) and $L_{K}^{\infty}$ (bottom panel). For the sake of visualization, we provide two slices in the right column: the radial profiles of the fluxes at $t=150r_g/c$ (upper right panel) and the fluxes through the integration surface $\Sigma$ at $r=26r_g$ as a function of time (lower right panel). In the upper right panel, we also display with dashed lines the time-averaged radial profiles in order to smooth out the impact of the plasmoids. The solid, dashed and dotted yellow lines are the event horizon, the distance to the separatrix footpoint and the distance to the Y-point respectively. Near the event horizon, the electromagnetic flux is of $\sim 9\cdot 10^{-3}$ $B^2_0 r^2_g c$, which corresponds approximately to half\footnote{Since we computed fluxes only in the upper hemisphere.} of the sum between the jet power and the amount of electromagnetic energy deposited by the coupling magnetic field lines per unit time (see Figure\,\ref{fig:rdcd}). Since the latter fraction is accounted for due to the integration surface we consider, there is no significant loss of electromagnetic energy before we reach the separatrix footpoint (dashed line). Then, the particles falling onto the disk from the Y-point along the separatrix experience a net acceleration due to the electric field parallel to their motion. It explains the slight decrease (resp. increase) of the electromagnetic flux (resp. the particle kinetic flux) visible in the radial profiles between the separatrix footpoint (yellow dashed line) and the Y-point (yellow dotted line). Yet, the main dissipation occurs in the current sheet beyond the Y-point. The net conversion of electromagnetic energy into particle kinetic energy manifests as a steady decrease in the time-averaged radial profiles. Interestingly enough, we find that in the plasmoids, the electromagnetic and kinetic fluxes are locally enhanced due to the higher plasma densities, currents and electromagnetic energy they carry. The motion of the plasmoids propagating outwards can be followed in the space-time diagrams. They produce stripes with an initially low bulk Lorentz factor, which progressively increases up to $\Gamma\sim 2$ (\ie $\beta\sim 0.75$), corresponding to the asymptotic slope of the stripes. In the lower left panel, we see the impact on the plasmoids crossing the $r=26r_g$ surface and leading to transient increases of the fluxes over a few $r_g/c$.

It must be noticed that in the immediate vicinity of the event horizon, we measure negative net fluxes of particle energy at infinity. It is due to the process first described by \citet{Penrose1969,Parfrey2019} which enables to extract energy from a Kerr \bh when particles of negative energy at infinity cross its event horizon. The effect is largely sub-dominant though, with $|L_{K}^{\infty}|$ approaching only a tenth to a fifth of its value at $r=26r_g$. Most of the energy provided to the particles comes from the electromagnetic field.


\subsubsection{Coronal heating}
\label{sec:Coronal_heating}

\begin{figure}
\centering
\includegraphics[width=0.99\columnwidth]{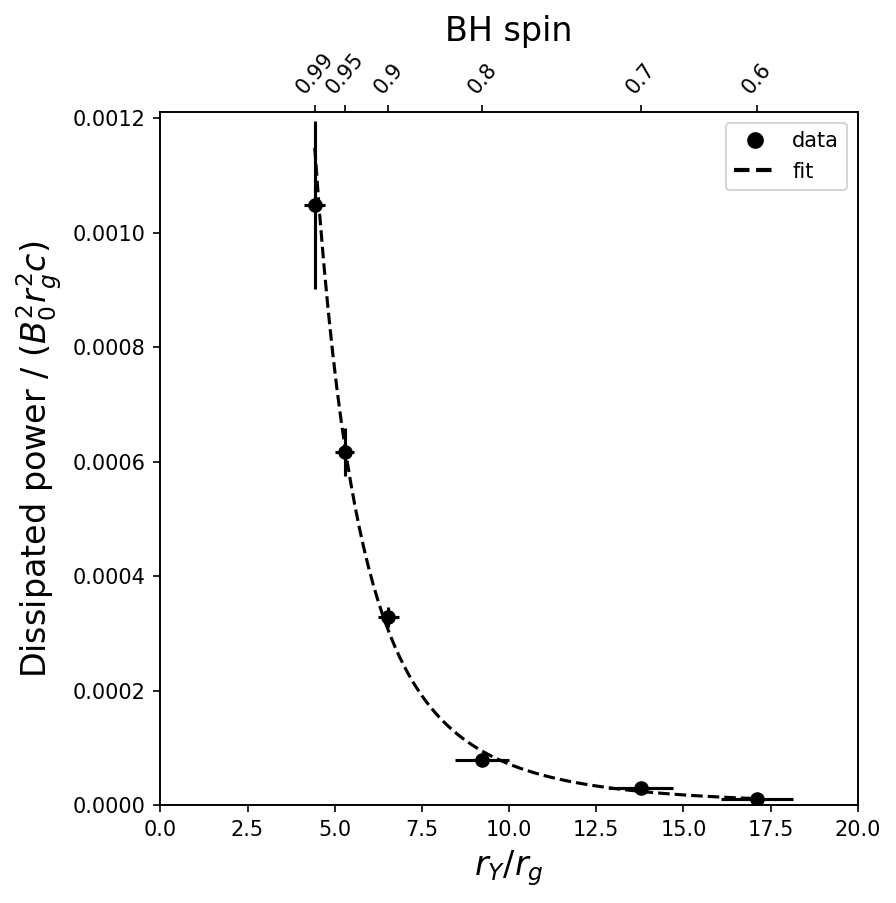}
\caption{Amount of electromagnetic energy dissipated per unit time in the current sheet, as a function of the distance of the Y-point to the \bh (bottom $x-$axis). The top $x-$axis indicates the corresponding \bh spin (analog to black points in Figure\,\ref{fig:rY}). The dashed line stands for the best semi-analytic fit (see text).}
\label{fig:dissipation}
\end{figure}

Magnetic reconnection has long be thought to be the main culprit for the heating of the corona above the disk \ileyk{\citep{Galeev1979}}. We thus examine the amount of electromagnetic energy dissipated by magnetic reconnection in the current sheet, from the separatrix footpoint up to $r=26r_g$. We isolate a narrow stripe of $1r_g$ thick around the current sheet and compute the volume integral of $\boldsymbol{J}\cdot\boldsymbol{E}$ within this stripe. While in force-free regions this quantity is negligible, it becomes significant in the current sheet and leads to the dissipated powers shown in Figure\,\ref{fig:dissipation}. With the intent of providing a semi-analytic interpretation of these values, we confront them to the distance $r_Y$ between the \bh and the Y-point previously given as a function of \bh spin in Figure\,\ref{fig:rY}. All the error bars show the $1\sigma$-variability of the computed values from a snapshot to another. For information, we plot the corresponding \bh spins on the top of the figure.

All dissipated powers are positive which means that overall, electromagnetic energy is tapped to accelerate particles \ileyk{(as visible also from the trend in the upper right panel in Figure\,\ref{fig:energy_budget})}. In order to understand the steep increase of the dissipated power with the \bh spin, we can first estimate the amount $P_{Y}$ of electromagnetic energy dissipated per unit time by continuous magnetic reconnection at the Y-point. If we assume that the surface through which magnetic reconnection occurs can be written $K r_Y^2$ with $K$ a multiplicative constant, we have:
\begin{equation}
    \label{eq:dissipation}
    P_{Y} \sim K \beta_{rec} \frac{B_Y^2r_Y^2c}{8\pi}
\end{equation}
with $B_Y$ the magnetic field near the Y-point. The magnetic reconnection rate $\beta_{rec}$ does not vary significantly from $a=0.6$ to $a=0.99$ in the highly magnetized regime we work in. We suppose that $B_Y\propto r_Y^{-p}$ with $p>0$ an unknown exponent. We fit $P_{Y}$ as a function of $r_Y$ for the fitting parameters $p$ and $K$ and we obtain the dashed line in Figure\,\ref{fig:dissipation} which corresponds to sensible values: $p\sim 2.7\pm 0.1$ and $K\sim 45\pm 13$. Indeed, a purely dipolar decay of the magnetic field would have given $p=3$. The lower value hints at the influence of the toroidal component, whose decay with the distance is slower than $r^{-3}$, in the reconnection process, consistent with the significant twisting of the magnetic field lines in Figure\,\ref{fig:3D_B}. The fact that $K$ is an order of magnitude higher than unity is probably due to magnetic reconnection in the current sheet beyond the Y-point. 

We can thus affirm that as the \bh spin increases, the Y-point gets closer \ileyk{to} the \bh and the regions probed by the magnetic reconnection sites host a much higher magnetic field, which yields a much stronger dissipation of electromagnetic energy. Since the dissipated power we obtain is similar to the Poynting and particle kinetic energy fluxes computed in Section\,\ref{sec:Energy_fluxes}, we conclude that most of the dissipation takes place in the current sheet and that the rest of the simulation space is essentially force-free.

Although qualitatively valid, the aforementioned interpretation is to be taken with caution since in reality, the dissipated power will strongly depends on the local conditions. It is also why we discarded a detailed discussion on the uncertainties and on the degeneracy between $p$ and $K$. For instance, $P_Y$ will change with the magnetic flux distribution on the disk which might not be dipolar: the results obtained by \cite{Chatterjee2021a} with GR-MHD simulations actually suggest a slower decrease of the magnetic field with the radius ($\propto r^{-1}$), which would lower the dependence between $P_Y$ and the spin.

\subsection{Light emission}
\label{sec:Observables}

\subsubsection{Ray tracing and emission process}
\label{sec:ray_tracing}

We now proceed to characterize the appearance and the time variability of the \bh magnetospheres we computed so far using a ray-tracer adapter from the \texttt{geokerr} code \citep{Dexter2009} and first used in \cite{Crinquand2021}. Once the initial conditions have relaxed, each particle has a certain probability to emit a photon at each time step. The photons are not advanced by \texttt{GRZeltron} but instead, we store the necessary information for ray-tracing in post-process among which the location, direction and time at emission, along with the energy of the photon (see Equation\,\eqref{eq:w_ph} below). When the photons reach $r=1000r_g$, we neglect the curvature of spacetime and the ray-tracing becomes straightforward. We refer the reader to Crinquand et al. (in prep.) for more information about the generation of synthetic synchrotron images from \texttt{GRZeltron} simulations.

\begin{figure}
\centering
\includegraphics[width=0.99\columnwidth]{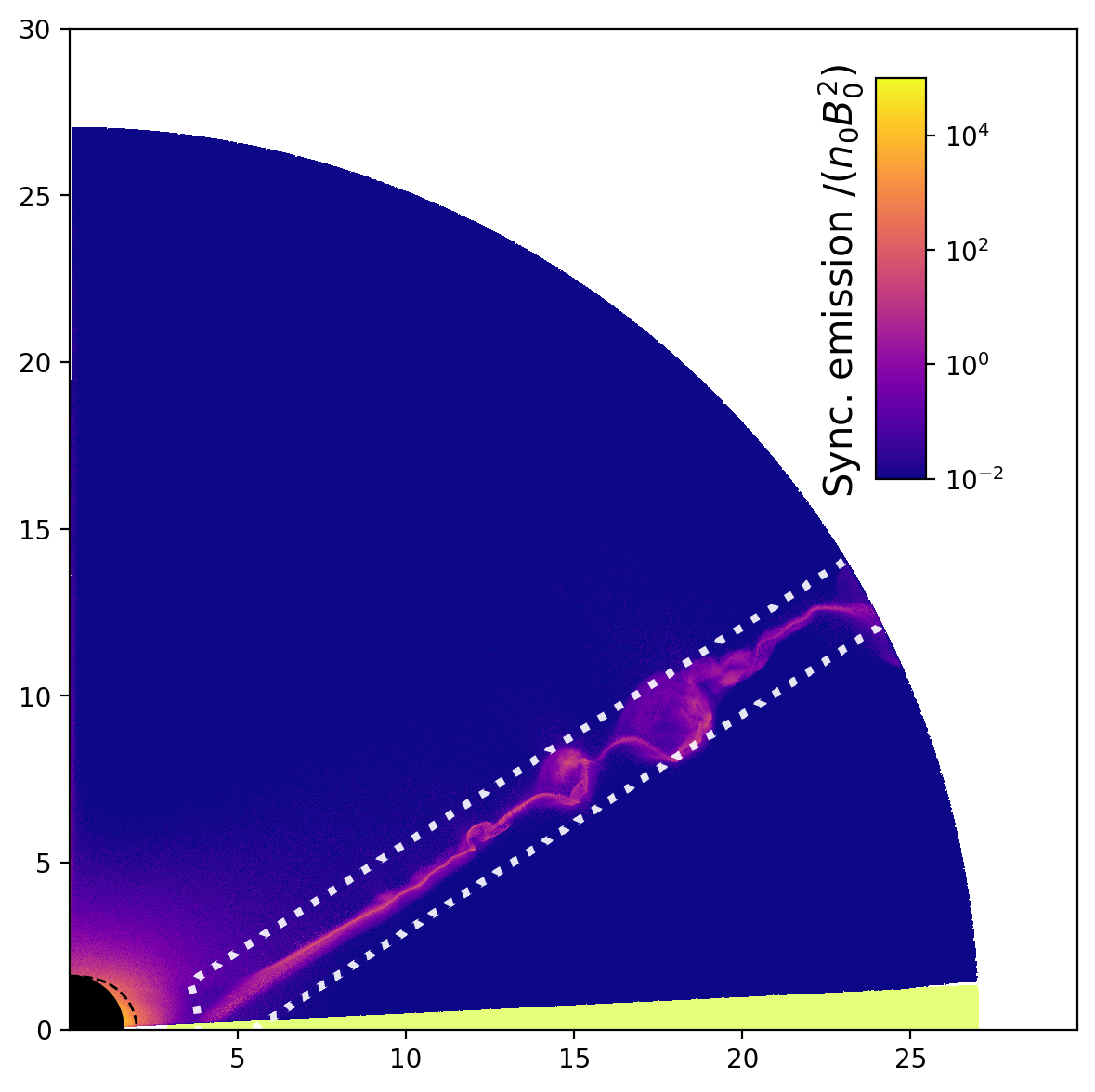}
\caption{\ileyk{Map of synchrotron emission power for a \bh spin $a=0.8$, multiplied by $r^2$ to highlight the contribution from the current sheet.} Photons emitted from the region delimited with a with dotted line are those shown as originating from the current sheet in the synthetic images in Figure\,\ref{fig:BH_intensity_unique} and \ref{fig:BH_intensity}.}
\label{fig:emission_map}
\end{figure}

We focus on synchrotron emission so the photons are emitted with a weight $w_{\gamma}$ defined by:
\begin{equation}
    w_{\gamma}=\frac{2e^4}{3m_e^2c^3}w\gamma^2 B_{\perp,\mathrm{eff}}^2,
    \label{eq:w_ph}
\end{equation}
where $w$ is the weight of the emitting particle, $\gamma$ is the particle Lorentz factor and $B_{\perp,\mathrm{eff}}$ is the magnitude of the effective magnetic field transverse to the particle motion after correcting for the drift velocity:
\begin{equation}
    \boldsymbol{B_{\perp,\mathrm{eff}}}=\boldsymbol{D}+\frac{\boldsymbol{v_p}\times\boldsymbol{B}}{c}-\frac{\left(\boldsymbol{v_p}\cdot\boldsymbol{D}\right)\boldsymbol{v_p}}{c^2},
\end{equation}
where $\boldsymbol{v_p}$ is the particle velocity. Both the corona and the disk are supposed to be optically thin to synchrotron radiation such as photons are not absorbed neither scattered. We account for the photons from the region $\theta\in[\pi/2;\pi]$ by associating to each photon emitted in the simulation space its symmetric with respect to the equatorial plane, with symmetric emission direction. In Figure\,\ref{fig:emission_map}, we represented the relativistic average synchrotron emission power $P_{sync}$ assuming an isotropic pitch-angle distribution. The current sheet displays an enhanced emissivity, especially where plasmoids are located: as is apparent in the upper right panel in Figure\,\ref{fig:energy_budget}, the plasmoids are regions of high plasma density and magnetic energy. On the other hand, the physical accuracy of the contribution from the ambient medium is uncertain since it might be influenced by our injection method. Its relative influence is artificially enhanced in Figure\,\ref{fig:emission_map} due to the isotropic pitch-angle assumption. In addition, the transverse motion of the particles responsible for synchrotron emission would be significantly damped in environments like \m{87} where synchrotron cooling is significant, which would lower the pitch angle along magnetic field lines and thus the emissivity in the ambient medium (Crinquand et al., in prep.). For these reasons, we discard the contribution of the ambient medium to the synchrotron emission and consider exclusively light emission from the current sheet delimited with a white dotted line in Figure\,\ref{fig:emission_map}. A comparison of the relative flux from both components is beyond the scope of our setup since due to the unrealistically low magnetization we use, the maximum Lorentz factor of the particles is artificially low and the emissivity of the current sheet is underestimated.



\subsubsection{Synthetic images}
\label{sec:Synthetic_images}

\ileyk{We computed synthetic synchrotron images for a variety of \bh spin and for different inclinations of the line-of-sight with respect to the \bh spin axis (see Crinquand et al. in prep. for a detailed analysis of synthetic images). The goal is to determine the apparent morphology of the current sheet and of the plasmoids. To do so, we compute the bolometric flux on a screen at infinity normal to the line-of-sight, with 400$\times$400 square pixels of equal size and a field-of-view extending from $-18r_g$ to $+18r_g$ in both directions. We limit the shot noise by applying a local median over the closest neighboring pixels. Then, to account for the point spread function of the instruments, we smooth the image with a Gaussian kernel of width one pixel.}

\begin{figure}
\centering
\includegraphics[width=0.99\columnwidth]{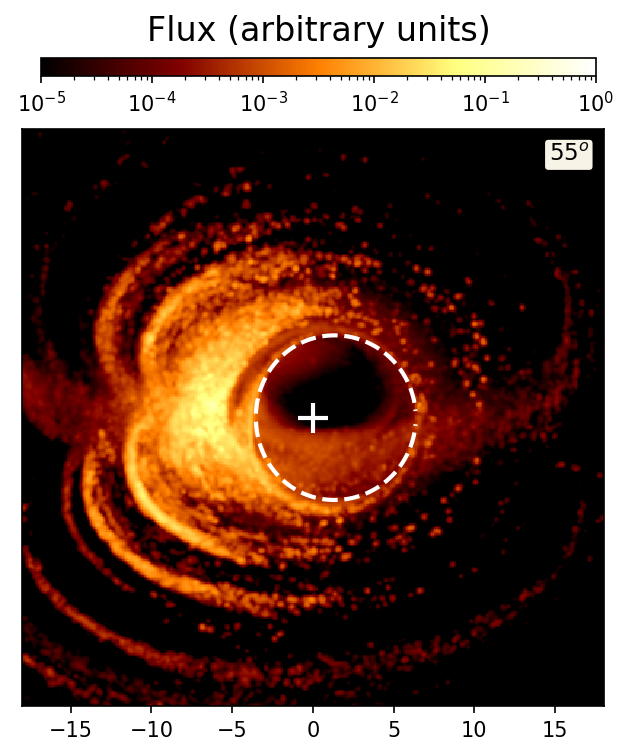}
\caption{Synthetic images of synchrotron emission from the current sheet for a simulation with a thin disk in prograde rotation around a Kerr \bh with a spin $a=0.8$.}
\label{fig:BH_intensity_unique}
\end{figure}

\ileyk{Figure\,\ref{fig:BH_intensity_unique} is a synthetic image for a \bh spin $a=0.8$ and an intermediate viewing angle, where the line-of-sight is almost aligned with the current sheet visible in Figure\,\ref{fig:emission_map}. In the innermost regions, we plotted the image of the photon ring (white dashed line) centered on the \bh position (white cross). First of all, we identify a bright crescent-shaped region near the photon ring. We interpret it as a manifestation of the significant \ileyk{relativistic beaming} for this inclination and \bh spin. Out of the lensed photon ring, bright arcs are visible with an intensity varying between 1 and 10\% of the brighter inner crescent. When observed under different inclinations (see panels in Figure\,\ref{fig:BH_intensity}), it appears that they correspond to plasmoids propagating along the conic current sheet. Due to the moderately relativistic bulk motion of the plasmoids ($\Gamma\sim1.2$ to 2), the emission is not strongly \ileyk{beamed}. The arc-shape of the plasmoids is an artifact induced by our axisymmetric assumption. In full 3D, the plasmoids would instead appear as tilted flux ropes of finite azimuthal extent moving on the envelope of the cone-shaped current sheet (see Section\,\ref{sec:hot_spot}). In addition to these features, for this grazing viewing angle, we get the cumulated photons emitted in the direction of outwards motion along the whole cone-shaped current sheet. They produce a faint stripe across the whole image, with an intensity which reaches a few 1\% of the one of the inner crescent. Contrary to the other sharper arcs, this arc is not due to our axisymmetric assumption and it is characteristic of the presence of a cone-shaped current sheet around the \bh. At high viewing angle inclinations, the emission is dominated by a hourglass shaped region corresponding to the projection of the conic current sheet.} 

\subsection{Retrograde disk}
\label{sec:retro_disk}

When the directions of the \bh and the disk’s angular momentum are opposite, the dynamics of the corona is very different from the case when the disk is in prograde rotation. \kyle{First of all, the \isco of a retrograde disk lies at a larger distance from the \bh, especially for the $a=-0.8$ case we study where $r_{\iscoo}\sim 8.4 r_g$ (while $r_{\iscoo}\sim 2.9 r_g$ for $a=0.8$)}. It limits the extent of the region on the disk where magnetic field loops anchored in the disk could close inside the event horizon. However, the very existence of these loops if the disk is in retrograde rotation is unlikely, even for low absolute spin values. Indeed, no steady coupling magnetic field line can sustain between the disk and the \bh since in the ergoregion, the field lines have to rotate in the same direction as the \bh. Consequently, the field line is necessarily twisted and a toroidal component quickly grows \ileyk{but contrary to the prograde case, there is no region where the differential rotation is low enough that the field line can remain closed.}

\begin{figure}
\centering
\includegraphics[width=0.99\columnwidth]{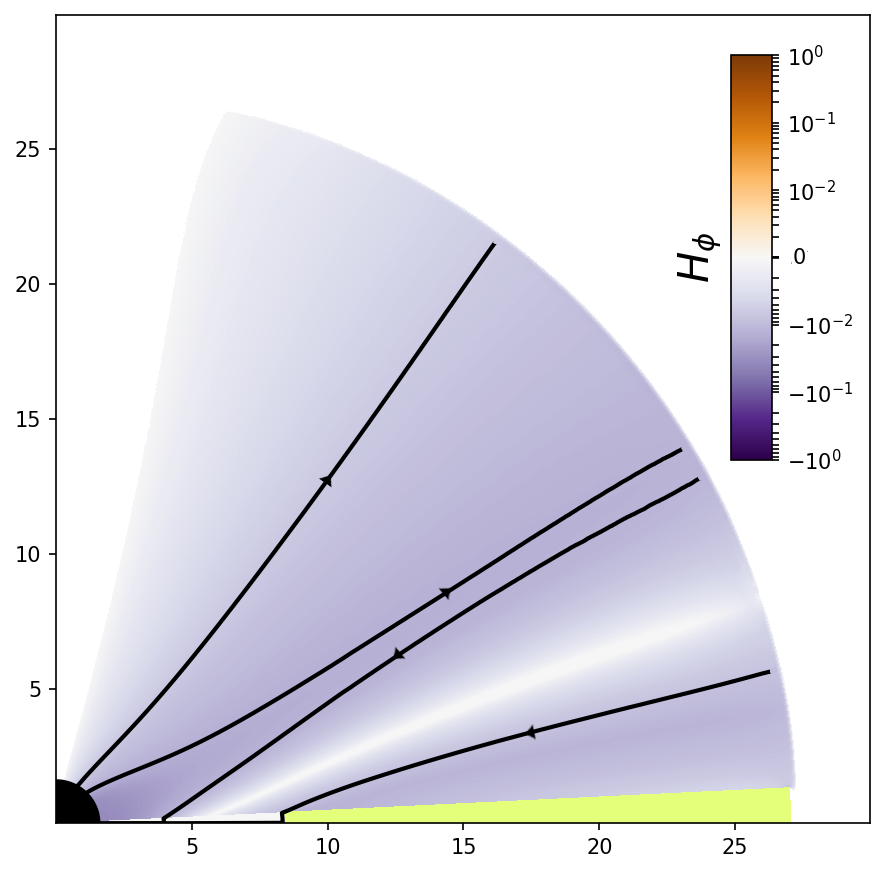}
\caption{Map of the toroidal component $H_{\phi}$ for a simulation with $a=-0.8$ (retrograde) and $\epsilon=5\%$. The magnetic field lines are those surrounding the non-reconnecting current sheet and the pair corresponding to the potential at the \isco. The Keplerian disk is in light green and extends down to $r_{\iscoo}\sim 8.4 r_g$.}
\label{fig:hphi_retro}
\end{figure}

Our results are in agreement with this analysis. After the initial relaxation phase, all magnetic field lines are open, except some in the region where $r<r_{\iscoo}$ and $\theta\in[\theta_d;\theta_{max}]$. We must keep in mind though that in the plunging region, the field lines do not rotate and their physical accuracy is questionable, as explained in Section\,\ref{sec:ICs}. \kyle{This numerical limitation would be alleviated in simulations where inwards advection of the magnetic field lines' footpoints is accounted for, beyond the scope of this paper.} Elsewhere, the field lines are either anchored in the disk or thread the event horizon. Between these two regions, a current sheet forms but no magnetic reconnection is observed. In Figure\,\ref{fig:hphi_retro}, we represented the toroidal component of the magnetic field. The same color scale as in Figure\,\ref{fig:wheel} is used in order to emphasize that in this configuration, $H_{\phi}$ does not change sign at the current sheet \ie where the radial component of the magnetic field reverses. The currents are essentially radial and close in the \bh equatorial plane.

The lack of magnetic reconnection in the current sheet confirms the preponderant role played by the toroidal component of the magnetic field, as suggested in Section\,\ref{sec:Toroidal_fields}. In contrast with the prograde configuration, the jump in magnetic field at the current sheet is exclusively produced by the poloidal component $B^{\parallel}$ which is much smaller than the toroidal component beyond a few gravitational radii. The direction of the magnetic field lines on both sides of the current sheet are thus too similar to trigger magnetic reconnection. In these conditions, plasmoids, if present, could not have been formed by a purely kinetic mechanism like the one at work in the case of a disk in prograde rotation around the \bh. Particles could not be steadily accelerated in the corona by magnetic reconnection and the Y-point would not be present. \kyle{We can not draw firm conclusions for the retrograde case since in this configuration, the corona could still be activated by the accretion of magnetic field loops of varying polarity \citep{Parfrey2015}. Furthermore, magnetic reconnection could still take place for a retrograde disk but in the equatorial region between the \isco and the event horizon \citep{Ripperda2020,Crinquand2021}.}


\section{Discussion}
\label{sec:discussion}

\subsection{Hot spot}
\label{sec:hot_spot}

\begin{figure*}
\centering
\includegraphics[width=1.9\columnwidth]{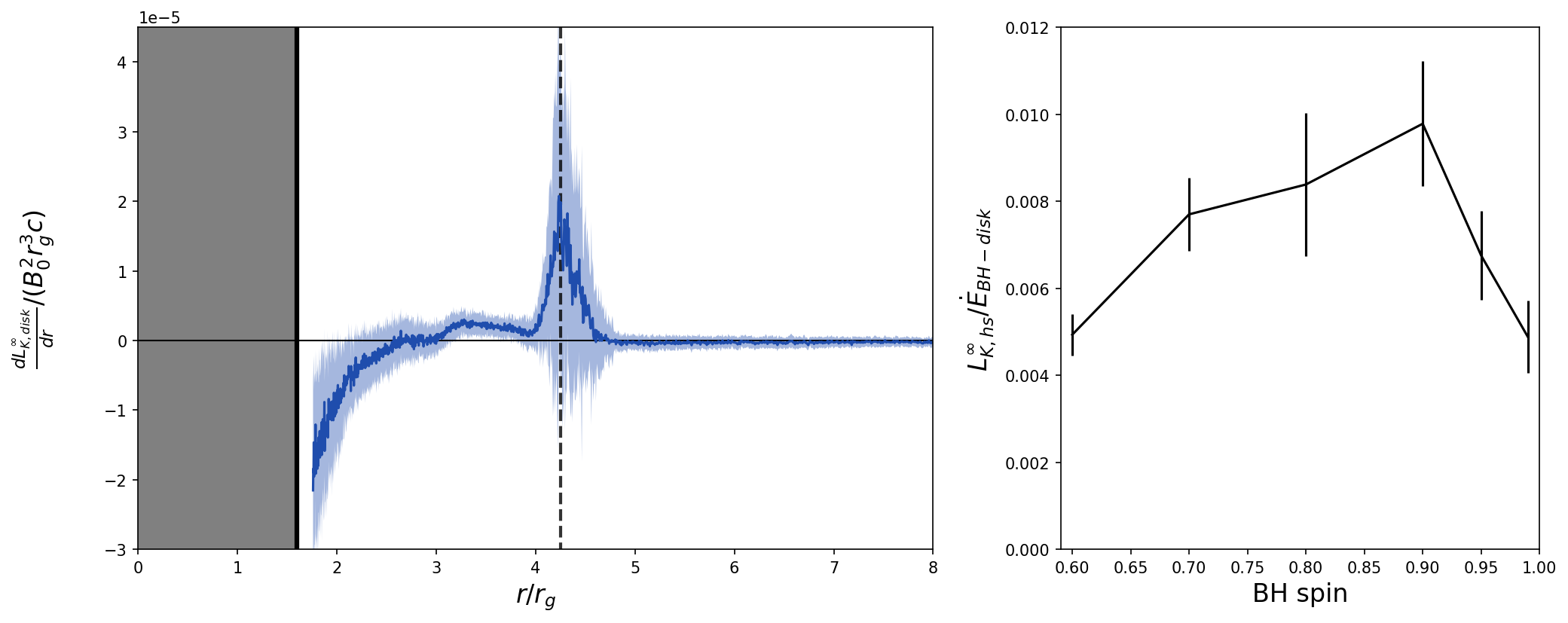}
\caption{(Left panel) Flux on the disk of particle kinetic energy per unit time per ring of infinitesimal width $\diff r$, as a function of the distance $r$ to the \bh, for $a=0.8$. The solid blue curve is the median profile while the light blue shaded region is the 1$\sigma$-variability. Vertical lines are the event horizon (solid) and the separatrix footpoint (dashed line). (Right panel) Flux of particle kinetic energy per unit time near the separatrix footpoint compared to the rate $\dot{E}_{BH-disk}$ of energy deposited onto the disk per unit time via the coupling magnetic field lines.}
\label{fig:energy_kin_slices}
\end{figure*}

\cite{Abuter2018b} reported on a centroid shift near \sgr during two flares that they interpreted as synchrotron emission from a hot spot orbiting \ileyk{around the \smbh. \cite{Baubock2020} reproduced the astrometric data with a compact hot spot of less than $5r_g$ located at 9$r_g$ from the \smbh.} The origin of the hot spots is still unknown and the uncertainties concerning the orientation of their trajectory with respect to the line-of-sight and the very center of the putative orbit prevent us from determining whether it was a structure inside a disk surrounding \sgr. \benoit{In the magnetically-coupled configuration considered in this paper, if the \bh is surrounded by a disk in prograde rotation, a hot spot could either be produced on the disk or correspond to a plasmoid ejected along the current sheet. In the former case, we identified two mechanisms susceptible to form a hot spot: either a deposit of electromagnetic energy onto the disk between the \isco and the separatrix footpoint, or a beam of relativistic particles accelerated by magnetic reconnection at the Y-point and hitting the disk at the separatrix footpoint.}

In Section\,\ref{sec:Coupling_field_lines}, we showed that the electromagnetic energy deposited per unit time via the coupling magnetic field loops is comparable to the jet power. \cite{Yuan2019b} evaluated the impact of this contribution on the radial profile of the disk emission. They showed that for a maximally spinning \bh and a separatrix footpoint similar to the one we find self-consistently, \ileyk{the thermal emissivity is significantly enhanced with respect to the standard model by \cite{Novikov1973} up to $\sim10r_g$, a distance approximately 3 to 4 times larger than the location of the separatrix footpoint. Since the electromagnetic energy deposit rates we compute match very well those estimated in the purely force-free regime, we do not repeat the derivation of thermal emissivity profiles performed in \cite{Yuan2019b}.}

We can however explore another source of energy deposit onto the disk, the one associated to the particle kinetic energy. In our simulations, we observe an important flow of relativistic particles accelerated by magnetic reconnection at the Y-point and flowing down to the separatrix footpoint. In the left panel in Figure\,\ref{fig:energy_kin_slices}, we show in blue the radial profile of the particle kinetic energy deposited per unit time and per radial bin by the particles onto the disk. The solid lines is the median profile over 50$r_g/c$ and the blue shaded region is the 1$\sigma$-variability. The negative values for the particle kinetic energy at infinity measured near the event horizon (vertical solid line) are a manifestation of the Penrose process, as mentioned in Section\,\ref{sec:Energy_fluxes}. Beyond, the profile is close to 0 everywhere except in a region centered on the separatrix footpoint (vertical dashed line). This positive peak corresponds to the particles cascading onto the disk at relativistic speed and we show in the right panel in Figure\,\ref{fig:energy_kin_slices} the energy deposited per unit time onto the disk obtained from integrating the radial profiles near the separatrix footpoint for each \bh spin. After normalization with the amount of electromagnetic energy deposited onto the coupled disk region per unit time, $\dot{E}_{BH-disk}$, we conclude that this source of energy is approximately 2 orders of magnitude lower than $\dot{E}_{BH-disk}$ for any spin above 0.6. Although this component seems minor, it must be noticed that it is more localized since energy is essentially deposited at the footpoint of the separatrix. 

Last, it is still unclear whether the hot spots observed were moving in a plane containing the \smbh itself. \cite{Matsumoto2020} found that their motion was super-Keplerian which could indicate that they are associated to structures flowing away from an underlying disk. \ileyk{\cite{Baubock2020} emphasized that their best-fit was obtained for near face-on orbits and an out-of-plane velocity component of 0.15$c$.} \cite{Ball2021} designed a model where synchrotron emission from plasmoids moving on the hull of an outflowing jet reproduces the features associated to the hot spots and in particular the offset between the center of the trajectory of the centroid and the position of the black hole. The presence of these plasmoids in GR-MHD simulations, produced by magnetic reconnection in current sheets induced by physical \citep{Ripperda2020} or numerical resistivity \citep{Nathanail2020}, brings strong support to these types of models where the apparent orbital motion of hot spots is the result of their slowly expanding helicoidal trajectory and of our viewing angle. In our PIC framework, these plasmoids are those moving along the current sheet with mildly relativistic speed ($v_d\sim 0.2$ to 0.5$c$) and with a rotation speed close from the one of the separatrix footpoint given by Equation\eqref{eq:omega_GR}, hence the super-Keplerian motion. If the \bh spin axis is aligned (or anti-aligned) with the line-of-sight, a hot spot formed at the Y-point  or farther would be at a projected distance $x_Y\sim6-10r_g$ for a \bh spin of at least 0.7 (see blue dashed curve in Figure\,\ref{fig:rY}). For lower spins, the Y-point is located too far from the \bh and a hot spot could not form within $10r_g$ via the mechanism we have exhibited. When we compute the outward projected drift $\Delta x$ over a full orbit, we obtain:
\begin{equation}
\frac{\Delta x}{x_Y}\sim 2\pi\frac{v_{d}}{v_{orb}}
\end{equation}
with $i_{cs}$ the inclination of the current sheet with respect to the equatorial plane and $v_{orb}=R_S\Omega_K(r=R_S)$ the azimuthal speed of the hot spot. With the inclinations $i_{cs}\sim30^{\circ}$ we measure, we find a relative shift of 2 to 7 times the initial distance projected on the \bh equatorial plane, with lower values for lower \bh spins. We notice that these values are marginally consistent with the motion of the main hot spot observed around \sgr\footnote{The large uncertainties on the trajectories measured for the other flares are not yet constraining enough to conclude}, which is first located at a projected distance of $\sim 8r_g$ and is found at a projected distance of $\sim 16r_g$ after two-third of an apparent orbit. Among the spins we explored, $a=0.7$ is thus the favored value provided the spin axis is aligned with the line-of-sight. We notice that for this spin, the amount of electromagnetic energy deposited onto the disk via the coupling magnetic field lines per unit time is $\dot{E}_{BH-disk}\sim7\cdot 10^{-4}B_0^2r_g^2c$ (see bottom panel in Figure\,\ref{fig:rdcd}), comparable to the dissipated electromagnetic power in the current sheet, $P_{Y}\sim3\cdot 10^{-4}B_0^2r_g^2c$ (see Figure\,\ref{fig:dissipation}).



\subsection{Flares and illuminating spectrum}
\label{sec:flares}

Daily non-thermal flares are observed from \sgr in NIR \citep{Eckart2006} and in X-rays \citep{Baganoff2003}. They are associated \ileyk{with} a drop of the magnetic field amplitude by almost an order of magnitude \citep{Ponti2017} which points at magnetic reconnection as a possible culprit for particle acceleration and subsequent non-thermal emission. In all models, NIR flares comes from optically thin synchrotron emission \citep{Witzel2020}. The origin of X-ray flares though is still a matter of debate. \cite{Ponti2017} ascribe it to optically thin synchrotron emission with a cooling break induced by a high-energy cut-off in the distribution of the underlying non-thermal population of electrons. Alternatively, it has been argued in favor of an inverse Compton \citep{Dodds-Eden2009} or a synchrotron self-Compton scattering \citep{Mossoux2016} origin.

\subsubsection{Particle energy distribution and scaling correction}
\label{sec:flares_spectrum}

\begin{figure}
\centering
\includegraphics[width=0.99\columnwidth]{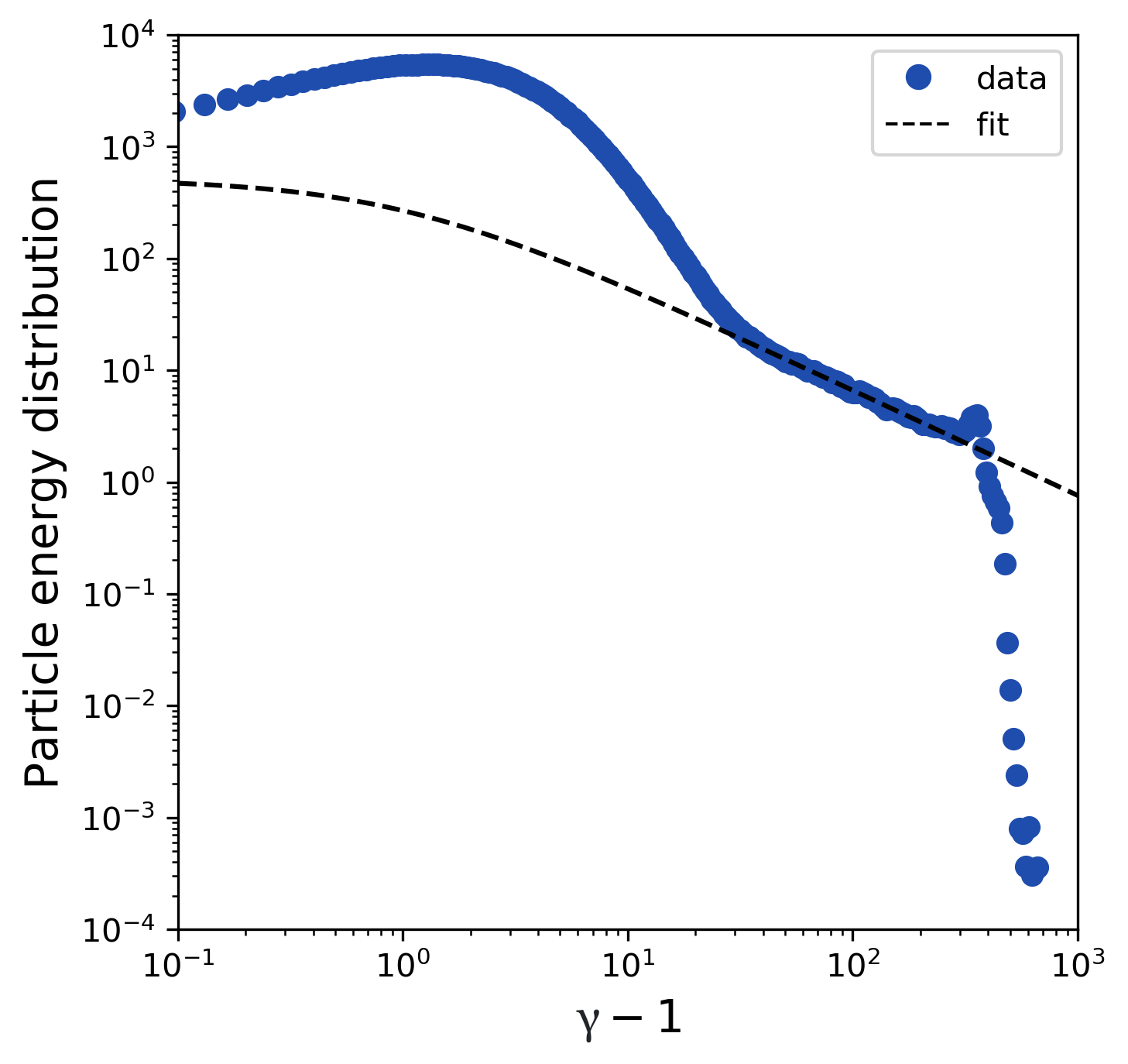}
\caption{Distribution of the electrons and positrons in the current sheet as a function of their Lorentz factor.}
\label{fig:spectrum_pcls}
\end{figure}

The results we presented in Section\,\ref{sec:Coronal_heating} show that a significant fraction of the electromagnetic energy is tapped to accelerate particles, with dissipated powers up to 10\% of the jet power: acceleration at magnetic reconnection sites produces a non-thermal population of relativistic electrons and positrons. 

In Figure\,\ref{fig:spectrum_pcls}, we show the energy spectrum of the electrons and positrons computed for a fiducial snapshot in the relaxed state of a simulation with $a=0.8$ and $\epsilon=5\%$. We only account for particles inside the current sheet (as defined by the region in Figure\,\ref{fig:emission_map}). We see that in addition to a thermal peak at $\gamma$ of a few, corresponding to the bulk motion of the plasma, a non-thermal component emerges and dominates beyond $\gamma\sim40$. The cutoff at $\gamma_{c}\sim200$ indicates that particle acceleration is limited. In our setup, this limitation arises from the magnetization of the plasma around the current sheet, upstream the reconnection region, since in the relativistic regime, we expect $\gamma_{c}$ to be of the order of $\sigma$ \citep{Kagan2018,Werner2015}. 

Between $\gamma=40$ and $\gamma=200$, we fit a power-law $N(\gamma)\propto\gamma^{-p}$ and find that the best fit is achieved for an exponent $p\sim 0.9$ (black dashed line in Figure\,\ref{fig:spectrum_pcls}). In the collisionless, relativistic and strongly magnetized regime, \cite{Werner2015} showed that particles accelerated by magnetic reconnection without guide field in a symmetric current sheet have an energy distribution which is close from a power-law with an exponent of $p\sim1.2$. This proximity with the slope index we measure, in a more realistic environment, suggests that this component we find in the particle energy distribution corresponds to particles accelerated up to relativistic Lorentz factors by magnetic reconnection in the conic-shaped current sheet. 

\benoit{An additional argument in favor of the accuracy of this exponent is that although our dimensionless magnetic fields $\tilde{B}_0$ are strongly underestimated compared to the typical values around accreting \bhs, our magnetization $\sigma$ is of the right order of magnitude. Indeed, in \m{87}, the analysis of the polarized synchrotron emission leads to a (non-relativistic) magnetization of the plasma in the innermost regions between 1 and up to $10^4$ for a magnetic field of 30G and a plasma density of $10^4$cm$^{-3}$ \citep{Akiyama2021}. In \sgr, the multiwavelength campaign carried by \cite{GravityCollaboration2021} indicates that the density of non-thermal electrons is of the order of $10^6$cm$^{-3}$ for an ambient magnetic field comparable to \m{87}. This would correspond to a magnetization of the order of 10-100. Finally, in \cyg, the estimated magnetization ranges from $10^3$ to $10^5$ for a plasma density of $10^{12}$cm$^{-3}$ and a magnetic field between $10^5$ and $10^6$G \citep{Cangemi2021}. These values are coherent with the maximum particle Lorentz factors. In \cyg, \m{87} and \sgr, the Lorentz factors of non-thermal particles responsible for NIR synchrotron emission are generally considered \benoit{to be $\lesssim 10^3$} \citep{Cangemi2021,Akiyama2021}. In \sgr, \cite{GravityCollaboration2021} found that in order to account for the spectral energy distribution and time evolution of the X-ray and NIR flares, sustained particles acceleration was necessary with particle Lorentz factors as high as a few $10^4$. In our simulations, near the Y-point, we reach a plasma magnetization of the order of a few 200 (accounting for a bulk Lorentz factor of 1.5), which corresponds to the cutoff $\gamma_c\sim 200$ we find. It is approximately an order of magnitude smaller than observations. Direct comparison to observed values is difficult though since radiative models are based on a one-zone approach, but we conclude that our magnetization lies within the range of possible values for \m{87} and \sgr, and is somewhat lower than the values in \cyg. Consequently, we can assume that the slope $p$ of the particle energy distribution beyond $\gamma\sim40$ does not depend on the values of $\tilde{B}_0$ and $\sigma$ in the regime we explored.} 


Alternatively, \cite{Yuan2019c} showed that significant amounts of the \bh rotational energy could be dissipated per unit time in magnetospheres containing small scale poloidal magnetic field loops. If magnetic confinement is strong enough, the magnetic field lines funneling the jet are disrupted by the kink instability. They suggest that most of the electromagnetic energy is dissipated through this magnetic reconnection process close to the \bh and that it could be a significant heat source for the corona. \benoit{Another candidate mechanism was found by GR-MHD simulations in the disk mid-plane, where magnetic reconnection happens in current sheets located on the edge of magnetic bubbles episodically formed in MADs \citep{Porth2019,Dexter2020,Ressler2021,Ripperda2021}. This mechanism corresponds to the current sheet found in the plunging region by PIC simulations \citep{Crinquand2021}.} Recently \cite{Scepi2021} computed the flaring and quiescent X-ray emission produced by these regions and retrieved fluxes and spectral shapes analogous to the observed ones.

\subsubsection{Illuminating spectrum}
\label{sec:lamppost}

The consequences of magnetic reconnection in the \bh corona also manifest in terms of irradiation of the underlying disk. Part of the hard X-ray emission produced by the particles accelerated in the current sheet illuminates the disk and produces a typical X-ray reflection spectrum characterized by, for instance, fluorescence lines \citep[in particular the Iron K$\alpha$ line at 6.4keV][]{Reynolds2020}. X-ray reflection spectrum models are used to determine the spin of \bhs accreting material from a geometrically thin and optically thick disk. However, in these semi-empirical models, the geometry and the location of the emission site above the disk where the hard X-ray illuminating photons come from is a degree-of-freedom of the problem, and so is the spectrum of this emission.

\begin{figure}
\centering
\includegraphics[width=0.99\columnwidth]{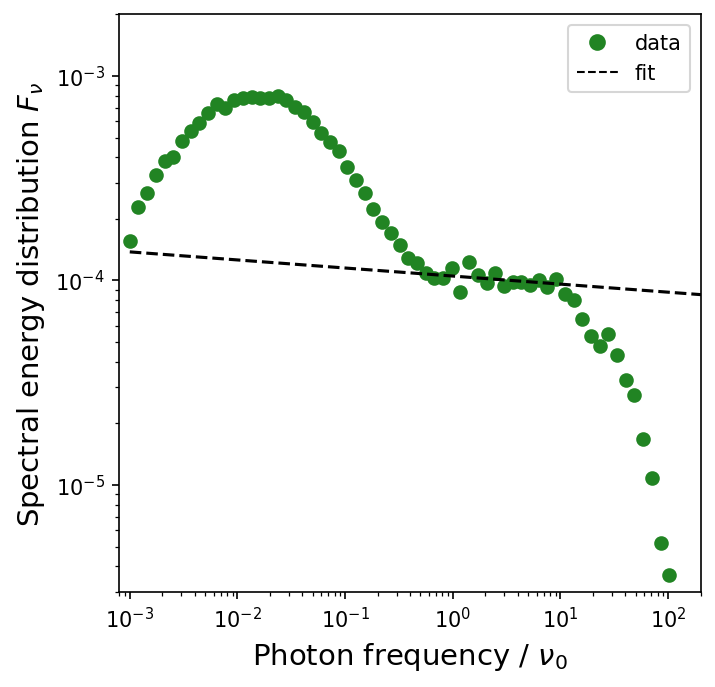}
\caption{Spectral energy distribution of synchrotron emission from the current sheet.}
\label{fig:spectrum_sync}
\end{figure}

We computed the spectral energy distribution of the synchrotron photons emitted by the leptons in the current sheet (Figure\,\ref{fig:spectrum_sync}). The photon frequency unit is given by the reference synchrotron frequency, $\nu_0$:
\begin{equation}
    \nu_0=\frac{3eB_0}{4\pi m_e c},
    \label{eq:nu0}
\end{equation}
that we could express as a function of the dimensionless magnetic field $\tilde{B}_0$:
\begin{equation}
    \nu_0=\frac{3}{4\pi}\tilde{B}_0\left(\frac{r_g}{c}\right)^{-1}.
    \label{eq:nu0_dimensionless}
\end{equation}
We retrieve the thermal peak at low energy and a non-thermal component at higher energy. We fitted this component for photon frequencies between $\nu_0$ and 10$\nu_0$ with a power-law $F_{\nu}\propto \nu^{p'}$ with $p'$ the spectral index. The best-fit is obtained for a spectral index $p'\sim -0.04\pm 0.03$. This component is approximately what we would expect for the synchrotron emission in a uniform magnetic field from a power-law distribution of positrons and electrons energies with an exponent $1-2p'=1.1\sim p$, so we can be confident that it is produced by the non-thermal population of particles accelerated by magnetic reconnection that we identified in Section\,\ref{sec:flares_spectrum}. It must be acknowledged that these values differ from the one derived in NIR by \cite{Ponti2017} for \sgr. They found a slope of the spectrum which corresponds to a spectral index $p'\sim -0.7$ (i.e. a photon index of $\sim$1.7), coherent with a nearly flat spectrum. One of their model ascribes this emission to a non-thermal population of particles with a power-law energy distribution of exponent $p\sim2.2$. The less steep particle energy distribution we find might be due to the limited photon frequency extent of the power-law component, over only one decade.

Notice that for magnetic field magnitudes $B\sim 100G$ and particle Lorentz factors $\gamma =100$ of \sgr, the synchrotron frequency $\gamma^2 \nu_0$ would correspond to infrared according to Equation\,\ref{eq:nu0}. We would retrieve this frequency range with Equation\,\ref{eq:nu0_dimensionless} for higher and more realistic values of $\tilde{B}_0$ than the one we use which is limited by the necessity to spatially resolve the Larmor radius (see Section\,\ref{sec:Parameters}).

In this section, we studied the properties of a simulation with a \bh spin $a=0.8$ surrounded by a thin disk ($\epsilon=5\%$). However, the more general exploration of the properties of the coronal emission illuminating the disk as a function of the \bh spin is now possible. \benoit{Most importantly, this result provides a physically-motivated conic-shaped geometry and emission spectrum for X-ray reflection spectroscopy models. It reduces the number of degrees-of-freedom compared to the more empiric lamppost model where a point source is located above the \bh on its spin axis.}

\subsubsection{Timing}
\label{sec:flares_timing}

Is the episodic detachment of plasmoids at the Y-point susceptible to yield flares with the appropriate timing properties? In \sgr, the flares occur on an approximately daily basis and last an hour or so. Using a mass of the \smbh of $4\cdot 10^6\text{\msun}$, we find that the time scales reported in the left panel in Figure\,\ref{fig:dY} is of the order of a few minutes. To our knowledge, there is no variability reported with a recurrence time of the order of a few minutes in \sgr. \benoit{Intrinsic variability of this mechanism is thus too fast to account for the flares from \sgr, although it could be the origin of the non-thermal population responsible for the flare emission. The triggering of a flaring episode itself is more likely due to changes in the accretion flow properties.} Time-dependent force-free simulations of magnetic loop accretion by \cite{Parfrey2015} show that for loops of a few $r_g$ and realistic accretion speeds, a periodicity of a few $100r_g/c$ (\ie a few days for \sgr) appears. Alternatively, in the GR-MHD simulations of the MAD regime performed by \cite{Chashkina2021}, the magnetic flux and accretion onto the \bh spike every few $\sim 100r_g/c$. This cycle is due to the formation of a transient current sheet in the \bh equatorial plane, similar to the one found by \cite{Ripperda2020} with GR-MHD simulations and studied by \cite{Crinquand2021} in the PIC framework. \benjamin{More generally, models \citep{Gutierrez2019} and numerical simulations \citep{Dexter2020,Porth2021,Chatterjee2021a} suggest that the MAD regime is prone to transient magnetic reconnection events.} Consequently, if the mechanism we describe in this paper is active during the flares, it could lead to quasi-periodic variability during \sgr's flares on a time scale of a few minutes. During a flare of \m{87}, this time scale would translate into a quasi-periodicity of a few days.

In \cyg, the plasma dynamics at the Y-point might well be responsible for the millisecond flares reported by \cite{Gierlinski2003c}. They observed short X-ray flares during which the flux increases by an order of magnitude, with stronger flares in the high/soft state when the standard accretion disk extends very close to the \bh, probably all the way down the \isco. \ileyk{\cite{Beloborodov2017} suggested that these flares might come from chains of plasmoids formed near the \bh.} In the high/soft state, the topology of the magnetosphere could be similar to the one we considered, with a conic-shaped current sheet forming above and below the disk. Unfortunately, the very short duration of the flares precluded any conclusive results on the spectrum during the flares, although \cite{Gierlinski2003c} suggested that a hybrid emission model with a black body disk component and a non-thermal tail from a population of relativistic electrons is possible. Alternatively, \cite{Mehlhaff2021} proposed that these flares might be due to kinetic beaming based on their PIC simulations of inverse Compton scattering by very high energy particles in the corona.

\subsection{Jet power}
\label{sec:jet_power}

In Section\,\ref{sec:BH_Opened_magnetic_field}, we computed the electromagnetic power carried by the Poynting vector through the open field lines threading the \bh at its poles. We identified this value to the power of the jet which, downstream the region we probe, will gain mass and where electromagnetic energy will eventually be converted into particle kinetic energy and radiation. These values can thus be interpreted as upper limits on the unbeamed jet luminosity, independently of the emission mechanism in the jet. In Figure\,\ref{fig:rdcd}, we show the maximum jet power we obtained for a \bh with spin $a=0.99$, approximately:
\begin{equation}
    P_{jet,99}=0.006 B_0^2 r_g^2 c.
    \label{eq:P_99}
\end{equation}
It would be misleading to use for $B_0$ which is, in our model, the amplitude of the magnetic field at $1r_g$, the values derived from polarization measurements or spectral modeling. Indeed, the values derived from one-zone models are averaged over extended emission regions where the magnetic field amplitude varies quickly. Instead, we consider the maximal value $B_0$ can reach by evaluating the dynamical importance of the magnetic field with the dimensionless ratio $\Phi_H/\sqrt{\dot{M}cr_g^2}$, with $\dot{M}$ the mass accretion rate in the innermost region. When this ratio reaches approximately 50, the magnetic field saturates and the accretion proceeds via a MAD \citep{Tchekhovskoy2011,Porth2019}. In these conditions, the magnetic field in the innermost regions must reach:
\begin{equation}
    B_{MAD}=\frac{25}{\pi}\sqrt{\frac{\dot{M}c}{r_g^2}},
    \label{eq:B_MAD}
\end{equation}
which is to be understood as a maximum value.

\cite{Collaboration2021b} performed a broad exploration of the parameter space with GR-MHD simulations in an attempt to model the polarized synchrotron emission reported from \m{87} in \cite{Collaboration2021a}. They showed that MAD were systematically better at matching the observations and found a mass accretion rate of $\dot{M}\sim 3-20\cdot 10^{-4}M_{\odot}\cdot$yr$^{-1}$, in agreement with the upper limit derived by \cite{Kuo2014} from Faraday rotation and by \cite{Russell2015} from X-ray analysis. With the parameters of \m{87}, we find a magnetic field $B_{MAD}=200-500G$ and, using Equation\,\eqref{eq:P_99}, a near maximal jet power of $P_{jet,99}\sim6\cdot 10^{42}$ $\text{erg}\cdot s^{-1}$ to $P_{jet,99}\sim3\cdot 10^{43}$ $\text{erg}\cdot s^{-1}$, in agreement with the jet kinetic power deduced by \cite{Lucchini2019a} from observations. 

In \cyg, the X-ray accretion luminosity is of the order of 1\% of the Eddington luminosity in the low-hard state, when jet emission is detected. With a conversion efficiency of 10\%, it corresponds to a mass accretion rate $\dot{M}\sim 5\cdot10^{-9}M_{\odot}\cdot$yr$^{-1}$. The corresponding MAD magnetic field amplitude is $B_{MAD}\sim 10^8$G and the maximal \benoit{jet power is $P_{jet,99}\sim 10^{37}\text{erg}\cdot s^{-1}$, approximately one order of magnitude higher than what the time-averaged estimate derived by \cite{Russell2007} from interaction with the interstellar medium. This discrepancy might be due to the fact that the jet is absent during the high-soft state, although \cyg is in the latter state for only 10\% of its lifetime \citep{Gallo2005}. Alternatively, it could mean that the magnetic field never reaches the MAD saturation value, contrary to \m{87}. A magnetic field amplitude in the jet launching region of $\sim10^7$G \citep{Cangemi2021} would give a jet power consistent with observations.}

For \sgr, the existence of a jet is more elusive. \cite{Li2013} identified a parsec-scale jet based on what they interpreted as a radio shock front due to the jet from \sgr crossing the local gas. Diagnostics such as Faraday rotation measurements grants us access to estimates for the mass accretion rate onto the \smbh, ranging from $10^{-9}$ to $10^{-7}$ $M_{\odot}\cdot$yr$^{-1}$ \citep{Bower2003,Marrone2006,Wang2013}. The simulations of stellar wind capture performed by \citep{Ressler2018} reproduce these values, with an extrapolated mass accretion rate at the event horizon of the order of $10^{-8}$ $M_{\odot}\cdot$yr$^{-1}$. It would yield a maximum magnetic field $B_{MAD}=600-6,000G$, several orders of magnitude above the measured values, and a near maximal jet power $P_{jet,99}\sim2\cdot 10^{37}$ $\text{erg}\cdot s^{-1}$ to $P_{jet,99}\sim2\cdot 10^{39}$ $\text{erg}\cdot s^{-1}$. This range is compatible with the values deduced from the ram pressure of the jet on the ambient medium \citep{Li2013}, from emission models \citep{Markoff2007} and from numerical simulations \citep{Yuan2012}. Even in the most optimistic case though, where \sgr would host a near maximally spinning \smbh accreting in a MAD regime, the putative jet for a mass accretion rate of $10^{-8}$ $M_{\odot}\cdot$yr$^{-1}$ is orders of magnitude less powerful than in \m{87}. We also notice that a Blandford-Znajek jet powered by a MAD magnetic field at this mass accretion rate would not be enough to fuel the Fermi bubbles observed in $\gamma$-rays \citep{Su2010} or the X-ray bubbles recently discovered by eROSITA \citep{Predehl2020}.



\subsection{Spinning up torques from the \bh}
\label{sec:spin_up}

\begin{figure}
\centering
\includegraphics[width=0.99\columnwidth]{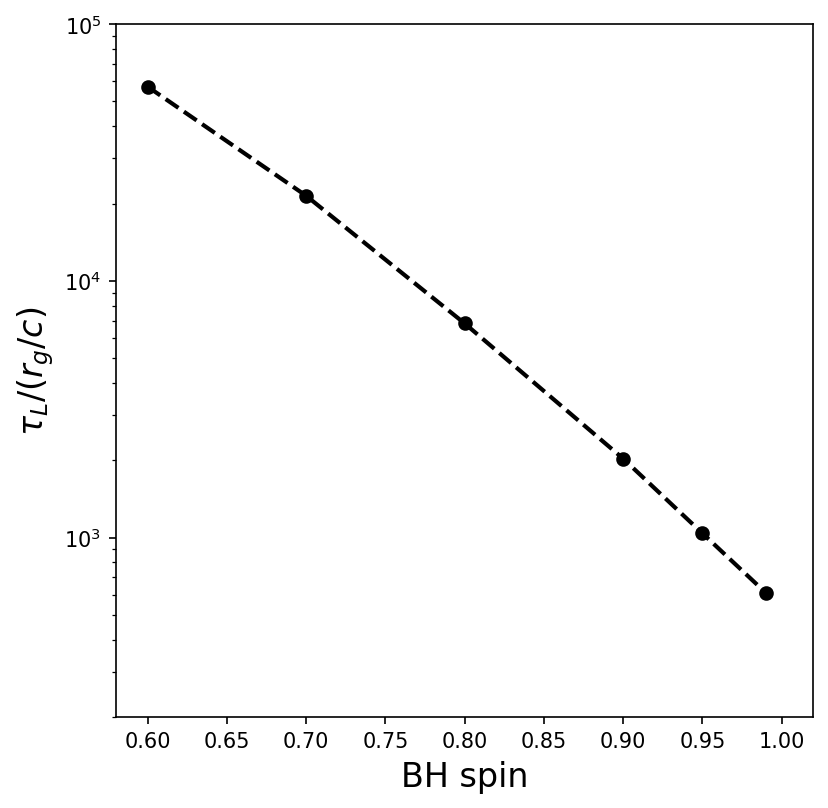}
\caption{Characteristic time scale $\tau_L$ of angular momentum deposit onto the coupled part of the disk as a function of the \bh spin.}
\label{fig:tauL}
\end{figure}

We want to evaluate the impact of the deposit of angular momentum from the \bh onto a disk in prograde rotation. We do so by extracting a characteristic time scale $\tau_L$ over which the angular momentum of the coupled section of the disk, between the \isco and the footpoint of the separatrix, changes significantly due to the angular momentum supplied by the \bh via the coupling magnetic field lines. An estimate for the angular momentum $L_{disk}$ of the coupled disk section is given by:
\begin{equation}
    \label{eq:Ldisk}
    L_{disk}=\int_{r_{\iscoo}}^{R_s}r^2\Omega_K \diff m,
\end{equation}
with $\Omega_K$ the angular velocity given by Equation\,\eqref{eq:omega_GR} and $\diff m$ the amount of mass in an infinitely thin annulus in the disk. Using mass conservation, we can express $\diff m$ in function of the mass accretion rate $\dot{M}$ and the accretion speed $v_r$ and we get:
\begin{equation}
    \label{eq:Ldisk}
    L_{disk}=a\frac{\dot{M}cr_g^2}{v_r}\int_{x_{\iscoo}}^{x_s}\frac{x^2\diff x}{1+x^{3/2}} \quad \text{with} \quad x=\left(r/r_g\right)^{3/2}/a.
\end{equation}
For convenience, we write this integral $I$. On the other hand, we measured $\dot{L}_{BH-disk}$ that we write $\lambda B_0^2 r_g^3$, with $\lambda$ a number dependent on the spin and taken from the bottom plot in Figure\,\ref{fig:rdcd}. The characteristic spin-up time scale $\tau_L$ of the disk is thus:
\begin{equation}
    \label{eq:tauL1}
    \tau_L=\frac{L_{disk}}{\dot{L}_{BH-disk}}=\frac{ac}{\lambda v_r}\frac{r_g}{c}\frac{\dot{M}c^2}{B_0^2r_g^2c}I.
\end{equation}
Assuming the \bh accretes in the MAD regime, in the innermost regions, the magnetic field is maximal and given by Equation\eqref{eq:B_MAD}. \benoit{Then, the characteristic time scale $\tau_L$ is given by:}
\begin{equation}
    \label{eq:tauL2}
    \tau_L=\left(\frac{\pi}{25}\right)^2\frac{aI}{\lambda}\frac{c}{v_r}\frac{r_g}{c}.
\end{equation}
We plotted this time scale as a function of the \bh spin from 0.6 to 0.99 in Figure\,\ref{fig:tauL} for a representative accretion speed $v_r=c/200$ \citep{Jacquemin-Ide2021}. They are always longer than the duration of our simulations but it turns out that $\tau_L$ can be as short as a few 100$r_g/c$ for $a=0.99$. \benoit{In comparison,} \cite{Chashkina2021} suggest that to activate the corona and trigger major reconnection events, the accretion of successive loops of opposite polarity and radial extent larger than 10$r_g$ is required. Such loops would need a couple of 1,000$r_g/c$ to be fully accreted once the inner parts of the loop cross the horizon. During this time lapse, feedback through angular momentum is significant if the system is in the MAD regime and if the \bh has a spin $a\gtrsim 0.9$. In this case, tremendous consequences on the structure of the innermost regions of the disk are to be expected. \benoit{This mechanism would manifest on time scales of a few 100$r_g/c$ which would be compatible with hour-long flares in \sgr. Whether it could play a role in the flare dynamics deserves further investigation.} Otherwise, if the system is not in the MAD regime or if the \bh spin is lower than 0.9, the magnetic coupling between the disk and the \bh is not sustained for an amount of time long enough to significantly modify the structure of the innermost regions of the disk via angular momentum deposit.

\section{Summary and perspectives}
\label{sec:Summary}


In this paper, we performed global PIC simulations of a collisionless and strongly magnetized electron-positron pair plasma embedded in a magnetic field sustained by a steady, aligned and perfectly conducting disk in Keplerian rotation around a Kerr \bh. We showed that in this environment, often called the corona or the \bh magnetosphere, magnetic field loops coupling the disk to the \bh maintain up to a maximal distance set by the \bh spin, provided the disk rotation is prograde. Beyond the separatrix (\ie the outermost closed magnetic field line), magnetic field lines open due to the strong shearing between the rotation speed at their footpoint on the disk and the one induced in the ergosphere by the Lense-Thirring dragging. \benoit{For higher spins, the coupling region is smaller.} The open magnetic field lines are twisted and are either anchored in the disk or thread the event horizon. The former ones are very inclined, which could favor the launching of a magneto-centrifugal disk outflow. The latter ones funnel an electromagnetic jet whose power dependence on the \bh spin and magnetic flux are in agreement with the force-free estimates of Blandford-Znajek jets by \cite{Tchekhovskoy2011}, for spins ranging from 0.6 up to 0.99. Provided we rely on the magnetic field of a MAD disk rather than the one derived from one-zone models, we reproduce the observed jet power in \m{87}. The two regions of open field lines are separated by a conic-shaped current sheet where the magnetic field reconnects and where plasmoids form and flow away from the Y-point, the outermost point on the separatrix (which would be a Y-ring in 3D). We find a quasi-periodic motion of the separatrix which progressively stretches until a plasmoid detaches, with typical characteristic time scales of the order of 20 to 5$r_g/c$ for \bh spins from 0.6 to 0.99. For the accreting stellar-mass \bh in \cyg, this quasi-periodicity corresponds to a few milliseconds. These episodes of sudden magnetic dissipation could thus be responsible for the millisecond flares observed by \cite{Gierlinski2003c} during the high/soft accretion states. 

We identified a highly relativistic component in the energy distribution of particles from the current sheet. Its slope close from $-1$ indicates that these particles have been accelerated by magnetic reconnection. We retrieve the corresponding flat synchrotron component in the spectral energy distribution. This mechanism can qualitatively reproduce part of the non-thermal emission from accreting \bhs where the cooling time scale is long compared to the light crossing time (e.g. \sgr). It could help to interpret their spectral properties, although the estimated dissipated powers and maximum Lorentz factor of the particles we derive might be underestimated by one to two orders of magnitude. This work \benoit{may also provide} an answer to the question of the origin of the coronal heating: in the current sheet, electromagnetic energy is tapped to continuously feed a relativistic and non-thermal population of electrons and positrons responsible for upscattering of X-ray photons from the disk up to $\gamma$-ray energies. Thanks to the \ileyk{plausible} magnetization we manage to reach in the corona, we are confident that these results provides reliable location of the emission sites and photometric and spectral properties as a function of the \bh spin: the higher the spin, the closer from the \bh the Y-ring is and the higher the dissipated electromagnetic power to fuel non-thermal emission. This physically-motivated coronal emission could be used in X-ray reflection spectroscopy models instead of the semi-empirical lamppost geometry where the height of the point source above the \bh is a degree-of-freedom. In our simulations, the current sheet co-exists with the jet launching region that it engulfs, which leaves the door open for a contribution of a compact radio jet to the NIR excess observed in \sgr for instance \citep{Moscibrodzka2013}. 

Finally, we computed the synthetic images of synchrotron emission within $\sim$10$r_g$ around the \bh and characterized the features due to the current sheet and the plasmoids it contains, for different viewing inclinations. If the mechanism we studied in this paper is at play around \m{87}, plasmoids could manifest in the images as transient and fainter sources of emission moving at mildly relativistic bulk speeds. These results also suggest that the hot spot reported by \cite{Abuter2018b} around \sgr could be due to synchrotron emission from a large plasmoid propagating outward in the conic-shaped current sheet along a helicoidal trajectory. 

Coupling magnetic field lines loops are anchored in the disk and close within the event horizon. We find that for the spins we explored ($a>0.6$), they deposit electromagnetic energy onto the disk surface, between the \isco and the footpoint of the separatrix, at a rate such that the associated power is comparable to the jet power. They also enable the extraction of angular momentum from the \bh. For $a>0.9$, this deposit of angular momentum in the innermost regions of the disk could dramatically alter its structure over a few 100$r_g/c$. Relativistic particles accelerated at the Y-ring are also a source of energy for the disk as they cascade onto the footpoint of the separatrix whose location depends on the \bh spin.


Our results remains qualitatively similar for thin ($\epsilon=5\%$) and slim ($\epsilon=30\%$) disks in prograde rotation, although magnetic reconnection is less vivid for thicker disks. In contrast, we find that for retrograde rotation (\ie negative effective \bh spin), the corona is not magnetically active. Reconnection is inhibited and no coupling magnetic field line can maintain. However, this conclusion strongly depends on the distribution of the magnetic flux provided by the disk and on the speed at which magnetic field lines are advected. It highlights the main limitation of our model where the disk is a fixed background whose dynamics we do not solve. Instead, we assume that over the few 100$r_g/c$ of our simulations, the disk and the magnetic field it brings are frozen. In order to capture the longer time scales of accreting \bhs, the computationally expensive kinetic approach we followed will have to be coupled to GR-MHD simulations which are much more suitable to describe the dynamics of the disk \citep{Ripperda2020}.

Another aspect of our model to be questioned is the physical motivation for the magnetic topology we considered. Small scale magnetic loops have been suggested to be a frequent outcome of the magneto-rotational instability \citep[MRI][]{Balbus1991}. Indeed, the MRI fuels turbulence and leads to buoyantly emerging magnetic field loops which rise above the disk surface, as shown in shearing box and global MHD simulations \citep{Davis2010,Zhu2018}. It confirms the semi-analytic stochastic picture envisioned by \citep{Galeev1979} and \cite{Uzdensky2008}. Depending on the extent of the loop and on its propensity to reconnect with neighboring loops, accretion could bring the innermost part of the loop into the event horizon while the outermost footpoint remains anchored in the disk. In our model and for the sake of simplicity, we only considered a single loop by starting with an initially dipolar distribution of the magnetic field on the disk. The main motivation was to explore the dynamics of this configuration on time scales short enough to neglect the accretion of such a large loop which would typically take a few 1,000$r_g/c$ while we run our simulations over a few 100$r_g/c$. In future PIC simulations, we intend to consider the effect of successive smaller scale loops on the corona but also on the formation of current sheets in a striped jet, analog to the GR force-free simulations of \cite{Parfrey2015} and \cite{Mahlmann2020}.

We could not draw conclusions from synthetic light curves due to the axisymmetric assumption our 2D model relies on. The latter artificially enhances the peak-to-peak variability while lengthening peak duration due to the gravitational delay induced by the curved spacetime. Timing information is also flawed by the hypothesis we make about the symmetry of the corona with respect to the equatorial plane. For these reasons, we postpone an analysis of the synthetic light curves to future work where the corona will be studied with fully 3D simulations. It will grant the possibility to study the fragmentation of the current sheet into plasmoids of finite azimuthal extent and to investigate whether magnetic reconnection can propagate along the Y-ring and mimic an orbiting hot spot.

\begin{acknowledgements}
This project has received funding from the European Research Council (ERC) under the European Union’s Horizon 2020 research and innovation programme (grant agreement No 863412). Computing resources were provided by TGCC and CINES under the allocation A0090407669 made by GENCI. The authors thank Ma\"{i}ca Clavel, Antoine Strugarek, Guillaume Dubus and Geoffroy Lesur for fruitful discussions and constructive feedback. 
\end{acknowledgements}


\bibliographystyle{aa} 
\begin{tiny}
\bibliography{all_my_library}
\end{tiny}

\begin{appendix}

\section{Numerical convergence}
\label{sec:num_conv}

\subsection{Numerical relaxation}
\label{sec:Numerical_relaxation}

\begin{figure}
\centering
\includegraphics[width=0.99\columnwidth]{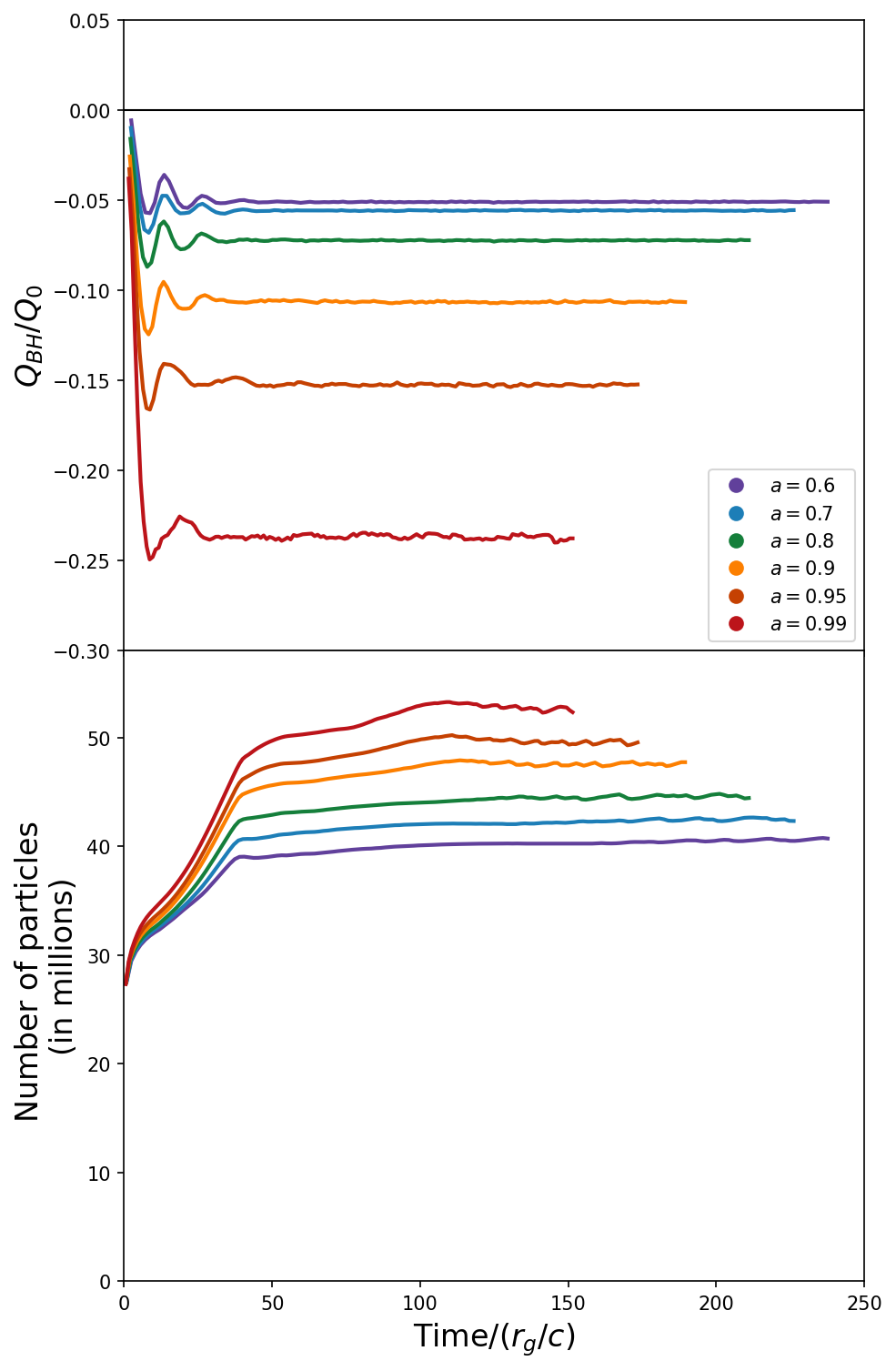}
\caption{(top panel) Net amount of electric charge $Q_{BH}$ within the event horizon as a function of time for different spin values and for a thin disk ($\epsilon=5\%$) in prograde rotation. (bottom panel) Total number of particles within the simulation space as a function of time.}
\label{fig:q_relaxing}
\end{figure}

We monitor the relaxation of the initial state by following reduced quantities representative of the dynamics of the fields and of the particles in the corona. For the former, we measure the \bh net electric charge $Q_{BH}$ defined based on the charges crossing the event horizon since the beginning of the simulation. Since the innermost regions within the event horizon are cropped ($r_{min}>0$), charges entering the horizon can exit the simulation space. Instead of counting the charges, we thus apply Maxwell-Gauss equation integrated over the event horizon:
\begin{equation}
    Q_{BH}=\frac{1}{4\pi}\oiint_{EH} \boldsymbol{D}\cdot\boldsymbol{\diff \Sigma}. 
\end{equation}
The electric field $\boldsymbol{D}$ is symmetric with respect to the equatorial plane such as we can integrate over the upper hemisphere and multiply by 2. We then compare $Q_{BH}$ to the electric charge $Q_0$ for which the time-component of the 4-potential cancels out \citep{Wald1974b}:
\begin{equation}
    Q_0=2aB_0.
\end{equation}
In the top panel in Figure\,\ref{fig:q_relaxing}, we see that the net charge eventually plateaus, which indicates that numerical relaxation is achieved in the innermost regions of the simulation space after approximately 50$r_g/c$. The net charge of the \bh is negative and for $a=0.99$, its magnitude reaches almost a fourth of $Q_0$. It was recently alleged by \cite{King2021} that if the \bh was to acquire an electric charge $Q_0$, the Blandford-Znajek mechanism would not be able to operate because the 4-potential no longer contains a term function of the \bh spin. However, the correspondence of the 4-potential with the Wald solution for a non-spinning \bh does not mean that the structure of the electromagnetic fields is the same due to the different underlying metrics. \cite{Komissarov2021} showed that in contrast with a Schwarzschild \bh, a maximally spinning \bh with an electric charge $Q_0$ has a magnetosphere where the component of the electric field parallel to the magnetic field does not cancel: particles are susceptible to be accelerated along the magnetic field lines.

Due to the injection procedure, the relaxation of the particle distribution takes more time. After a steep increase of the number of particles visible in the bottom panel in Figure\,\ref{fig:q_relaxing}, the number of particles keeps increasing slower, especially for simulations with higher \bh spin values. A plateau is eventually achieved after $\sim 100r_g/c$, once the rate at which particles exit the simulation space equals the injection rate. The relaxation of the outermost regions of the simulation space takes more time than the relaxation of the innermost parts, so we typically work between $t=100r_g/c$ and $t=150r_g/c$ to guarantee full time convergence. The total number of particles correspond to approximately 20 particles per grid cell, with more particles in the current sheet. It ensures a good signal-to-noise ratio by lowering the particle shot noise.

\subsection{Resolution and scale separation}
\label{sec:Resolution}

\begin{figure}
\centering
\includegraphics[width=0.99\columnwidth]{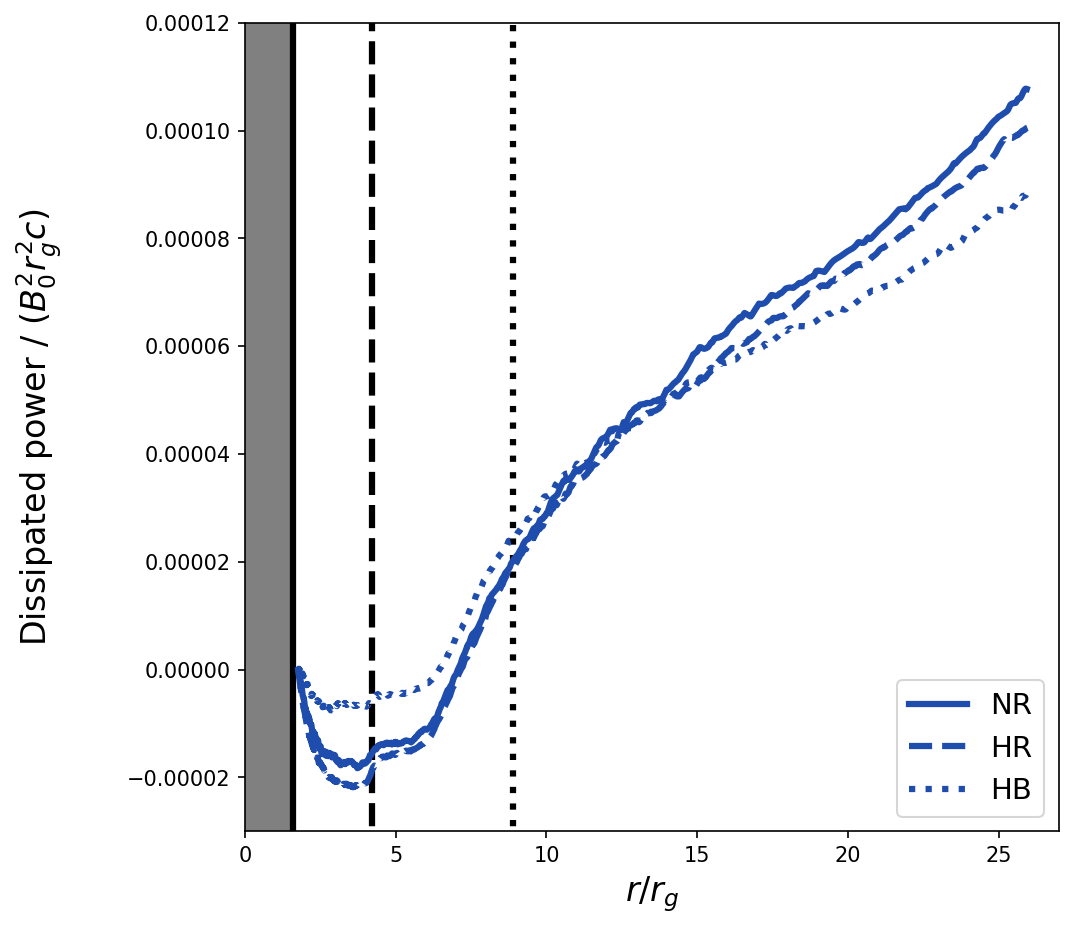}
\caption{Time-averaged dissipated electromagnetic energy between the event horizon and a given radius for simulations with normal resolution (NR), high resolution (HR) and high magnetic field (HB). The dashed and dotted black lines show the distance to the separatrix footpoint and to the Y-point respectively.}
\label{fig:ene_CV}
\end{figure}

We evaluated the impact of the resolution and of the scale separation on the dissipation by computing the radial profiles for simulations with $a=0.8$, $\epsilon=5$\% and a disk in prograde rotation. Figure\,\ref{fig:ene_CV} shows the integral of $\boldsymbol{J}\cdot\boldsymbol{E}$ over a volume bounded by the pole and the disk in the longitudinal direction, and between the event horizon (vertical black solid line) and a given distance $r$ to the \bh in the radial direction. Apart from the innermost regions, energy is progressively transferred from the electromagnetic field to the particles until reaching a value of approximately 0.001$B_0^2r_g^2c$ at 26$r_g$, consistent with Figure\,\ref{fig:dissipation}. We compare a simulation with the standard resolution we used in this paper (NR in Figure\,\ref{fig:ene_CV}, with 2,128 and 1,120 cells in the radial and longitudinal directions respectively) to a simulation with a resolution twice higher (HR, 4,256$\times$2,240 cells) and a high-resolution simulation like HR but with a dimensionless magnetic field scale $\tilde{B}_0$ higher by a factor 4 (HB). As emphasized by Equation\,\eqref{eq:parameters}, the Larmor radius is resolved by the same number of cells in all 3 simulations. The profiles are averaged over 40$r_g/c$ after the simulations have relaxed.

The normal and high resolution cases NR and HR yield very similar profiles and prove that the number of grid cells we use is sufficient to resolve the magnetic reconnection process and the subsequent particle acceleration. The simulation HB at higher magnetic field gives a profile within 15\% of the NR and HR ones. Due to our injection method (see Section\,\ref{sec:Particle_injection}), the higher magnetic field produces a denser corona. The magnetization is 4 times higher than in cases NR and HR but in the relativistic reconnection regime we explore, this does not significantly modify the reconnection rate. Instead, the slightly lower dissipation might be due to the higher plasma inertia.

\section{Y-point location}
\label{app:best_fit}

We provide fitting formulas for the coordinates $(x_Y,y_Y)$ and the distance $r_Y$ to the \bh of the Y-point as a function of the spin value. We fitted the measured median positions with the following power law:
\begin{equation}
    r=c_1a^{c_2}
    \label{eq:fit}
\end{equation}
where $c_1$ and $c_2$ are the fitting parameters. The best fit parameters for the range of spins we explore are given in Table\,\ref{tab:best_fit}. Although reasonable in the range of \bh spin values we sample, these values might lead to significant departure from the physical ones when the power law\,\eqref{eq:fit} is extrapolated to spins lower than 0.6.

\begin{table}[h!]
\caption{Parameters of the best fit power laws to locate the Y-point as a function of the \bh spin based on Equation\,\eqref{eq:fit}.}
\label{tab:best_fit}
\begin{tabular}{l|ccc}
 & $r_Y$ & $x_Y$  & $y_Y$ \\ 
\hline 
$c_1$ & 5.21 & 4.88 & 1.81 \\
$c_2$ & -2.39 & -2.34 & -2.70
\end{tabular}
\end{table}

\section{Synthetic images}
\label{app:synthetic_images}

\begin{figure*}
\centering
\includegraphics[width=1.9\columnwidth]{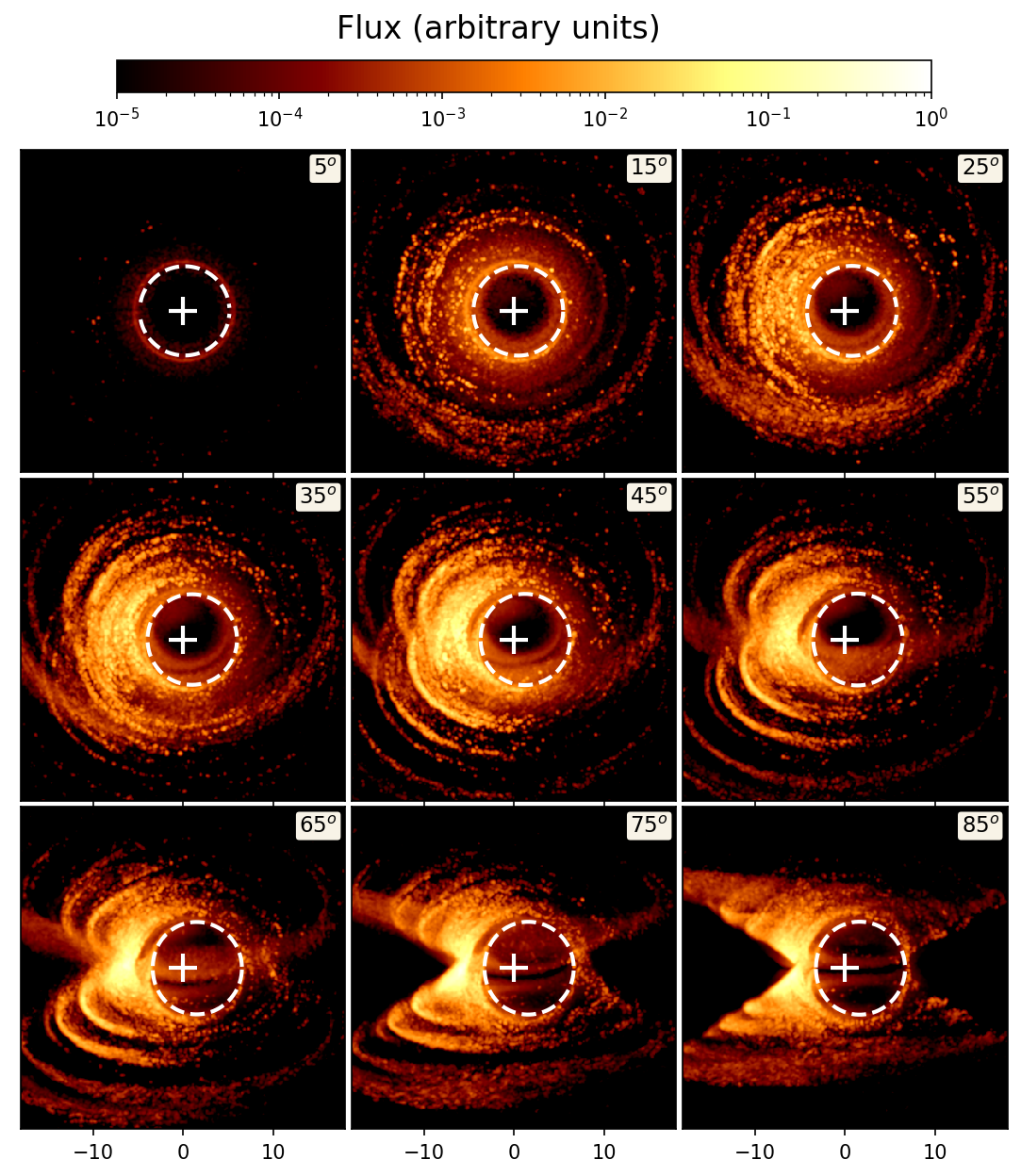}
\caption{Intensity maps of synchrotron emission from the current sheet for different viewing angles.}
\label{fig:BH_intensity}
\end{figure*}

In Figure\,\ref{fig:BH_intensity}, we show the synthetic images of synchrotron emission from the current sheet for different inclinations of the line-of-sight with respect to the \bh spin axis. The \bh spin is $a=0.8$ and the disk is thin and in prograde rotation around the \bh. The inclination $i=55^{\circ}$ corresponds to Figure\,\ref{fig:BH_intensity_unique}.

\end{appendix}

\end{document}